 \documentclass[pre,amsmath,amssymb,twocolumn, showpacs, superscriptaddress,10pt]{revtex4-1}

\usepackage{amsmath}
\usepackage{hyperref}
\usepackage{graphicx}
\usepackage{amsfonts}
\usepackage{amsthm}
\usepackage{cases}
\usepackage{bm}

\usepackage[normalem]{ulem}

\usepackage{color}
\definecolor{Blue}{rgb}{0.00, 0.00, 1.00}
\definecolor{Red}{rgb}{1.00, 0.00, 0.00}
\definecolor{Green}{rgb}{0.00, 0.70, 0.00}

\hypersetup{
    colorlinks=true,       
    linkcolor=red,          
    citecolor=blue,        
    filecolor=magenta,      
    urlcolor=cyan           
}

\newcommand{\nn}{\nonumber}
\newcommand{\be}{\begin{equation}}
\newcommand{\ee}{\end{equation}}
\newcommand{\bea}{\begin{eqnarray}}
\newcommand{\eea}{\end{eqnarray}}

\newcommand{\eqrefMT}{\eqref}
\newcommand{\generalab}{generalab}
\newcommand{\thetaxZERO}{thetax0}
\newcommand{\varianceNddimHO}{variance_N_ddimHO}
\newcommand{\Ad}{Ad}
\newcommand{\defNab}{defNab}
\newcommand{\eqrelationKH}{eq:relation_K_H}
\newcommand{\resWKBTWO}{resWKB2}
\newcommand{\aTWO}{a2}
\newcommand{\mrONE}{mr1}
\newcommand{\FTWO}{F2}
\newcommand{\DMGeneral}{DMGeneral}
\newcommand{\Haa}{Haa}
\newcommand{\varBessel}{var_Bessel }
\newcommand{\eqsumofcumulantsangularsectors}{eq:sum_of_cumulants_angular_sectors}
\newcommand{\integrate}{integrate}
\newcommand{\final}{final}
\newcommand{\cumulevenGUEgen}{cumuleven_GUE_gen}
\newcommand{\cumulfree}{cumul_free}
\newcommand{\entropy}{entropy}

\newcommand{\x}{{\bf x}}
\newcommand{\y}{{\bf y}}
\newcommand{\p}{{\bf p}}
\newcommand{\z}{{\bf z}}

\newcommand{\beq}{\begin{equation}}
\newcommand{\eeq}{\end{equation}}
\newcommand{\beqn}{\begin{eqnarray}}
\newcommand{\eeqn}{\end{eqnarray}}

\DeclareMathOperator{\Tr}{Tr}

\begin{document}

\title{Counting statistics for noninteracting fermions in a $d$-dimensional potential}

\author{Naftali R. Smith}
\affiliation{LPTMS, CNRS, Univ. Paris-Sud, Universit\'e Paris-Saclay, 91405 Orsay, France}
\author{Pierre Le Doussal}
\affiliation{CNRS-Laboratoire de Physique Th\'eorique de l'Ecole Normale Sup\'erieure, 24 rue Lhomond, 75231 Paris Cedex, France}
\author{Satya N. \surname{Majumdar}}
\affiliation{LPTMS, CNRS, Univ. Paris-Sud, Universit\'e Paris-Saclay, 91405 Orsay, France}
\author{Gr\'egory \surname{Schehr}}
\affiliation{LPTMS, CNRS, Univ. Paris-Sud, Universit\'e Paris-Saclay, 91405 Orsay, France}
\date{\today}

\begin{abstract}
We develop a first-principle approach to compute the counting statistics in the ground-state of $N$ noninteracting spinless fermions in a general potential in arbitrary dimensions $d$ (central for $d>1$).
In a confining potential, the Fermi gas is supported over a {bounded} domain. In $d=1$, for specific potentials, this system is related to standard random matrix ensembles. We study the quantum fluctuations of the number of fermions ${\cal N}_{\cal D}$ in a domain $\cal{D}$ of macroscopic size in the bulk of the support.
We show that the variance of ${\cal N}_{\cal D}$ grows as $N^{(d-1)/d} (A_d \log N + B_d)$ for large $N$, and obtain the explicit dependence of $A_d, B_d$ on the potential {and on the size of ${\cal D}$} (for a spherical domain in $d>1$). This generalizes the free-fermion results for microscopic domains, given in $d=1$ by the Dyson-Mehta asymptotics from random matrix theory. This leads us to conjecture similar asymptotics for the entanglement entropy of the subsystem $\cal{D}$, in any dimension, supported by exact results for $d=1$.

\end{abstract}



\maketitle


An important concept to study quantum noise and correlations in many body fermionic systems is the counting statistics (CS), which characterises the fluctuations of the number of particles ${\cal N}_{\cal D}$ inside a domain ${\cal D}$. Applications include 
shot noise \cite{Levitov}, quantum transport \cite{Lev96,PGM},
quantum dots \cite{Been06,Gus06}, 
spin 
and fermionic chains \cite{EiserRacz2013,AbanovIvanovQian2011,IvanovAbanov2013,Caux2019},
trapped fermions \cite{Eisler1,DeanPLDReview}. In the {related} context of random matrix theory (RMT), the statistics of the number of eigenvalues in an 
interval also generated a lot of interest \cite{MehtaBook,CL95,FS95,AbanovIvanovQian2011,DIK2009,MMSV14,MMSV16,CharlierSine2019,MNSV09,MNSV11,MV12,MSVV13,FyodorovPLD2020}.
The CS is particularly important for 
noninteracting fermions 
because of its connection 
\cite{Kli06,KL09,CalabreseMinchev2,Hur11} to the
bipartite entanglement entropy (EE) of the subsystem ${\cal D}$ 
with its complement $\overline{\cal D}$.  The EE is a highly non local quantity, much studied in the context of quantum information
\cite{Nahum2017, Cornfeld2019}, conformal field theory
\cite{CalabreseCardyDoyon,CC04,DubailStephanVitiCalabrese2017}, topological phases~\cite{SRev}, 
quantum phase transitions \cite{RM04,MFS09}, or quantum spin chains 
\cite{JinKorepin2004,KeatMezzadri}. Both the CS and the EE 
are difficult to compute analytically, {in particular in the presence of an external potential}. There exist
however {standard} results for free fermions, in the absence of external potential.
In this case, at zero temperature, both the variance of ${\cal N}_{{\cal D}}$ and the EE grow as
$\sim \! R^{d-1} \log R$ with the typical size $R$ of the domain ${\cal D}$
\cite{Widom1,Widom2,Widom3,Klitch,CalabreseMinchev1,Torquato, Fraenkel2020,TanRyu2020}.

In 
cold Fermi gases \cite{BDZ08}, the quantum microscopes \cite{Fermicro1,Fermicro2,Fermicro3} allow to take 
an instantaneous ``picture'' and measure the counting statistics. In experiments the fermions are in a trapping potential, of tunable shape and interaction
\cite{BDZ08,flattrap}. It is thus important to calculate both the CS and the EE 
in an inhomogeneous background, for which very few analytical results exist even for noninteracting fermions, apart from
the $d\!=\!1$ harmonic oscillator 
\cite{CalabresePLDEntropy,DubailStephanVitiCalabrese2017,V12},
and the rotating harmonic trap in $d \! = \! 2$ \cite{LMG19}.

There has been recent progress to describe noninteracting 
spinless fermions in traps in $d$ dimensions \cite{DeanPLDReview}.
In $d=1$, for a single particle Hamiltonian $\hat H=\frac{p^2}{2} + V(x)$
(in units $\hbar=m=1$),
there 
is a useful connection with random matrices for 
a few specific potentials $V(x)$. 
The many body ground state wavefunction $\Psi_0$ of $N$ fermions is 
a Slater determinant with all energy levels of $\hat H$ 
occupied up to the Fermi energy $\mu$, 
a function of $N$. The quantum joint probability $|\Psi_0|^2$ of the positions $\{ x_j \}$ of the $N$ fermions,
maps onto the joint 
probability for the eigenvalues $\{\lambda_j \}$ of random matrices of size $N \times N$. For the harmonic oscillator (HO), $V(x) \! = \! \frac{x^2}{2}$, the random matrix is Hermitian from the Gaussian unitary ensemble (GUE).
At large $N$, the mean fermion density, i.e., the quantum average $\rho(x) \! = \! 
 \langle \sum_i \delta(x-x_i) \rangle$, has support $[x^-,x^+]$, with $x^\pm \! \simeq \! \pm \sqrt{2N}$. In the bulk, i.e., away from the edges $x^\pm$, it takes the semi-circle form $\rho(x) \! \simeq \!  \rho^{\rm bulk}(x) \! = \! k_F(x)/\pi$, where $k_F(x) \! = \! \sqrt{2 \mu- x^2}$ is the local Fermi momentum, and in this case $\mu \! \simeq \! N$. There are two natural length scales, the microscopic one of order the inter-particle distance
$\sim \! 1/k_F(x)$, and the macroscopic one of order $x^+ \! -x^-$. 
For an interval ${\cal D} \! = \! [a,b]$ of microscopic size, it is well known from {standard} results of RMT \cite{Dyson,DysonMehta} that for $\sqrt{N}|b-a| \! = \! O(1) \! \gg \! 1$ the variance behaves as
\cite{MehtaBook,CL95,FS95,AbanovIvanovQian2011,DIK2009,MMSV14,MMSV16,CharlierSine2019}
\be
\label{DM}
{\rm Var}\,{\cal N}_{\left[a,b\right]}\simeq\frac{1}{\pi^{2}}\left[\log\left(\sqrt{2N  -a^{2}}\,|b-a|\right)+c_{2}\right]
\ee
with $c_2  =  \gamma_E +1+\log 2$, where $\gamma_E$ is Euler's constant. The fermion/eigenvalue correlations can be expressed as determinants of a central object called the kernel,
which depends on $V(x)$, see below. At microscopic scales, the kernel takes a universal scaling form, called
the sine-kernel, independent of the (smooth) potential, which leads to \eqref{DM}. 
However, except for free fermions on the infinite line, it does not apply when both $a,b$ are well separated in the bulk. For the HO, some results in that regime were obtained 
in \cite{MMSV14,MMSV16} using a Coulomb gas method, and for the GUE in the math literature \cite{Borodin1,BaiWangZhou,Charlier_hankel,JohanssonLS}.

Despite recent advances a general framework is
still lacking for computing the counting statistics and entanglement
entropy for noninteracting fermions in general potential and arbitrary dimension.
In this Letter we 
provide a first principle approach to compute these 
quantities
in $d=1$ for a general potential $V(x)$, and in $d>1$ 
for a general central potential. Our method recovers the existing results in various special cases, see below.

Let us summarize our main results.
For a confining potential in $d \! = \! 1$, such that the bulk density $k_F(x)/\pi$, $k_F(x) \! = \! \sqrt{2 (\mu-V(x))}$, 
has a single support $[x^-,x^+]$, we obtain an explicit
formula for ${\rm Var} {\cal N}_{[a,b]}$, with $a,b$ well separated in the bulk, $|a-b| \! \gg \! 1/k_F(a)$. In the limit $N \! \gg \! 1$ (i.e., $\mu \! \gg \! 1$)
where $N \! \simeq \! \int_{x^-}^{x^+} \frac{dx}{\pi} k_F(x)$
\bea \label{generalab} 
&& (2 \pi^2) {\rm Var} {\cal N}_{[a,b]}  = 2 \log \left( 2 k_F(a) k_F(b) 
\int_{x^-}^{x^+} \frac{dz}{\pi k_F(z)} \right) \nonumber  \\
&& \qquad +  \log \left( \frac{\sin^2\frac{\theta_a-\theta_b}{2}}{\sin^2\frac{\theta_a+\theta_b}{2}}|\sin \theta_a \sin \theta_b| \right) + 2 c_2  + o(1)  \\
&& \text{where} ~~~~~\label{thetax0} 
\theta_x =
\pi  \frac{\int_{x^-}^{x} dz/k_F(z) }{\int_{x^-}^{x^+} dz/k_F(z) }
\quad , \quad \begin{cases} \theta_{x^-}=0 \\ \theta_{x^+}=\pi \end{cases}  .
\eea 

\begin{figure}[ht]
\centering
\includegraphics[angle=0,width=1.0\linewidth]{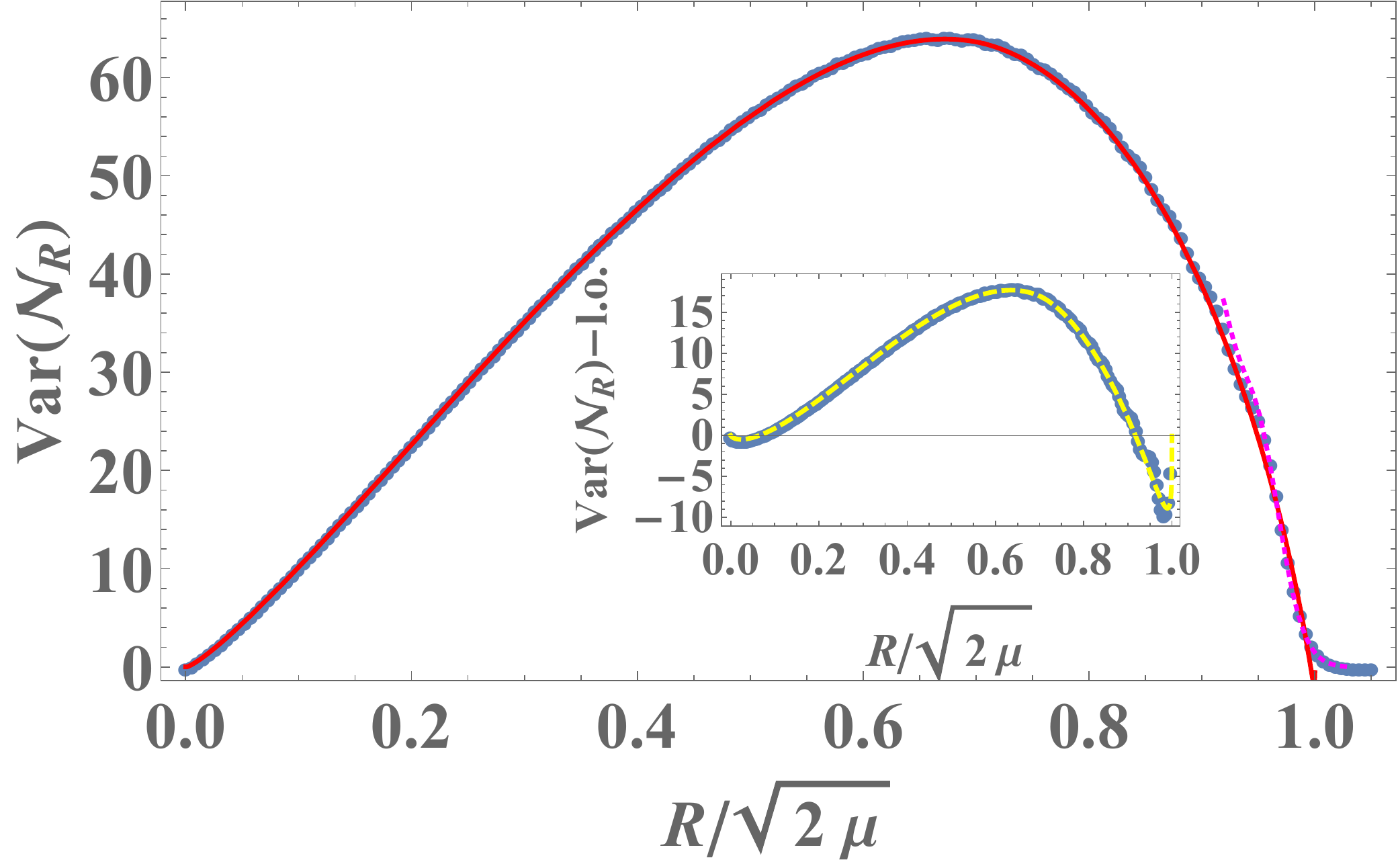}
\caption{Variance of ${\cal N}_{\cal D} = {\cal N}_R$ for a disk of radius $R$ in $d  =  2$, plotted vs $\tilde R  =  R/  \sqrt{2 \mu}$ for $\mu  =  100$ {corresponding to $N=\mu\left(\mu+1\right)/2=5050$.}
The simulations (symbols) \cite{SM} show excellent agreement with our predictions: In the bulk, with 
(\ref{variance_N_ddimHO}) (solid line), where $A_2(\tilde R)$ is given in (\ref{Ad}) and 
$B_2(\tilde R)$ in 
{\eqref{B2}}, and near the edge $\tilde R =  1$, with the scaling form (\ref{final})  (dotted line).
{\bf Inset:} the sub-leading term $B_2(\tilde R)$ 
plotted vs $\tilde R$ (dashed line), 
compared to the simulations (symbols), the leading term $A_2(\tilde R) \mu \log \mu$ being subtracted from the variance.}
\label{Fig_density}
\end{figure}

We then consider noninteracting fermions in a general central potential in $d$ dimension, with
single particle Hamiltonian $\hat H \! = \! \frac{{\bf p}^2}{2} + V(r)$,
where $r \! = \! |{\bf x}|$. We obtain the variance ${\rm Var} {\cal N}_{{\cal D}}$ for 
any rotationally invariant domain ${\cal D}$. For instance, for the HO, $V(r) \! = \! \frac{1}{2} r^2$, the support of the density is the ball of radius $\sqrt{2 \mu}$, and for a sphere of macroscopic radius $R \! = \! \tilde R \sqrt{2 \mu}$
we obtain for large $\mu$, with fixed $\tilde R \in [0,1[$
\bea \label{variance_N_ddimHO}
 {\rm Var} {\cal N}_{{\cal D}} &=& \mu^{d-1}\left[A_{d}\left(\tilde{R}\right)\log\mu+B_{d}\left(\tilde{R}\right)+o(1)\right]\\
\label{Ad} 
 A_d(\tilde R) &=& \frac{1}{ \pi^2 \Gamma(d)} \left(2 \tilde R \sqrt{1- \tilde R^2}\right)^{d-1} 
\eea 
and $B_d(\tilde R)$ is given below for $d=2$ in~(\ref{B2}) and for $d = 3$ in~(\ref{B3}). As 
seen from the comparison to simulations 
in Fig.~\ref{Fig_density} (see \cite{SM} for details on the simulations), the prediction in (\ref{variance_N_ddimHO})
for a disk in $d \! = \! 2$ is already excellent for $\mu \! = \! 100$ {(it is crucial to include the sub-leading term $B_d(\tilde R)$)}.
In the microscopic limit $\tilde R \to 0$ we obtain 
\be \label{var_ff}
{\rm Var}{\cal N}_{{\cal D}}\!\simeq\!\frac{1}{\pi^{2}\Gamma(d)}\left(k_{F}R\right)^{d-1}\left[\log\left(k_{F}R\right)+b_{d}\right]\;,
\ee
where $k_F \! = \! \sqrt{2\mu}$. The leading term reproduces the free fermion result
\cite{Klitch,CalabreseMinchev1,Torquato,Widom1,Widom2,Widom3} for a sphere 
and 
we further obtain 
the subleading term 
\be
b_{d}=2\log2-\frac{\gamma_{E}}{2}+1-\frac{3}{2}\psi^{(0)}\left(\frac{d+1}{2}\right),
\ee
$\psi^{(0)}(x)$ being the di-gamma function.
These results lead us to the conjecture \eqref{entropy} for the entanglement
entropy of the subsystem ${\cal D}$ in any dimension
for arbitrary smooth central potential, corroborated by 
exact results in $d \! = \! 1$. 

Let us start with fermions on the infinite line in $d=1$.
It is useful to introduce the height field $h(x)$ \cite{Haldane}, also called the ``index'' in 
RMT~\cite{MNSV09,MNSV11,MV12,MSVV13}, and its two-point covariance function $H(x,y)$, from which the variance of ${\cal N}_{\cal D}$ for any interval ${\cal D} \! = \! [a,b]$ is obtained as
\bea \label{defH}
&& h(x) = {\cal N}_{]-\infty,x]}  \quad , \quad H(x,y) = {\rm Cov}[ h(x), h(y)], \\
&& {\rm Var}{\cal N}_{\left[a,b\right]}=H\left(a,a\right)+H\left(b,b\right)-2H\left(a,b\right)
\label{defNab} 
\eea 
with ${\rm Var}{\cal N}_{\left]-\infty,a\right]} \! = \! {\rm Var}{\cal N}_{\left[a,+\infty\right[} \! = \! H(a,a)$,
for a semi infinite interval \cite{footnote1}.

For $N$ noninteracting fermions 
the correlation functions are obtained from 
the kernel
\be \label{kernel}
K_\mu(x,y) = \sum_{k=1}^N \psi_k^*(x) \psi_k(y) 
\ee
where the $\psi_k(x)$ are the eigenstates of $\hat H= \frac{p^2}{2} + V(x)$. 
We will denote $\{ \epsilon_k \}_{k=1,2,\dots}$ the eigenenergies in increasing order.
The mean density is $\rho(x)=K_\mu(x,x)$,
and the $n$-point correlation
is given by $\det_{n \times n} K_\mu(x_i,x_j)$ (see e.g. \cite{DeanPLDReview}).
This leads to the exact 
relation \cite{SM}
\be
\label{eq:relation_K_H}
K_\mu(x,y)^2= - \partial_x \partial_y H(x,y) + \delta(x-y)  \rho(x)
\ee
from which we 
determine 
the height field covariance 
\eqref{defH}.

We now obtain an estimate of $K_\mu(x,y)^2$, and of $H(x,y)$,
valid anywhere in the bulk in the large $N$ limit. In this regime,
the sum over $k$ in \eqref{kernel} is dominated by $k \gg 1$ \cite{footnote:large_k}. 
One can thus use the 
WKB asymptotics \cite{LandauLifshitz,ValleeBook}
\be \label{WKB1} 
\psi_{k}(x)\simeq\frac{C_{k}}{\left[2\left(\epsilon_{k}-V(x)\right)\right]^{1/4}}\sin\left(\phi_{k}(x)+\frac{\pi}{4}\right)
\ee
where $\phi_k(x) \! = \! \int_{x^-}^x dz
\sqrt{2( \epsilon_k - V(z))}$ and $C_k^2 \! = \! \frac{2}{\pi} \frac{d\epsilon_k}{dk}$
is a normalization \cite{Furry,footnote10}. Inserting \eqref{WKB1} in \eqref{kernel},
we relabel $k \! = \! N \! - \! m$ around the Fermi energy
$\mu \! = \! \epsilon_N$. Noting that the phase $\phi_N(x)$ at large $N$ is also very large,
we can expand 
$\phi_{N-m}(x) \! = \! \phi_N(x)  -  m \frac{d\phi_N(x)}{dN} \! + \! o(1)
\! = \! \phi_N(x)  -  m \theta_x \! + \! o(1)$,
where $\theta_x$ is given in \eqref{thetax0},
using 
$\frac{dN}{d\mu} \simeq \int_{x^-}^{x^+} \frac{dx}{\pi k_F(x)}$.
Performing the geometric sum over $m$ we obtain
\be \label{resWKB2} 
K_{\mu}(x,y)\simeq\frac{d\mu/dN}{2\pi\sqrt{k_{F}(x)k_{F}(y)}}\sum_{\sigma=\pm1}\frac{\sin(\tilde{\phi}_{N}(x)-\sigma\tilde{\phi}_{N}(y))}{\sin\left(\left(\theta_{x}-\sigma\theta_{y}\right)/2\right)}
\ee 
with $\tilde \phi_N(x) \! = \! \phi_N(x) + O(1)$. In Eq.~\eqref{resWKB2} 
the sine terms oscillate on microscopic scales. For 
$\left|x-y\right| \! \sim \! 1/k_{F}\left(x\right)$ the term $\sigma \! = \! 1$
dominates \cite{footnote3}. Using $\tilde \phi_N'(x) \! \simeq \! k_F(x)$ and
$\frac{d \theta_x}{dx} \! = \! \frac{d\mu}{dN} \frac{1}{k_F(x)}$, one
recovers the sine-kernel 
{
\be \label{sinekernel} 
K_{\mu}(x,y)\simeq\frac{\sin\left(k_{F}(x)|x-y|\right)}{\pi|x-y|}
\ee}
valid on microscopic scales. On the other hand, for $x,y$ 
well separated on macroscopic scales in the bulk $]x^-,x^+[$,
taking the square of \eqref{resWKB2}, one can neglect the cross term and replace
the $\sin^2$ by $1/2$, leading to
\be \label{a2} 
K_{\mu}(x,y)^{2}\simeq\frac{\left(d\mu/dN\right)^{2}}{2\pi^{2}k_{F}(x)k_{F}(y)}\frac{1-\cos\left(\theta_{x}\right)\cos\left(\theta_{y}\right)}{\left(\cos\theta_{x}-\cos\theta_{y}\right)^{2}}
\ee
up to fast oscillating terms averaging to zero on scales larger than microscopic.
Note that Eq.~\eqref{a2} is valid for any smooth potential: for the HO we also derived these estimates using the Plancherel-Rotach asymptotics for the Hermite polynomials \cite{SM}. Having obtained $K_\mu(x,y)^2$ in the two regimes, 
we use \eqref{eq:relation_K_H}
to compute the height correlator.

(i) For $x,y$ well separated in the bulk, i.e., $\left|x-y\right|\gg1/k_{F}\left(x\right)$, the 2-point
height covariance is given by
\be \label{mr1} 
\!\!\!\! H \! \left(x,y\right) \! \simeq \! \frac{1}{2\pi^{2}} \! \left( \! \log\left|\sin\frac{\theta_{x}+\theta_{y}}{2}\right|-\log\left|\sin\frac{\theta_{x}-\theta_{y}}{2}\right|\right)
\ee 
up to $o(1)$ terms at large $\mu$. One 
checks that 
\eqref{mr1} is consistent with \eqref{a2} and \eqref{eq:relation_K_H} (in this regime the $\delta$ function does not contribute). Using \eqref{thetax0}, the right hand side (r.h.s.)
in \eqref{mr1} vanishes when $x$ is in the bulk and $y$ reaches an edge $y=x^\pm$,
and for 
$y \notin ]x^-,x^+[$, $H(x,y) \simeq o(1)$ \cite{SM}.
The r.h.s. in \eqref{mr1} coincides with
the correlator 
of the 2D Gaussian free field (GFF)
in the upper-half plane (with Dirichlet boundary conditions) along part of 
a circle $z=e^{i \theta_x}$, thus extending the result of \cite{Borodin1} for
the GUE/HO \cite{footnote4}. Similar connections to the GFF also emerge in recent approaches using inhomogeneous
bosonization \cite{DubailStephanVitiCalabrese2017, BrunDubail2018, Unterberger, RuggieroBrunDubail2019}.

(ii) On microscopic scales, $\left|x-y\right| \! \sim \! 1/k_{F} \left(x\right)$, 
one uses the sine kernel \eqref{sinekernel} 
in the left hand side of \eqref{eq:relation_K_H}.
The integration constants are fixed so that
$H(x,y)$ for $\left|x-y\right| \! \gg \! 1/k_{F}\left(x\right)$ matches
with the limit $y \to x$ in \eqref{mr1} leading to
\be \label{F2} 
\!\!\!\!H(x,y) \! \simeq \! \frac{1}{2\pi^{2}} \! \left[U \! \left(k_{F}(x)|x-y|\right)+\log\frac{2k_{F}(x)\sin\theta_{x}}{d\theta_{x}/dx} \! \right]
 \ee
where 
{
\be
U(z) \! = \! \text{Ci}(2z)+2 z \text{Si}(2 z)-\log z +1 -2 \sin ^2(z) - \pi z,
\ee}
 with $U(z  \! \gg \! 1) \! = \! - \log z \! + \! o(1)$
and
$U(z \! \ll \! 1) \! = \! 1+\gamma_E + \log 2 - \pi z + z^2 + o(z^2)$. 
One checks, using $U''(z)=2 \sin^2 z/z^2$, that \eqref{F2} is consistent with \eqref{eq:relation_K_H}
(including the delta function) and $K_\mu$ given by the sine kernel \eqref{sinekernel}, since 
$k_F(x) \simeq k_F(y)$ on microscopic scales. Using \eqref{defNab}, it leads to the Dyson Mehta behavior 
{
\bea
\label{DMGeneral}
\pi^2 {\rm Var} {\cal N}_{[a,b]} & \simeq & U(0)-U(k_F(a)|a-b|) \nn\\
&\simeq& \log k_F(a)|a-b| + c_2.
\eea}

From \eqref{mr1}, \eqref{F2} and \eqref{defNab}, we obtain our result \eqref{generalab}
as well as, for any $a$ in the bulk
\be \label{Haa}
H(a,a)={\rm Var}{\cal N}_{[a,+\infty[}\simeq\frac{1}{2\pi^{2}}\left(\log\frac{2k_{F}(a)^{2}\sin\theta_{a}}{d\mu/dN}+c_{2}\right).
\ee
Expanding \eqref{Haa} for $a \to x^+$, $a \! < \! x^+$, one 
obtains \cite{SM} $H(a,a) \! \simeq \! \frac{1}{2 \pi^2} ( 
 \frac{3}{2} \log (-\hat a)  + c_2 + 2 \log 2 )$
 for $- \hat a \! \gg \! 1$. Here the edge scaling variable 
 is $\hat a \! = \! (a-x^+)/w_N$, and
$w_N \! = \! (2 V'(x^+))^{-1/3}$, the width of the edge regime
\cite{DeanPLDReview}, appears naturally.
Inside the edge regime, i.e., for $\hat a \! = \! O(1)$, $H(a,a) \! \simeq \! \frac{1}{2} {\cal V}_2(\hat a)$,
where the scaling function ${\cal V}_2$ was defined in \cite{MMSV14, MMSV16}
for the HO, but is universal for a smooth potential \cite{SM}.
The matching with the bulk for $\hat a \to - \infty$ obtained above
agrees with known results for the HO/GUE \cite{footnote5,MMSV14,Gustavsson}.

For the HO
 $x^\pm \! = \! \pm \sqrt{2 \mu}$,
$\theta_x \! = \! \arccos (-x/\sqrt{2\mu})$ and \eqref{mr1}
agrees with the rigorous results for the GUE \cite{Borodin1}.
In this case \eqref{generalab} gives a general result \cite{SM}
which agrees with known results in special cases
\cite{MMSV14,MMSV16,Charlier_hankel,MNSV09,MNSV11}.

Another important example is the inverse square well
$V(x) \! = \! \frac{x^2}{2} + \frac{\alpha(\alpha-1)}{2 x^2}$
for $x \! > \! 0$ and $\alpha \! \geq \! 1/2$. It corresponds \cite{SM} to the 
Wishart-Laguerre unitary ensemble (LUE) of random matrices 
\cite{Forrester} with the correspondence between fermion positions $x_j$ and
eigenvalues $\lambda_j \! \sim \! x_j^2$ \cite{NadalMajumdar2009,Farthest,Hardwalls}.
One has $\mu \! = \! 2 N +\alpha + 1/2$, hence $d\mu/dN \! \simeq \! 2$
and $\cos \theta_x \! = \! \frac{\mu-x^2}{\sqrt{\mu^2-\alpha(\alpha-1)}}$.
We focus on the interval $[0,a]$ and scale both $a \! = \! O(\sqrt{\mu})$ and
$\alpha \! =\! O(\mu)$ in the large $\mu$ limit. This scaling, used below for
$d$-dimensional central potentials, is also the standard large-$N$ limit 
for Wishart matrices. Setting $\tilde a \! = \! a/ \! \sqrt{2\mu}$ and
$\lambda \! = \! \alpha/\mu$, one obtains from \eqref{Haa} in the
bulk $|2 \tilde a^2-1| \! < \! \sqrt{1- \lambda^2}$ 
\be  \label{var_Bessel } 
2\pi^{2}{\rm Var}{\cal N}_{[0,a]}^{{\rm LUE}}\simeq\log(\mu)+\log\left(4\tilde{a}\frac{\left(1-\tilde{a}^{2}-\frac{\lambda^{2}}{4\tilde{a}^{2}}\right)^{3/2}}{\left(1-\lambda^{2}\right)^{1/2}}\right)+c_{2}
\ee 
with the superscript LUE added for later convenience.
A similar result was recently reported in the mathematics literature
\cite{Charlier1,Charlier2}. The result \eqref{mr1} also 
agrees with rigorous GFF results for
the LUE \cite{BorodinGorin,Paquette}. We have extended these results to other cases
related to RMT \cite{SM}.


We now address a central potential $V(r)$ in $d \! > \! 1$
and focus on the number of fermions ${\cal N}_R$ in a spherical domain ${\cal D}$ of radius $R$ centered at the origin.
The single particle Hamiltonian $\hat H$ commutes with the angular momentum $\hat {\bm L}$,
and with $\hat {\bm L}^2$ of eigenvalues $\ell(\ell+d-2)$,
$\ell=0,1,\dots$, defining the sector
of angular momentum $\ell$. The eigenstates of $\hat H$ are
obtained from those of a collection of 1D radial problems
$\hat H_\ell \! = \! - \frac{1}{2} \partial_r^2  + V_\ell(r)$, $r \! \geq \! 0$, with potentials
\cite{MoshinskyBook,Farthest} 
\be \label{Vl} 
V_{\ell}(r)=V(r)+\frac{\left(\ell+\frac{d-3}{2}\right)\left(\ell+\frac{d-1}{2}\right)}{2r^{2}}
\ee
and eigenenergies $\epsilon_{n,\ell}$, each with degeneracy
$g_d(\ell)$, which behaves as 
$g_{d}(\ell\gg1)\simeq\frac{2\ell^{d-2}}{\Gamma(d-1)}$.
We consider the $N$ fermion ground state 
where all levels with $\epsilon_{n,\ell} \leq \mu$ are filled.
In each sector $\ell$, the levels $n=1,\dots,m_\ell$ are
occupied, with $N =  \sum_{\ell} g_d(\ell) m_\ell$, with
$m_\ell  =  0$ for $\ell  >  \ell_{\max}(\mu)$.
Remarkably, we show \cite{SM} that the quantum joint probability of the
radial positions $\{ r_i \}_{i=1,\dots,N}$ of the fermions
decouples into a symmetrized product over the 
angular sectors. As a consequence, the cumulants
$\langle {\cal N}_R^p \rangle^c$ for $p  \geq  1$ are simply sums over the angular sectors
as
\be
\label{eq:sum_of_cumulants_angular_sectors}
\left\langle {\cal N}_{R}^{p}\right\rangle ^{c}=\sum_{\ell=0}^{\ell_{\max}(\mu)}g_{d}(\ell)\left\langle {\cal N}_{[0,R]}^{p}\right\rangle _{\ell}^{c}
\ee
where $\langle {\cal N}_{[0,R]}^p \rangle^c_{\ell}$ are the cumulants
of 
${\cal N}_{\left[0,R\right]}$ for the 1d potential
$V_\ell(r)$ in \eqref{Vl} with $m_\ell$ fermions. In the large $\mu$ limit, the 
sum in \eqref{eq:sum_of_cumulants_angular_sectors} is dominated
by large values of $\ell$ and $m_\ell$, and, for $p > 1$, is effectively cut-off at 
$\ell_{c}(\mu,R)  \simeq  R k_F(R)  \leq  \ell_{\max}$, where $k_F(r) = \sqrt{2 (\mu-V(r))}$.
This allows us to use our results in 1d
and to obtain the variance of ${\cal N}_R$ 
for a general central potential, see \cite{SM}. 

We discuss here the HO $V(r) =  r^2  /2$,
for which the density has a spherical support, 
with $\rho^{{\rm bulk}}(r)\sim\left(2\mu-r^{2}\right)^{d/2}$,
and an edge at $r =  r_e  =  \sqrt{2 \mu}$  \cite{DeanEPL2015}. In this case $V_\ell(r)$ in \eqref{Vl} is the inverse square well 
studied above with $\alpha  =  \ell + \frac{d-1}{2}$. For large $\mu$, the occupation
numbers $m_\ell$ are determined by $\epsilon_{m_\ell,\ell}  \simeq  2 m_\ell + \ell
 \simeq  \mu$. Hence, defining $\lambda =  \ell/\mu$, one has $m_\ell \simeq \frac{\mu}{2} (1- \lambda)$ for $\lambda < 1$ and $m_\ell  = 0$ for $\lambda >  1$. The total number of
fermions is thus 
{
\be
N \! \simeq \! \frac{\mu^d}{\Gamma(d-1)} \! \int_0^1 d\lambda (1- \lambda) \lambda^{d-2} \! = \! \frac{\mu^d}{\Gamma(d+1)}.
\ee}
Substituting the result \eqref{var_Bessel }
with $a  =  R$, i.e., $\tilde a  =  \tilde R  =  R/\sqrt{2 \mu}$,
into \eqref{eq:sum_of_cumulants_angular_sectors} with $p  = 2$,
and approximating the 
sum by an integral, one obtains,
using $\ell_{c}(\mu,R)/\mu  =  2 \tilde R \sqrt{1- \tilde R^2}$
\be \label{integrate} 
{\rm Var} {\cal N}_{R} \simeq \frac{2 \mu^{d-1}}{\Gamma(d-1)} \int_0^{2 \tilde R \sqrt{1- \tilde R^2}} \!\! d\lambda \, \lambda^{d-2} 
{\rm Var} {\cal N}^{\rm LUE}_{[0,R]} \; .
\ee
Performing the integral over $\lambda$ 
yields the result in \eqref{variance_N_ddimHO} and
\eqref{Ad} for the HO in the large $\mu$ limit.
The coefficient $A_d(\tilde R)$ has a maximum at $\tilde R  = 1/\sqrt{2}$ for any $d  >  1$, and vanishes at the edge
as $A_d(\tilde R) \sim  (1-\tilde R)^{(d-1)/2}$.
The $O(\mu^{d-1})$ term $B_d$ is obtained in \cite{SM} 
for general $d$. 
{For $d=2$ and $d=3$ it reads
\bea
\label{B2} 
&&2\pi^{2}B_{2}\left(x\right)=\log\left(\frac{\left|1-2x\sqrt{1-x^{2}}\right|}{1+2x\sqrt{1-x^{2}}}\right)\nn\\
&&+2x\sqrt{1-x^{2}}\left\{ \log\left[\left(\frac{64x}{1-2x^{2}}\right)^{2}\left(1-x^{2}\right)^{3}\right]+2\gamma_E-2\right\} \nn\\
\eea 
and
\bea \label{B3}
&&2\pi^{2}B_{3}(x)=(1-2x^{2})^{2}\log|1-2x^{2}|\nn\\
&&\qquad +4x^{2}(1-x^{2})\left\{ \log\left[8x(1-x^{2})^{3/2}\right]+\gamma_{E}\right\} .
\eea
respectively.}
 $B_d(x)$ has a 
singularity $(1-x)^{\frac{d-1}{2}} \log(1-x)$ near the edge at $x  =  1$.
As in $d  =  1$, there is an edge region 
of width $w_N = (2 V'(r_e))^{-1/3}$ where the variance becomes 
a universal function of 
$\hat R =  (R-r_e)/w_N$
\be \label{final}
 {\rm Var} {\cal N}_R \simeq \left(\frac{r_e}{w_N}\right)^{d-1} 
\int_0^\infty \frac{d\xi\, \xi^{\frac{d-3}{2}}}{2 \Gamma(d-1)}
{\cal V}_2(\hat R+ \xi) \; .
\ee
Here $(r_e/w_N)^{d-1}$ is the typical number of
fermions in the edge region \cite{DeanPLDReview} and 
${\cal V}_2$ is the above scaling function for $d=1$, defined in
\cite{MMSV14, MMSV16}. For the HO, 
Eq.~\eqref{final} matches, for $\hat R \to -\infty$, the behavior of
$B_d(x)$ for $x \to 1^-$ \cite{SM}. Finally, 
the small $R$ limit corresponding to free fermions,
given in the introduction,
can also be obtained directly \cite{SM} using the sine-kernel analog  
in $d$ dimensions \cite{Torquato,DeanPLDReview}.


One can ask about higher cumulants of ${\cal N}_{\cal D}$. In $d = 1$,
for potentials related to RMT 
they can be extracted from known Fisher-Hartwig asymptotics
of Hankel and Toeplitz determinants 
\cite{DIK2009,Charlier_hankel,Charlier1,CharlierJacobi}. In all cases we find
for $n \geq 2$ \cite{footnote6} 
\be
\!\!\!\! \left\langle \! {\cal N}_{\left[a,b\right]}^{2n} \right\rangle ^{\!c}\!= \! \kappa_{2n}\!+ o\!\left(1\right),~\kappa_{2n}\!=\!\left(-1\right)^{n+1} \! \left(2n\right)!\frac{2 \zeta \! \left(2n \! -\! 1\right)}{n\left(2\pi\right)^{2n}}
 \label{cumuleven_GUE_gen} 
\ee
and $\langle {\cal N}^{2n+1}_{[a,b]}\rangle^c  = o(1)$, where 
$\zeta(x)$ is the Riemann zeta function. This leads to two important
observations. First, from very recent results
\cite{Charlier3},
Eq.~\eqref{cumuleven_GUE_gen} also holds
for the potential $V_\ell(r) \simeq \frac{\ell^2}{2 r^2}$ even when $\ell \sim \mu$.
Using our Eq.~\eqref{eq:sum_of_cumulants_angular_sectors} 
we obtain \cite{SM} the cumulants 
of ${\cal N}_R$ for free fermions in dimension $d>1$, with $k_F R \gg 1$
\be \label{cumul_free}
\left\langle {\cal N}_{R}^{2n}\right\rangle ^{c} \! = \! \frac{(k_{F}R)^{d-1}}{\Gamma(d)}\! \left(\kappa_{2n}+o(1)\right), \quad n  \geq  2.
\ee
Second, since \eqref{cumuleven_GUE_gen} coincides with the results 
from the sine-kernel (and the Circular Unitary Ensemble) \cite{DIK2009,AbanovIvanovQian2011,CharlierSine2019, CalabreseMinchev2},
it is natural to conjecture that these higher cumulants
arise solely from fluctuations on microscopic scales and that \eqref{cumuleven_GUE_gen} 
actually holds in $d = 1$ for any smooth potential $V(x)$
\cite{footnote7}.
For $d > 1$, 
using $\ell_{c}(\mu,R) \simeq R k_F(R)$ in 
Eq.~\eqref{eq:sum_of_cumulants_angular_sectors},
our conjecture leads to 
$\left\langle {\cal N}_{R}^{2n}\right\rangle ^{c}\!=\!\frac{(k_{F}(R)R)^{d-1}}{\Gamma(d)}\left(\kappa_{2n}+o(1)\right)$,
a natural extension of our 
result for free fermions
\eqref{cumul_free}, where $k_F(R)$ now depends on $R$. 
In fact, for $V(r) = \frac{1}{2} r^2$ the argument is already close to 
being rigorous \cite{thank_Charlier}.

We now apply our results to the calculation of the bipartite R\'enyi entanglement entropy of a $d$-dimensional domain ${\cal D}$ with its complement $\overline{\cal D}$. It is defined for $q  \geq  1$ as
{
\be
S_q({\cal D}) = \frac{1}{1-q}\ln\Tr[\rho_{\cal D}^q],
\ee
}
 where
$\rho_{\cal D}  =  \Tr_{\overline{\cal D}}[\rho]$ is obtained by tracing out 
the density matrix $\rho$ of the system over $\overline{\cal D}$.
For noninteracting fermions 
$S_q({\cal D})$ can be expressed
as a series, 
{
\be
S_{q}({\cal D})=\sum_{n\geq1}s_{n}^{(q)}\left\langle {\cal N}_{{\cal D}}^{2n}\right\rangle ^{c},
\ee}
 in
the cumulants of ${\cal N}_{\cal D}$, 
where 
 $s_n^{(q)}$ are given in \cite{CalabreseMinchev2}
and $s_1^{(q)} =  \frac{\pi^2}{6}(1+ \frac{1}{q})$. In $d = 1$ this relation leads to the well known result for the entropy of free fermions,
{
\be
S_{q}^{{\rm ff}}\left([a,b]\right)\simeq\frac{q+1}{6q}\log\left(2k_{F}|a-b|\right)+E_{q}
\ee}
where $E_q$ is given in Eq.~(11) in \cite{CalabreseEntropyFreeFermions} (see also \cite{JinKorepin2004}). 
Our conjecture for the the higher cumulants in an arbitrary potential (central for $d \! > \! 1$) leads to
\be \label{entropy}
\!\!\! S_{q} \! \left({\cal D}\right) \! = \! \frac{\pi^{2}}{6}\frac{q+1}{q}{\rm Var}{\cal N}_{{\cal D}}  +  \frac{\left(k_{F}(R)R\right)^{d-1}}{\Gamma(d)} \! \left( \! \tilde{E}_{q}+o(1) \! \right)
\ee 
with $\tilde E_q =  E_q   -  \frac{q+1}{6 q} (1 + \gamma_E)$. It holds in $d > 1$ 
for the sphere centered at the origin, and in $d = 1$ for any interval ${\cal D} = [a,b]$
with both $a,b$ in the bulk \cite{footnote9}.
In \eqref{entropy} the simple form of the second term 
arises from the common $R$ dependence of the 
cumulants of order $4$ and higher.
This conjecture is corroborated by the rigorous results leading to 
\eqref{cumuleven_GUE_gen} in $d = 1$ for the HO, the inverse square well 
and the hard box \cite{SM}. It also agrees with existing results 
for $d = 1$ 
\cite{CalabresePRLEntropy,CalabreseEntropyFreeFermions,CalabresePLDEntropy,DubailStephanVitiCalabrese2017}.
Thanks to \eqref{cumul_free}, Eq.~\eqref{entropy} is exact for free fermions 
in $d > 1$. The leading term $S_q({\cal D})  \propto R^{d-1} \log R$ at large $R$ is consistent with
the result obtained using the Widom conjecture applied to a spherical domain \cite{Widom1,Widom2,Widom3,Klitch,CalabreseMinchev1}, and also with the rigorous proof in \cite{ProofKlitch}. Here, in addition, we obtain the first
correction $O(R^{d-1})$. 

In conclusion we obtained analytically the counting statistics and the entanglement entropy for $N \gg 1$ noninteracting fermions at 
temperature $T = 0$ in a general 
potential in $d = 1$, 
and a central potential in $d > 1$. 
They depend non trivially on the shape of the potential, 
already at leading order in $d > 1$, e.g. in \eqref{variance_N_ddimHO}.
These results can be extended to finite 
$T$ \cite{TBP} and it would 
be interesting to extend them to interacting particles,
as was done recently for bosons \cite{Calabrese_etal}.

{\it Acknowledgments:} 
We thank A. Borodin and C. Charlier for interesting discussions.
NRS acknowledges support from the Yad Hanadiv fund (Rothschild fellowship). This research was supported by ANR grant ANR-17-CE30-0027-01 RaMaTraF.

{}


\newpage
.
\newpage

\begin{widetext} 

\setcounter{secnumdepth}{2}

\begin{large}
\begin{center}
Supplementary Material for\\  {\it Counting statistics for non-interacting fermions in a $d$-dimensional potential }
\end{center}
\end{large}

\bigskip
We give the principal details of the calculations described in the main text of the Letter. 
\bigskip

\tableofcontents

\renewcommand{\theequation}{S\arabic{equation}}
\setcounter{equation}{0}

\section{Generalities}

Here we consider noninteracting spinless fermions in $d=1$ with single particle Hamiltonian
$\hat H = \frac{p^2}{2} + V(x)$, working in units such that $\hbar=1$, and fermion mass $m=1$.
For a confining potential $V(x)$, the orthonormal eigenfunctions of $\hat H$, denoted $\psi_k(x)$, are labeled by integers $k=1,2,\dots$ such that their associated eigenenergies $\epsilon_k$ form an increasing sequence. The ground state wave function is the Slater determinant
$\Psi_0(x_1,\dots,x_N)= \frac{1}{\sqrt{N}} \det_{1 \leq i,j \leq N} \psi_j(x_i)$, and the joint probability distribution function (JPDF) of the fermion positions takes a determinantal form, $|\Psi_0(x_1,\dots,x_N)|^2 = \frac{1}{N!} \det_{1 \leq i,j \leq N} K_\mu(x_i,x_j)$ in terms of the kernel $K_\mu$ \citep{DeanPLDReview}
\be \label{kernel_app} 
K_\mu(x,y) = \sum_k \theta(\mu-\epsilon_k) \psi^*_k(x) \psi_k(y) = \sum_{k=1}^N \psi^*_k(x) \psi_k(y)  .
\ee
Here $\mu$ is the Fermi energy, related to $N$ via $N=\sum_k \theta(\mu-\epsilon_k)$. 
Using the orthonormalization of the eigenfunctions $\int\psi_{k}^{*}\left(x\right)\psi_{k'}\left(x\right)=\delta_{k,k'}$, it is straightforward to show that the kernel satisfies
\be
\label{eq:reproducibility}
\int K_{\mu}\left(x,z\right)K_{\mu}\left(z,y\right)dz=K_{\mu}\left(x,y\right),
\ee
a useful property which is called the ``reproducibility'' of the kernel. In particular, it implies \cite{MehtaBook} that
each $n \leq N$ point correlation function of the fermion positions also take a determinantal form in terms of $K_\mu$ (see below for $n=2$), in particular the mean fermion 
density (normalized to $\int dx \rho(x)=N$) is $\rho(x)=K_\mu(x,x)$. 

These properties extend to the case of a non confining potential $V(x)$, with a continuum spectrum, e.g. free fermions $V(x)=0$, with the kernel
given by a continuum limit of \eqref{kernel_app}. In this case $N$ can be infinite and 
the control parameter is $\mu$. This is extended to $d>1$ starting from 
Section \ref{sec:decoupling}.

\section{Fermions in special potentials in $d=1$ and random matrix ensembles}

\label{sec:special} 

As mentioned in the text, for specific potentials $V(x)$ and geometries, the JPDF of the fermion positions $\vec x=\{x_i\}$ in the ground state 
can be mapped, for any $N$, to the JPDF of the eigenvalues $\vec \lambda = \{ \lambda_i \}$ of some random 
matrices.

\begin{itemize}

\item Free fermions on a circle of perimeter $L$ with $V(x)=0$, map to the eigenvalues 
$\lambda_j = e^{i 2 \pi x_j/L}$ of random matrices from the circular unitary ensemble (CUE)
\cite{MehtaBook,Forrester,Cunden2018,FyodorovPLD2020}.
The eigenfunctions are plane waves $\sim e^{2 \pi i p x/L}$, $p=0,\pm 1,\dots$, and
$|\Psi_0(\vec x)|^2 \propto P_{\rm CUE}(\vec \lambda) \propto |\Delta_N(\lambda)|^\beta$,
where $\Delta_N(z)= \prod_{1\leq i<j \leq N}(z_i-z_j)$ and $\beta=2$. The mean density is uniform $\rho(x)=k_F/\pi=\frac{N}{L}$ with $k_F=\sqrt{2\mu}$. 

\item Fermions on $\mathbb{R}$  in the harmonic oscillator (HO) potential, $V(x)=\frac{x^2}{2}$, map to the eigenvalues $\lambda_i = x_i$ of Hermitian random matrices from the Gaussian unitary ensemble (GUE)
\cite{Eisler1,MMSV14,DeanPLDReview,DeanReview2019}.
The eigenfunctions are $\sim e^{-x^2/2} H_n(x)$, $n=0,1,\dots$, 
where $H_n$ are the Hermite polynomials, with energies (with the labeling used here) $\epsilon_{n+1}=n+ \frac{1}{2}$, and $|\Psi_0(\vec x)|^2 = P_{\rm GUE}(\vec \lambda)
\propto |\Delta(\lambda)|^\beta e^{-\sum_i \lambda_i^2}$ with $\beta=2$. For large $N$, the mean density is 
the semi-circle of support $[x^-,x^+]$ with $x^\pm \simeq \pm \sqrt{2 N}$, i.e.,   
$\rho(x) \simeq \rho^{\rm bulk}(x)=k_F(x)/\pi$, where $k_F(x)=\sqrt{(2 \mu- x^2)_+}$ 
and $\mu \simeq N$. We denote here and below $(x)_+=\max(x,0)$. 


\item Fermions on $\mathbb{R}^+$ in the inverse square potential, $V(x) = \frac{x^2}{2} + \frac{\alpha(\alpha-1)}{2x^2}$, $\alpha \geq 1/2$,
map to eigenvalues $\lambda_i = x_i^2$ of Wishart-Laguerre random matrices \cite{NadalMajumdar2009,Farthest,Hardwalls,DeanReview2019}.
The eigenfunctions are $\sim x^\alpha L_n^{\alpha - 1/2}(x^2) e^{-x^2/2}$, where 
$L_n^\gamma(z)$ are the generalized Laguerre polynomials, with
energies $\epsilon_n=2 n + \alpha + \frac{1}{2}$, $n=0,1,\dots$. One has 
$|\Psi_0(\vec x)|^2 d^N \vec x \propto \prod_{i=1}^N x_i^{2 \alpha} e^{- x_i^2} dx_i
\prod_{1 \leq j,k \leq N} (x_j^2 -x_k^2)^2 \propto P_{\rm LUE}(\vec \lambda) d^N \vec \lambda$
where $P_{\rm LUE}(\vec \lambda) \propto e^{- \frac{\beta}{2} \sum_{i=1}^N \lambda_i} \prod_{i=1}^N \lambda_i^{\frac{\beta}{2} (M-N+1)-1} |\Delta_N(\vec \lambda)|^\beta$ with $\beta=2$ is the 
JPDF of the eigenvalues $\lambda_j$ of the
Laguerre-Wishart complex random matrices (also called LUE)
of the form $X^T X$ where $X$ is a $M \times N$ rectangular random
matrix with i.i.d. unit complex Gaussian entries with $M \geq N$
(see \cite{VivoMajumdar} for the case $M<N$) and $\alpha - \frac{1}{2} = M - N$. The mean fermion density is $\rho^{{\rm bulk}}(x)=\frac{1}{\pi}\sqrt{\left(2\mu-x^{2}-\frac{\alpha(\alpha-1)}{x^{2}}\right)_{+}}$ which maps to the 
Marcenko-Pastur density for the $\lambda_i$. The case $\alpha=0$ corresponds to the half-harmonic oscillator
with a half-semi-circle density for the fermions $x_i$. In the other limiting case, $V(x) = \frac{\alpha(\alpha-1)}{2x^2}$, the spectrum of $\hat H$ is continuous and the fermions are described by the Bessel kernel
(see below and \cite{Hardwalls}).

\item Fermions in the "Jacobi box" with $x \in [0,L]$ and potential $V(x)= \frac{\pi^2}{L^2} (
\frac{{\sf a}^2-\frac{1}{4}}{8 \sin^2(\pi x/2 L)} + \frac{{\sf b}^2-\frac{1}{4}}{8 \cos^2(\pi x/2 L)} )$, map to the eigenvalues
$\lambda_i = \frac{1- \cos (\pi x_i/L)}{2}$ of the Jacobi unitary ensemble (JUE) of RMT.
The eigenvectors can be expressed in terms of Jacobi polynomials and the energies are
$\epsilon_n = \frac{\pi^2}{2 L^2} (n + \frac{{\sf a}+{\sf b}+1}{2})^2$. One has $|\Psi_0(\vec x)|^2 d^N \vec x
= P_J(\vec \lambda) d^N \lambda$ where $P_J(\vec \lambda) \propto \prod_i \lambda_i^{\sf a} (1-\lambda_i)^{\sf b} \Delta_N(\vec \lambda)$. For more details see Appendix C in \cite{Hardwalls}. In the case ${\sf a}={\sf b}=\frac{1}{2}$ one obtains the hard box 
with $V(x)=0$ and Dirichlet boundary conditions at $x=0,L$ (note that the eigenfunctions
vanish at $x=0,L$ for any ${\sf a},{\sf b}$).

\end{itemize} 


\section{Counting statistics and kernel in $d=1$}
\label{sec:counting} 

For $N$ fermions on the infinite line one defines the height field observable 
$h(a)=\int_{-\infty}^a \hat \rho(x) dx
=  {\cal N}_{]-\infty,a]}=N-{\cal N}_{[a,+\infty[}$,
with $\hat \rho(x) = \sum_{i=1}^N \delta(x-x_i)$, where ${\cal N}_{\cal D}$ denotes the number of fermions in the subset ${\cal D}$. The number of fermions in the interval $[a,b]$ is thus
${\cal N}_{[a,b]}=h(b)-h(a)$, and its quantum average is $\langle {\cal N}_{[a,b]} \rangle= \int_a^b \rho(x) dx$ where $\rho(x)=\langle \hat \rho(x) \rangle$ is the mean density and here $\langle \dots \rangle$ denotes averages w.r.t. the ground state quantum JPDF $|\Psi_0|^2$. 

Next one defines the two point covariance of the height field, i.e., the function $H(x,y) = {\rm Cov}(h(x),h(y))= \langle h(x) h(y) \rangle - \langle h(x) \rangle \langle h(y) \rangle$.
If $H(x,y)$ is known, then the variance of the number of fermions in any interval, or any collection
of intervals is also known as
\bea \label{HH1} 
&&  {\rm Var}{\cal N}_{\left[a,b\right]}=H\left(a,a\right)+H\left(b,b\right)-2H\left(a,b\right) 
\quad , \quad {\rm Var}{\cal N}_{\left]-\infty,a\right]}={\rm Var}{\cal N}_{\left[a,+\infty\right[}=H(a,a), \\
&& {\rm Var} {\cal N}_{\cup_i [a_i,b_i]} = \sum_{ij} H(a_i,a_j)+H(b_i,b_j)-2 H(a_i,b_j).
 \label{HH2} 
\eea
The height field covariance is related to the two point correlation function defined as
\be \label{R2} 
R_{2}(x,y):=\left\langle \sum_{1\leq i\neq j\leq N}\delta(x-x_{i})\delta(y-x_{j})\right\rangle =\langle\hat{\rho}(x)\hat{\rho}(y)\rangle-\delta(x-y)\langle\hat{\rho}(x)\rangle .
\ee 
Indeed, from its definition one has $\partial_x \partial_y H(x,y) = \langle \hat \rho(x) \hat \rho(y) \rangle 
- \rho(x) \rho(y) = R_2(x,y) + \delta(x-y) \rho(x) - \rho(x) \rho(y)$, using \eqref{R2}. For noninteracting fermions
the correlation functions are given by determinants \cite{MehtaBook,DeanPLDReview} for instance
\be
R_2(x,y) = N(N-1) \int_{-\infty}^{+\infty} dx_3 \dots dx_N |\Psi_0(x,y,x_3,\dots,x_N)|^2 = K_\mu(x,x) K_\mu(y,y) - K_\mu(x,y)^2  \label{R22} 
\ee
where we used that $K_\mu$ is symmetric in its arguments in the cases of interest here. 
Eq. \eqref{R22} then implies $\partial_x \partial_y H(x,y) = \delta(x-y)  \rho(x) - K_\mu(x,y)^2$, which is the equation 
\eqrefMT{\eqrelationKH}
 of the text. In the text we provide solutions to this equation, but one can 
equivalently compute the height covariance from the kernel by integrating 
\eqrefMT{\eqrelationKH} of the main text twice
which yields 
\be \label{HK} 
H(a,b)=\int_{-\infty}^{a}dx\int_{-\infty}^{b}dy\left(-K_{\mu}(x,y)^{2}+\rho(x)\delta(x-y)\right)=\int_{-\infty}^{\min(a,b)}dx\int_{\max(a,b)}^{+\infty}dy\,K_{\mu}(x,y)^{2}
\ee 
where in the second equality we used that $\rho(x)=K_\mu(x,x)= \int_{-\infty}^{+\infty}
dy \, K_\mu(x,y)^2$, which follows from the reproducibility \eqref{eq:reproducibility} of the kernel. 

For fermions on $\mathbb{R}^+$, e.g. for the inverse square well, 
one should replace everywhere above $-\infty$ by $0$. More generally,
for fermions defined in an interval $[c^-,c^+]$ one should replace everywhere above $-\infty$ by $c^-$ and
$+\infty$ by $c^+$, e.g. for the Jacobi box $c^-=0$ and $c^+=L$. In all cases we call for convenience below $[c^-,a]$ and $[a,c^+]$ a "semi-infinite" interval.

\medskip

\section{Calculation of $K_\mu(x,y)^2$}
\label{sec:K2} 

In this section we provide more details on the calculation of the kernel $K_\mu(x,y)$ using the WKB asymptotics, leading to \eqrefMT{\resWKBTWO} in the main text, as well as the evaluation of
$K_\mu(x,y)^2$ on macroscopic scales 
given in \eqrefMT{\aTWO} in the text, later used to obtain the result \eqrefMT{\mrONE} in the text
for the height correlator. We discuss separately the case of a confining potential 
presented in the text, and the case of fermions on the circle.

{\bf Fermions in a confining potential}. In the large $N$ limit the sum over $k$ in \eqref{kernel_app} 
is dominated by $k \gg 1$, i.e., semi-classical eigenstates.
Consider first a confining potential $V(x)$ such that there
are exactly two turning points at the Fermi energy, i.e., two roots to $V(x^\pm)=\mu$.
Energy levels are then non-degenerate. The semi-classical eigenstates obey the quantization condition 
$\int dz \sqrt{2( \epsilon_k - V(z))_+} \simeq \pi k$.
This leads to $d\epsilon_k/dk = (\frac{1}{\pi} \int \frac{dz}{ \sqrt{2( \epsilon_k - V(z))_+}})^{-1}$,
where here and below 
we introduce the standard convention that $1/(x)^\gamma_+=0$ for $x<0$
and $1/(x)_+^\gamma=1/x^\gamma$ for $x>0$ (used below for the values 
$\gamma=1/2$ and $\gamma=1/4$). For $k=N$, $\epsilon_N=\mu$, 
this leads to $dN/d\mu = \int_{x^-}^{x^+} dz/(\pi k_F(z))$. For $x$ in the bulk we can use the 
WKB asymptotics \cite{LandauLifshitz,ValleeBook}
\be \label{WKB1_app} 
\psi_{k}(x)\simeq\frac{C_{k}}{\left[2(\epsilon_{k}-V(x))_{+}\right]^{1/4}}\sin\left(\phi_{k}(x)+\frac{\pi}{4}\right)
\ee 
where $\phi_k(x)= \int_{-\infty}^x dz \sqrt{2( \epsilon_k - V(z))_+}$ is a fast oscillating function. 
The normalization is
$C_k^2 = \frac{2}{\pi} \frac{d\epsilon_k}{dk}$ \cite{Furry}. It naturally allows to
recover the usual formula for the bulk density as
\be
\rho(x) = \sum_{k=1}^N |\psi_N(x)|^2 \simeq \frac{1}{\pi} \int_0^\mu 
\frac{d\epsilon}{\sqrt{2 (\epsilon - V(x))_+} } = \frac{k_F(x)}{\pi}  \quad , \quad 
k_F(x)=\sqrt{2 (\mu- V(x))_+}  \label{rhosemi} 
\ee 
where we have replaced $\sum_k \frac{d\epsilon_k}{dk} \simeq \int d\epsilon$
and $\sin^2 \to 1/2$ up to fast oscillating terms which are neglected.
Note that \eqref{WKB1_app} is approximated by zero in the classically forbidden region, 
$\epsilon_k < V(x)$, which was essential to recover the bulk mean density in \eqref{rhosemi}.
Eq. \eqref{WKB1_app} thus does not accurately describe terms in the sum \eqref{kernel_app} such that
$\epsilon_k \simeq V(x)$, but for $x$ in the bulk it is safe to assume that these contributions are subdominant (as they are
in \eqref{rhosemi}). Note also that we do not need to assume that there are also 
only two roots to $V(x^\pm(\epsilon))=\epsilon$ for all $\epsilon<\mu$, the formula
below are also correct, to the same order, for e.g. a double well potential 
(whenever $\mu>V_0$ where $V_0$ is the local maximum). 

Below we need the limit of large $N$ at fixed $m$, i.e., an expansion near the Fermi energy,
for which one can write, 
\be \label{expansion} 
\phi_{N-m}(x)= \int_{x^-}^x dz \sqrt{2 (\epsilon_{N-m}  - V(z))_+}
= \phi_N(x) - m \frac{d \epsilon_N}{dN} \int_{x^-}^x \frac{dz}{\sqrt{2 (\epsilon_{N}  - V(z))_+}} + o(1)
= \phi_N(x) - m \theta_x + o(1)
\ee
where, as in the text, we define 
\be
\theta_x = \frac{d \mu}{dN} \int_{x^-}^x \frac{dz}{k_F(z)} = \pi \frac{\int_{x^-}^x \frac{dz}{k_F(z)}}{\int_{x^-}^{x^+} \frac{dz}{k_F(z)}} \, .
\ee 

As an example, for the HO one has $\epsilon_k=k-\frac{1}{2}$ and 
$\phi_{N}(x)=\frac{1}{2}x\sqrt{2\epsilon_{N}-x^{2}}+\epsilon_{N}\left(\arcsin\left(x/\sqrt{2\epsilon_{N}}\right)+\frac{\pi}{2}\right)$,
 and 
$\theta_x = \arccos(-x/\sqrt{2 \mu})$ with $\mu \simeq N$. 
We can compare 
the formulae \eqref{WKB1_app} and \eqref{expansion} with the Plancherel-Rotach formula
as given in \cite{ForresterFrankelGaroni2006} Eq. (3.10),
with $k=N-m$, $n=k-1$, setting $x=\sqrt{2 N} X$, for large $N$ and $m$ fixed
\bea \label{HOWF} 
&& \psi_{k=N-m}(x=\sqrt{2N}X)=\left(\frac{1}{\sqrt{\pi}2^{n}n!}\right)^{1/2}e^{-x^{2}/2}H_{n}(x)
\\
&& \qquad\qquad = 2^{1/4}\sqrt{\frac{1}{\pi}}\frac{1}{(1-X^{2})^{1/4}}N^{-\frac{1}{4}-\frac{m+1}{2}}\frac{(N!)^{1/2}}{((N-m-1)!)^{1/2}}\cos\left(\Phi_{N}(X)+(m+1)\arccos X\right)\left(1+O\left(\frac{1}{N}\right)\right) \nonumber
\eea
with $\Phi_N(X)= N X \sqrt{1-X^2} + (N + \frac{1}{2}) \arcsin X - \frac{N \pi}{2}$.
Using in \eqref{WKB1_app} that $\frac{C_{N-m}}{((2 \epsilon_{N-m} - x^2)_+)^{1/4}}
\simeq \frac{C_{N}}{((2 \epsilon_{N} - x^2)_+)^{1/4}}$ 
and $d \epsilon_N/dN=1$ we see that the prefactors of the oscillating term
agree in formulas \eqref{WKB1_app} and \eqref{HOWF}. In addition one can check that 
\be
\Phi_N(X) +  (m+1) \arccos X = \phi_N(x) - m \arccos(-X) + \frac{3 \pi}{4} - (N-m) \pi + o(1),
\ee
hence 
\be
\cos\left(\Phi_{N}(X)+(m+1)\arccos X\right)\simeq(-1)^{N-m+1}\sin\left(\phi_{N}(x)-m\arccos(-X)+\frac{\pi}{4}\right)
\ee
for $x=\sqrt{2 N} X$.
Hence up to a factor $(-1)^{k+1}$, which has no effect in our calculation below, the Plancherel-Rotach formula as given in \cite{ForresterFrankelGaroni2006} coincides with
the WKB approximation \eqref{WKB1_app}, together with our estimate \eqref{expansion}.

We can now insert the WKB asymptotics \eqref{WKB1_app}-\eqref{expansion} for the eigenstates into the formula for the kernel 
\eqref{kernel_app}. We will see that, in the limit of large $N$ and for the observable of interest, the sum over $k$ is dominated 
by $k=N-m$ with $m=O(1)$. Hence we can use the same approximations as in
the previous paragraph, and take for the WKB eigenstates the form
$\psi_{N-m}(x) \simeq \frac{C_{N}}{(2( \mu - V(x)))^{1/4}} \cos(\phi_N(x) - m \theta_x - \frac{\pi}{4} )$,
recalling that $\epsilon_N=\mu$. Using that $C_N^2 = \frac{2}{\pi} \frac{d\mu}{dN}$
and the identity $\cos a \cos b = \frac{1}{2} (\cos(a+b) + \cos(a-b))$ we obtain 
\be
K_{N}(x,y)\simeq\frac{d\mu/dN}{\pi\sqrt{k_{F}(x)k_{F}(y)}}\sum_{m\geq0}\sum_{\sigma=\pm1}\cos\left(\phi_{N}(x)-\sigma\phi_{N}(y)-m(\theta_{x}-\sigma\theta_{y})-\frac{\pi}{4}(1-\sigma)\right) .
\ee
Now we perform the geometric sum over $m$, i.e., we write 
\be
\sum_{m \geq 0} \cos(A - m B) = {\rm Re} \sum_{m \geq 0} e^{i A - i m B}
= {\rm Re} \frac{e^{i A }}{1 - e^{- i B}}= \frac{\sin (A + \frac{B}{2})}{2 \sin \frac{B}{2}}
\ee
with $A= \phi_N(x) - \sigma \phi_N(y) - \frac{\pi}{4} (1- \sigma)$ and $B=\theta_x - \sigma \theta_y$
and we obtain the Eq. \eqrefMT{\resWKBTWO} of the main text with
$\tilde \phi_N(x) = \phi_N(x) + \frac{1}{2} \theta_x - \frac{\pi}{4}$.

We have checked numerically, in the case of the harmonic oscillator,
that \eqrefMT{\resWKBTWO} of the main text provides an excellent approximation 
not only of the amplitude but also of the phase of the (rapid) oscillations.
Note that in that case (and more generally for the potentials related to
RMT) one can use the equivalent Christoffel-Darboux form of the kernel \cite{MehtaBook,Forrester,DeanPLDReview}
(a consequence of the recurrence relations of the Hermite polynomials) 
\be
\label{eq:KHarmonic_exact}
 K_{\mu}\left(x,y\right)=\sqrt{\frac{N}{2}}\,\frac{\psi_{N+1}\left(x\right)\psi_{N}\left(y\right)-\psi_{N}\left(x\right)\psi_{N+1}\left(y\right)}{x-y}
= \frac{e^{-\left(x^{2}+y^{2}\right)/2}}{\sqrt{\pi}\,2^{N}\left(N-1\right)!}\frac{H_{N}\left(x\right)H_{N-1}\left(y\right)-H_{N-1}\left(x\right)H_{N}\left(y\right)}{x-y}
\ee
with our conventions (see \eqref{HOWF} and above). Using the Plancherel-Rotach formula \eqref{HOWF} one then arrives at the same result \eqrefMT{\resWKBTWO} of the main text.
Since that method circumvents the summation over the eigenstates, it provides 
an independent check of our results in some special cases.

The next step is to calculate $K_\mu(x,y)^2$ when the distance $|x-y|$ is macroscopic.
The direct square of \eqrefMT{\resWKBTWO} in the main text leads to the sum of two parts. The first part is obtained from
the $\sin^2(\tilde \phi_N(x) - \sigma  \tilde \phi_N(y))$ terms and the replacement of each of them by $1/2$
(ii) the second part is a linear combination of terms proportional to 
$\cos 2 (\tilde \phi_N(x) - \sigma  \tilde \phi_N(y))$ and 
(from the product of the sine) $\cos 2 \tilde \phi_N(x)$ and $\cos 2 \tilde \phi_N(y)$. 
These terms oscillate on microscopic scales $O(1/k_F(x),1/k_F(y))$, hence 
any local average of them on macroscopic scales (e.g. upon integration
over $x,y$ when computing the
height correlator in \eqref{HK}) will give negligible contributions 
\cite{footnote16}. Retaining thus only the first part, and using that $\sum_{\sigma=\pm 1} \frac{1}{\sin^2(\frac{\theta_x-\sigma\theta_y}{2})}=
4 \frac{1 - \cos(\theta_x) \cos(\theta_y)}{(\cos \theta_x - \cos \theta_y)^2}$ we obtain
Eq. \eqrefMT{\aTWO} of the text. {Note that we have performed a numerical check
of the formula for $K_\mu^2$ for the HO in 
Fig. \ref{Fig:Kernel} in Section \ref{sec:numerics}.}

\bigskip

{\bf Fermions on the circle}. Consider now fermions on the circle $x \in [0,L]$,
and a periodic potential of period $L$ such that $V(x) < \mu$ for all $x$, i.e., without turning points. 
Let us start with the case $V(x)=0$ (i.e., the CUE) which is quite pedagogical. The kernel reads $K_\mu(x,y) =\frac{1}{L}  \sum_{p=-p_N}^{p_N} e^{\frac{2 \pi i p (x-y)}{L}}
= \frac{\sin(k_F (x-y))}{L \sin \pi \frac{x-y}{L}} $ with $k_F = \frac{N \pi}{L}$ and $p_N=\frac{N-1}{2}$
({we restrict ourselves here to the case where $N$ is odd. In this case the many-body ground state is not degenerate}). For $x-y \ll L,$  {$K_\mu(x,y)$} reduces to the sine-kernel. For $x-y= O(L)$ one has,
discarding the fast oscillating term $\cos(2 k_F (x-y))$
\be \label{K2circle} 
K_{\mu}\left(x,y\right)^{2}\simeq\frac{1}{2L^{2}\sin^{2}\left(\pi\frac{x-y}{L}\right)}=\frac{1}{8\pi^{2}}\frac{d\theta_{x}}{dx}\frac{d\theta_{y}}{dy}\frac{1}{\sin^{2}\frac{1}{2}\left(\theta_{y}-\theta_{x}\right)}=\partial_{x}\partial_{y}\frac{1}{2\pi^{2}}\log\left|\sin\frac{\theta_{x}-\theta_{y}}{2}\right|
\ee
with $\theta_x= 2 \pi x/L$ for free fermions. We now show that the last two identities extend to a general potential $V(x)$ where 
$\theta_x$ is given below (by a different formula than the one for the confining well).


In the semi-classical approximation one can consider that the energy levels $\epsilon_k>\max_x V(x)$ are doubly degenerate on the circle \cite{footnote11}. The WKB states are $\psi_{k}^{\pm}(x)\simeq\frac{C_{k}}{\left[2(\epsilon_{k}-V(x))\right]^{1/4}}e^{\pm i\phi_{k}(x)}$ with $\phi_k(x)=\int_{0}^x du
\sqrt{2( \epsilon_k - V(u))}$. Their normalization implies
$C_k^2 = \frac{1}{2 \pi} \frac{d\epsilon_k}{dk}$, using the {quantization} condition $\int_{0}^L dx \sqrt{2( \epsilon_k - V(x))}  = 2 k \pi$. Denoting $n$ the highest fully occupied level, $\mu=\epsilon_n$, we have $N=2 n+1 \simeq 2 n$ for
$N \gg 1$, and  $\int_{0}^L dx \sqrt{2( \mu - V(x))}  \simeq 2 n \pi \simeq N \pi$ so we have 
$\rho(x) \simeq k_F(x)/\pi$ as usual 
and $dN/d\mu \simeq \int_0^L \frac{dx}{\pi k_F(x)}$. In particular 
$C_n^2 =  \frac{1}{2 \pi} \frac{d\epsilon_n}{dn} \simeq \frac{1}{\pi} \frac{d\mu}{dN}$, with
a factor of $2$ compared to the case of two turning points.
Inserting the WKB wavefunctions in the kernel 
$K_\mu(x,y) \simeq \sum_{\sigma=\pm 1} \sum_{m=0}^{n} 
\psi^\sigma_{n-m}(x)^* \psi^\sigma_{n-m}(y)$ one expands
$\phi_{n-m}(x)= \phi_n(x) - m \theta_x + o(1)$,
where $\theta_x=\frac{d\phi_n(x)}{dn}
= \frac{d \epsilon_n}{dn} \int_0^x \frac{du}{\sqrt{2( \mu - V(u))}}$.
Since $\frac{d \epsilon_n}{dn} \simeq 2 \frac{d\mu}{dN}$,
one thus obtains that for the circle 
$\theta_x = 2 \pi  \frac{\int_{0}^{x} \frac{dz}{k_F(z)} }{\int_{0}^{L} \frac{dz}{k_F(z)} }$.
Performing the same
manipulations as in the text we obtain
\be \label{appr2} 
K_{\mu}(x,y)\simeq\frac{2}{\pi}\frac{d\mu/dN}{\sqrt{k_{F}(x)k_{F}(y)}}\sum_{m=0}^{n}\cos\left(\phi_{n}(y)-\phi_{n}(x)+m(\theta_{x}-\theta_{y})\right)\simeq\frac{1}{\pi}\frac{d\mu/dN}{\sqrt{k_{F}(x)k_{F}(y)}}\frac{\sin\left(\tilde{\phi}_{n}(y)-\tilde{\phi}_{n}(x)\right)}{\sin\frac{1}{2}\left(\theta_{y}-\theta_{x}\right)}
\ee
where $\phi_n(x)=\int_{0}^x du \sqrt{2( \mu - V(u))}$ and
$\tilde \phi_n(x)=\phi_n(x) + \frac{1}{2} \theta_x$.
Using that $d \theta_x = 2 \frac{d \mu}{d N} \frac{dx}{k_F(x)}$ 
and $\sin^2 \to 1/2$ up to fast oscillating terms we arrive at 
\eqref{K2circle}.

\bigskip

\section{More details on the results for the counting statistics in $d=1$}

Consider fermions with Fermi energy $\mu$ in a general potential $V(x)$ in $d=1$,
defined on the interval $[c^-,c^+]$ (which may be infinite or semi-infinite).
In this section we explain the formula for the variance of the number of fermions 
${\cal N}_{\cal D}$ in a {\it macroscopic} interval in the large $N,\mu$ limit. These
formula will 
differ slightly depending on whether the bulk
density, $\rho(x)=k_F(x)/\pi$, has e.g. (i) a bounded support on a single interval 
$[x^-,x^+]$ with $V(x^\pm)=\mu$, (ii) a semi-infinite support $[x^-,+\infty[$
(iii) no edge such as fermions on the circle with $V(x)<\mu$ for all $x$.
Other cases, such as multiple interval supports can also be studied.

{\bf (i) Confining potentials: two turning points}.
The case (i) relevant for a confining trap was detailed in the main text, leading to formula 
\eqrefMT{\generalab} in the text for the variance of ${\cal N}_{[a,b]}$ and formula \eqrefMT{\mrONE}, \eqrefMT{\Haa} in the text
for the height field correlator. They are expressed in terms
of the semi-classical variable $\theta_x$ defined in \eqrefMT{\thetaxZERO} in the text,
which reaches values $0$ and $\pi$ at $x^-$ and $x^+$,
and has the interpretation of the time along the classical 
trajectories (normalized by the period).
This case corresponds to two turning points at energy $\mu$ at positions $x^-$ and $x^+$. 
For the potentials $V(x)$ related to RMT introduced in Section \ref{sec:special},
 $\theta_x$ has a simple expression. One finds, from the definition
in \eqrefMT{\thetaxZERO} in the text
\be \label{thetaRMT} 
\theta_{x}=\begin{cases}
\arccos(\frac{-x}{\sqrt{2\mu}})\quad, & V(x)=\frac{1}{2}x^{2}\quad,\quad x^{\pm}={\pm}\sqrt{2\mu}\\
\arccos(\frac{\mu-x^{2}}{\sqrt{\mu^{2}-\alpha(\alpha-1)}})\quad, & V(x)=\frac{x^{2}}{2}+\frac{\alpha(\alpha-1)}{2x^{2}}\quad,\quad(x^{\pm})^{2}=\mu\pm\sqrt{\mu^{2}-\alpha(\alpha-1)}\\
\arccos(\frac{\cos(\pi x/L)-A}{B})\quad, & V(x)=\frac{\pi^{2}}{L^{2}}(\frac{{\sf a}^{2}-\frac{1}{4}}{8\sin^{2}(\pi x/2L)}+\frac{{\sf b}^{2}-\frac{1}{4}}{8\cos^{2}(\pi x/2L)})\quad,\quad\cos(\pi x_{\pm}/L)=A\mp B
\end{cases}
\ee
with $A = \frac{{\sf b}^2-{\sf a}^2}{8 \mu}$ and $B= \sqrt{ 1 + A^2 - \frac{1-2 {\sf a}^2 - 2 {\sf b}^2}{8 \mu}}$. 
We now discuss each potential separately. The following formula are useful in all cases \cite{footnote12}
\be
\frac{d\mu}{dN} \simeq \frac{\pi}{\int_{x^-}^{x^+} dz/k_F(z) } 
~,~ \theta_x \simeq \frac{d\mu}{dN} \int_{x_-}^{x} dz/k_F(z)
\quad , \quad 2 (\sin \frac{1}{2} (\arccos p \pm \arccos q) )^2 
= 1- p q \pm \sqrt{(1-p^2)(1-q^2)} \; .
 \label{trigo} 
\ee
\begin{itemize}
\item
For the HO (first line in \eqref{thetaRMT}) one has $d\mu/dN \simeq 1$ and from \eqref{thetaRMT} 
$|\sin \theta_a|=  \sqrt{1- \frac{a^2}{2 \mu}} = \frac{k_F(a)}{\sqrt{2 \mu}}$.
Inserting in \eqrefMT{\Haa} in the text it leads to the explicit expression for the variance 
for the semi-infinite interval
\be \label{aaa} 
{\rm Var} {\cal N}_{[a,+\infty[} ={\rm Var} {\cal N}_{]-\infty,a]} 
= \frac{1}{2 \pi^2} \left(\log \mu + \frac{3}{2} \log (1 - \tilde a^2) + c_2 + 2 \log 2  + o(1) \right) .
\ee
Using in addition the trigonometric relation \eqref{trigo} one can check that the
general formula \eqrefMT{\generalab} in the text leads to the expression 
\be  \label{HOHO} 
2\pi^{2}{\rm Var}{\cal N}_{[a,b]}=2\log\mu+\frac{3}{2}\log\left[\left(1-\tilde{a}^{2}\right)\left(1-\tilde{b}^{2}\right)\right]+2\log\left|\frac{4|\tilde{a}-\tilde{b}|}{1-\tilde{a}\tilde{b}+\sqrt{(1-\tilde{a}^{2})(1-\tilde{b}^{2})}}\right|+2c_{2}+o(1)
\ee
in the limit $\mu \to +\infty$ with
$\tilde a = \frac{a}{\sqrt{2 \mu}}$, $\tilde b = \frac{b}{\sqrt{2 \mu}}$
fixed, $-1 < \tilde a \neq \tilde b < 1$.
For $a=-b$ the leading-order term in \eqref{HOHO} agrees with the Coulomb gas
calculations in \cite{MMSV14,MMSV16}. 
The $O(1)$ term also agrees with some exact results by other methods in the RMT context:
1) Eq. \eqref{aaa} for $a=0$ agrees with the calculation of the ``index'' in \cite{MNSV09,MNSV11},
see also \cite{ForresterWitte} where higher order corrections in $1/N$ where 
obtained using Painlev\'e equations. In particular the leading corrections to \eqref{aaa} are $O(\frac{\log N}{N})$; 2) a study of Fisher-Hartwig asymptotics using Riemann Hilbert methods
in \cite{Charlier_hankel} (for the comparison with this
work see Section 
\ref{sec:cum}). Our results \eqref{HOHO} and \eqref{aaa} are compared with numerical
simulations in Fig. \ref{Fig_1d_HO} in Section \ref{sec:numerics}.

\item
For the inverse square well (second line in \eqref{thetaRMT}) one has $d\mu/dN \simeq 2$ and from \eqref{thetaRMT},
$|\sin \theta_a| = \frac{a k_F(a)}{\sqrt{\mu^2 - \alpha (\alpha-1)} }$. Formula
\eqrefMT{\Haa} in the text then leads to 
\be
{\rm Var}N_{[0,a]}={\rm Var}N_{[a,+\infty[}=\frac{1}{2\pi^{2}}\left[\log\left(\frac{ak_{F}(a)^{3}}{\sqrt{\mu^{2}-\alpha(\alpha-1)}}\right)+c_{2}\right]+o(1)
\ee
which leads to the equation \eqrefMT{\varBessel} in the text
with $\lambda^2 = \frac{\alpha (\alpha-1)}{\mu^2}$ and $\tilde a=a/\sqrt{2 \mu}$. 
In addition, using the above relations we find that \eqrefMT{\generalab} in the text leads to 
\bea \label{Wish} 
&& (2 \pi^2) {\rm Var} N_{[a,b]}  =  2 \log \mu + \log(16 \tilde a\tilde b \kappa_{\tilde a}^3 
\kappa_{\tilde b}^3) 
+ 2 \log \frac{|\tilde a^2 - \tilde b^2|}{\tilde a^2 + \tilde b^2 - 2 \tilde a^2 \tilde b^2 - \frac{\lambda^2}{2} 
+ 2 \tilde a\tilde b \kappa_{\tilde a}
\kappa_{\tilde b}}
+ 2 c_2 + o(1) 
\eea
with $\kappa_{\tilde{a}}=\left(1-\tilde{a}^{2}-\frac{\lambda^{2}}{4\tilde{a}^{2}}\right)^{1/2}$ and 
$\kappa_{\tilde{b}}=\left(1-\tilde{b}^{2}-\frac{\lambda^{2}}{4\tilde{b}^{2}}\right)^{1/2}$. 
Eq. \eqref{Wish} is obtained in the limit of large $\mu$ with $\tilde a, \tilde b$ fixed
and $\lambda = O(1)$ fixed, i.e., $\alpha \sim \mu$. For $\lambda=0$ our result 
\eqref{Wish} agrees with the  
Fisher-Hartwig asymptotics obtained using Riemann Hilbert methods
for the LUE 
in \cite{CharlierJacobi}, which assumes $\alpha=O(1)$ (see also Section \ref{sec:cum}). Finally, one 
can check that as $\tilde a - \tilde b \ll 1$, Eq. \eqref{Wish} reproduces the result for a
microscopic interval,
$\pi^2 {\rm Var} {\cal N}_{[a,b]} = \log( 2 \mu |\tilde a-\tilde b| \kappa_{\tilde a}) + c_2
= \log(|a-b| k_F(a)) + c_2$.
{The result for ${\rm Var} \mathcal{N}_{\left[0,a\right]}$ is compared
with numerical simulations in Fig. \ref{Fig:hard_box_and_LUE}
in Section \ref{sec:numerics}.
}

\item
For the Jacobi box (third line in \eqref{thetaRMT}), one can check directly that $\frac{d\theta}{dx} = \frac{d\mu}{dN} \frac{1}{k_F(x)}$ with
$\frac{d\mu}{dN} = N = \sqrt{2 \mu}$, and $\theta_{x^-}=0$, $\theta_{x^+}=\pi$. 
Using \eqref{thetaRMT} and trigonometric relations, one obtains explicit formula (not displayed here)
for ${\rm Var} {\cal N}_{[a,b]}$ from \eqrefMT{\generalab} in the text
for $a,b$ in the bulk, and for ${\rm Var} {\cal N}_{[0,a]}={\rm Var} {\cal N}_{[a,L]}$
from \eqrefMT{\Haa} in the text, in the regime ${\sf a} \sim {\sf b} \sim \mu$. 
One can check that in the limit ${\sf a} \sim {\sf b} \ll \mu$ they agree
with the Fisher-Hartwig asymptotics obtained using Riemann-Hilbert methods
for the JUE in \cite{CharlierJacobi} (see also Section \ref{sec:cum}).
Here we only display the result for the 
{\it hard box} $V(x)=0$ for $x \in [0,L]$ and Dirichlet boundary conditions for the
wavefunction at $x=0,L$. It can be obtained as the limit of the JUE 
for ${\sf a}= {\sf b} = 1/2$, i.e., $A=0$ and $B=1$. One has $\mu = \frac{\pi^2}{2 L^2} N^2$,
$\frac{d \mu}{dN}= \frac{\pi^2}{L^2} N = \frac{\pi}{L} \sqrt{2 \mu}$ and $\theta_x= \pi \frac{x}{L}$
and from \eqrefMT{\Haa} and \eqrefMT{\generalab} in the text we obtain for the hard box 
(where $N= \frac{L}{\pi} \sqrt{2 \mu}$) 
\bea \label{hardboxresults} 
&& {\rm Var}{\cal N}_{[0,a]}={\rm Var}{\cal N}_{[a,L]}=\frac{1}{2\pi^{2}}\left(\log N+\log\left|\sin\pi\frac{a}{L}\right|+c_{2}+\log2+o(1)\right) , \\
&& {\rm Var}{\cal N}_{[a,b]}={\rm Var}{\cal N}_{[0,a]}+{\rm Var}{\cal N}_{[0,b]}+\frac{1}{\pi^{2}}\log\left|\frac{\sin\frac{\pi(a-b)}{2L}}{\sin\frac{\pi(a+b)}{2L}}\right|+o(1) .
\eea
{The result \eqref{hardboxresults} is compared with a numerical calculation in Fig. \ref{Fig:hard_box_and_LUE}
in Section \ref{sec:numerics}.
}
\end{itemize}

{\bf (ii) Non confining potential: single turning point}. Consider here a general potential $V(x)$, such that the support of the bulk density is the semi-infinite interval $[x^-,+\infty[$, $x^- \in \mathbb{R}$. One example for fermions in $[0,+\infty[$ is the repulsive inverse square wall, $V(x)= \frac{\alpha (\alpha-1)}{ 2 x^2}$, with $\alpha>1$, with $x^-=\sqrt{\frac{\alpha(\alpha-1)}{ 2 \mu}}$. Although the variable $\theta_x$ cannot be defined as in \eqrefMT{\thetaxZERO} in the text, one can still obtain the variance by taking the limit $x^+ \to +\infty$. Consider the formula \eqrefMT{\Haa} and \eqrefMT{\generalab} in the text. In that limit 
$\int_{x^-}^{x^+} \frac{dx'}{k_F(x')}  \to +\infty$ but $\int_{x^-}^{x} \frac{dx'}{k_F(x')}$ remains
fixed. Hence $\theta_x$ becomes small and one can Taylor expand in it. 
We see that $\frac{\sin \theta_a}{d\mu/dN} \to \int_{x^-}^{a} \frac{dx'}{k_F(x')}$ in \eqrefMT{\Haa} in the text 
leading to (for any $c^- < x^-$) 
\be
(2\pi)^{2}{\rm Var}{\cal N}_{[c^{-},a]}\simeq(2\pi)^{2}{\rm Var}{\cal N}_{[a,+\infty]}\simeq\log\left(2k_{F}(a)^{2}\int_{x^{-}}^{a}\frac{dx'}{k_{F}(x')}\right)+c_{2}=\log\mu+\log4\tilde{a}\left(1-\frac{\lambda^{2}}{4\tilde{a}^{2}}\right)^{3/2}+c_{2}
 \label{1turning} 
\ee
where the last equality is specialized to the inverse square wall, with $\lambda = \alpha/\mu$ and $\tilde a = a/\sqrt{2 \mu}$,
in which case
$\int_{x^-}^{a} \frac{dx'}{k_F(x')}= \frac{a}{2 \mu} k_F(a)$,
and it is valid in the limit $\alpha \to \infty$, $\mu \to \infty$ keeping $\lambda$ and $\tilde{a}$ fixed.
Similarly, performing the limit $x^+ \to +\infty$ on \eqrefMT{\generalab} in the text,
one obtains the variance for an interval in the bulk, for a general such potential
\be \label{generalab1t} 
\pi^{2}{\rm Var}{\cal N}_{[a,b]}=\log\left(2k_{F}(a)k_{F}(b)\int_{x^{-}}^{a}\frac{dz}{k_{F}(z)}\int_{x^{-}}^{b}\frac{dz}{k_{F}(z)}\right)+\log\left|\frac{\int_{a}^{b}\frac{dz}{k_{F}(z)}}{\int_{x^{-}}^{a}\frac{dz}{k_{F}(z)}+\int_{x^{-}}^{b}\frac{dz}{k_{F}(z)}}\right|+c_{2}+o(1) \, .
\ee
For the inverse square wall, it is known that the kernel is the Bessel kernel 
$K_\mu(x,y) = 2 k_F^2 \sqrt{x y} K_{Be,\alpha-1/2}(k_F^2 x^2, k_F^2 y^2)$
with $k_F=k_F(+\infty)=\sqrt{2 \mu}$, and the set of $k_F^2 x_i^2$ form a determinantal Bessel process
of index $\alpha-1/2$. In \cite{Charlier1} results are obtained for the Bessel process 
for fixed $\alpha$. The correspondence amounts to set in \cite{Charlier1} 
$x=1$ and $r=k_F^2 a^2=2 \mu a^2$, and $N_{[0,r]}$ there equal to our $N_{[0,a]}$.
The formula (1.16) in \cite{Charlier1} is obtained for $r \to +\infty$ at fixed $\alpha=O(1)$
and we see that it agrees with our result \eqref{1turning} for $\lambda=0$. 
Note however that, while this paper was in progress, a new result was obtained
\cite{Charlier2} in the limit $\alpha \sim \mu$. This new result, obtained by very different
methods, also agrees
with our formula \eqref{1turning} for generic $\lambda$. Similarly our formula
\eqref{1turning} for the interval $[a,b]$ can be compared with (1.19) in
\cite{Charlier1}.

{\bf (iii) Fermions on the circle: no turning point}. Consider fermions on the circle $x \in [0,L]$,
and a periodic potential of period $L$ such that $V(x) < \mu$ for all $x$, i.e., without turning points. 
Let us first display our result, and then sketch how it is obtained from the results in Section \ref{sec:K2}
(since it is slightly different from the other cases). 
We find that for any macroscopic interval $[a,b]$ the formula for the variance of ${\cal N}_{[a,b]}$ in Eqs. \eqrefMT{\generalab}, \eqrefMT{\thetaxZERO} in the text are
replaced, in the case of the circle, by 
\be \label{generalabcircle} 
\pi^2 {\rm Var} {\cal N}_{[a,b]} =  \log|\sin\frac{\theta_a-\theta_b}{2}| 
+  \log (k_F(a) k_F(b)
  \int_0^L \frac{dz}{\pi k_F(z)}
) + c_2  + o(1)  ~,~ \theta_x := 2 \pi  \frac{\int_{0}^{x} \frac{dz}{k_F(z)} }{\int_{0}^{L} \frac{dz}{k_F(z)} } ~,~ k_F(x) = \sqrt{2 (\mu - V(x))} .
\ee 
Note the factor of $2$ in the definition of $\theta_x$. In Section \ref{sec:K2}
we computed $K_\mu(x,y)^2$ for the circle, given by \eqref{K2circle}.
To obtain \eqref{generalabcircle} we use that
${\rm Var} {\cal N}_{[a,b]}$, considered as a symmetric function of $(a,b)$,
obeys (from Section \ref{sec:counting})
$\frac{1}{2} \partial_a  \partial_b {\rm Var} {\cal N}_{[a,b]}
= K_\mu(a,b)^2 - \delta(a-b) \rho(a)$. The Eq.
\eqref{K2circle} then determines ${\rm Var} {\cal N}_{[a,b]}$ up to 
a term $f(a)+f(b)$, and the function $f$ is then fixed using the matching
onto the microscopic result, leading to \eqref{generalabcircle}. Note that
if one knows ${\rm Var} {\cal N}_{D}$ for any interval ${\cal D}=[a,b]$, one knows it 
for ${\cal D}$ being any collection of intervals, since e.g. one can always define
a height function 
$h(a)={\cal N}_{[0,a]}$ and use \eqref{HH1},\eqref{HH2}.

Let us discuss some properties of the result \eqref{generalabcircle}.
For a microscopic interval $|a-b| = O(1/k_F(a))$ one has as usual,
$\pi^2 {\rm Var} {\cal N}_{[a,b]} \simeq 
U(0)-U(k_F(a)|a-b|) = \log[ k_F(a)|a-b| ] + c_2 + o(1)$ with the same function $U(z)$ as in the text.
One can check that \eqref{generalabcircle} matches with this microscopic result in two 
limits (i) $|a-b| \ll 1$ and (ii), due to the
periodicity of the circle, $a \to 0$ and $b \to L$, using 
$|\sin\frac{\theta_a-\theta_b}{2}| \simeq \frac{\pi}{ \int_{0}^{L} \frac{dz}{k_F(z)}}
( \int_0^a \frac{dz}{k_F(z)} + \int_b^L \frac{dz}{k_F(z)})$ which leads to $ {\rm Var} N_{[a,b]}  \simeq \log [k_F(0)(L-b+a)] + c_2+o(1)$ in that limit. 
Finally it is interesting to note that the formula
to be used may change as $\mu$ is varied, e.g. for the potential $V(x)=V_0 \cos 2 \pi x/L$ on the circle, 
for $\mu>V_0$ we must use \eqref{generalabcircle}, while for $\mu<V_0$ we must use \eqrefMT{\generalab} in the text.

\medskip

\section{Central potential for $d>1$: Generalities and decoupling}
\label{sec:decoupling}

Here we consider noninteracting fermions in their ground state in a central potential $V(r)$.
The $N$ body Hamiltonian is ${\cal H}_N = \sum_{i=1}^N \hat H_i$, where the single particle Hamiltonian 
$\hat H= \frac{{\bf p}^2}{2} + V(r)$. We show the decoupling between the
different angular momentum sectors which leads to Eq. \eqrefMT{\eqsumofcumulantsangularsectors}
 in the text.

We use the spherical coordinates ${\bf x} = (r,{\bm \theta})$ where ${\bm \theta}$ is a $d-1$ dimensional angular vector. In these coordinates, the single particle Hamiltonian can be written as {$\hat{H}=-\frac{1}{2}\Delta_{{\bf x}}+V(r)=-\frac{1}{2}r^{1-d}\partial_{r}\left(r^{d-1}\partial_{r}\right)+\frac{1}{2r^{2}}\hat{\bm{L}}^{2}+V(r)$.}
The eigenfunctions of $\hat H$, using spherical symmetry, are labeled by the quantum numbers $(n,{\bf L})$, where $n$ is a positive integer, and where ${\bf L}$ stands collectively for all the angular quantum numbers. They can be written as \citep{Farthest}
\be
\psi_{n,{\bf L}}(r,{\bm \theta})=r^{\frac{1-d}{2}}\chi_{n,l}(r)Y_{{\bf L}}(\bm{\theta}) \,.\label{eq: decomp}
\ee  
The $Y_{{\bf L}}(\bm{\theta})$ are the $d$-dimensional spherical harmonics, labeled by the set of angular quantum numbers ${\bf L}$. They are eigenfunctions of $\hat{\bm{L}}^2$ with eigenvalues $\ell(\ell+d-2)$ depending on a single {nonnegative} integer $\ell$.
The radial part $\chi_{n,\ell}(r)$ is the eigenfunction of a $1d$ effective Hamiltonian, $\hat H_\ell \, \chi_{n,\ell} = \epsilon_{n,\ell} \chi_{n,\ell}$, with an effective potential 
\be \label{VlSuppMat} 
V_{\ell}\left(r\right)=V\left(r\right)+\frac{\left(\ell+\frac{d-3}{2}\right)\left(\ell+\frac{d-1}{2}\right)}{2r^{2}}.
\ee

The ground state wavefunction is given by the Slater determinant 
$\Psi_0({\bf x_1}, \cdots, {\bf x_N}) = \frac{1}{\sqrt{N!}} \det_{1\leq i,j \leq N} \left[ \psi_{{\bf k}_i}({\bf x}_j) \right]$,
where ${\bf k}_i = (n_i, {\bf L}_i)$ labels the single particle eigenfunction of the occupied eigenstates. 
In the ground-state, the occupied eigenstates are all the energy levels $\epsilon_{n, \ell} \leq \mu$,
where $\mu$ is the Fermi energy {(we assume for simplicity that $N$ is such that the many-body ground state is not degenerate)}. Using standard methods, such as the Cauchy-Binet formula (see e.g. \cite{Farthest,CalabreseMinchev1}) we can write the generating function of the cumulants of the number of fermions ${\cal N}_R$ in a sphere of radius $R$
centered at the origin using the overlap matrix $\mathbb A$ 
\be \label{cdf_det_2}
\left\langle e^{-s{\cal N}_{R}}\right\rangle =\det_{1\leq i,j\leq N}\left[\delta_{ij}-(1-e^{-s})\mathbb{A}_{ij}\right]\quad,\quad\mathbb{A}_{ij}=\int_{r=|\x|\leq R}d^{d}{\bf x}\,\psi_{n_{i},{\bf L}_{i}}^{*}({\bf x})\psi_{n_{j},{\bf L}_{j}}({\bf x})\;\;,\;
\ee
where $\psi_{n,{\bf L}}({\bf x})$ is given in Eq. (\ref{eq: decomp}).
and $\left\langle \dots\right\rangle $ denotes the quantum expectation value with respect to $|\Psi_0|^2$. 
Using $d^d{\bf x} = r^{d-1} \,dr \, d{\boldsymbol \theta}$ and the orthogonality property of the spherical harmonics 
$\int d{\boldsymbol \theta}\ 
  Y_{\bf L}({\boldsymbol \theta})Y_{{\bf L}'}({\boldsymbol \theta}) = \delta_{{\bf L},{\bf L}'}$
  the angular integral gives 
\be
\label{integral_1}
\int_{r \leq R} d{\bf x} \, \psi^*_{n_i, {\bf L}_i}({\bf x}) \psi_{n_j,{\bf L}_j}({\bf x}) = \delta_{{\bf L}_i, {\bf L}_j} \int_{0}^{R} dr \, \chi_{n_i,\ell_i}(r) \chi_{n_j,\ell_i}(r)  
\ee
where we recall that $\int_0^\infty  dr\ \chi_{kl}(r)\chi_{k'l}(r)=\delta_{kk'}$. Hence the overlap
matrix $\mathbb{A}$ is diagonal in the variables ${\bf L}_i$ and the determinant factorises
over the different angular sectors. Therefore Eq. \eqref{cdf_det_2} takes the product form
\be
\left\langle e^{-s{\cal N}_{R}}\right\rangle = \prod_{\ell \geq 0}^{\ell_{\max}(\mu)} 
Z_\ell(s,m_\ell)^{g_d(\ell)} \quad , \quad Z_\ell(s,m_\ell) =  \langle e^{- s {\cal N}_{[0,R]}} \rangle_{\ell}  =
\det_{1 \leq i,j \leq m_\ell} \left[\delta_{ij} - (1- e^{-s}) \int_0^R dr \chi_{i,\ell}(r) 
\chi_{j,\ell}(r) \right] \label{product} 
\ee
where $g_d(\ell)=\frac{(2 \ell + d-2) \Gamma(\ell+d-2)}{\Gamma(\ell+1) \Gamma(d-1)} $ are the (angular) degeneracies with $g_d(0)=1$ for all $d \geq 1$ and $g_1(0)=g_1(1)=1$ together with
$g_1(\ell)=0$ for all $\ell \geq 2$. Here $Z_\ell(s,m_\ell)$ is the generating function of cumulants of the
number of fermions ${\cal N}_{[0,R]}$ in the interval $r \in [0,R]$ for the 1d system of $m_\ell$
fermions described by the single particle Hamiltonian $\hat H_\ell$. Taking the logarithm in 
\eqref{product} and expanding in $s$ one obtains the equation \eqrefMT{\eqsumofcumulantsangularsectors} in the text. Note that the above arguments extends to any domain with
radial symmetry, for instance the spherical shell $R_1<r<R_2$ which maps onto the study of the
interval $[R_1,R_2]$ in one dimension. Note that the proof given here of \eqrefMT{\eqsumofcumulantsangularsectors} in the text  assumes that the 
total number of fermions $N$ is finite (and the same for the $m_\ell$), which is natural for a confining potential. However it also
applies to the case where the potential is non confining, e.g. for free fermions $V(r)=0$, as
can be seen by taking a limit where the right edge tends to infinity with fixed $R$. 

This property is even more general, as can be understood by the following physical argument.
The single-particle angular momentum $\hat {\bm L}$ and the radial distance operator $\hat r$, commute and can therefore be measured simultaneously. The measurement of $\hat {\bm L}^2$ in the ground state
leads to the values $\left\{ \ell_{i}\right\} _{i=1,2,\dots,\mathcal{M}}$ of all of the angular sectors which have a nonzero number of particles, where each value of $\ell$ appears in this list $g_{d}\left(\ell \right)$ times
and $\mathcal{M}=\sum_{\ell=0}^{\ell_{\max}(\mu)} g_d(\ell)$. Since after the measurement, the Pauli exclusion principle only acts between the particles in the same angular sector we find that the radial JPDF decouples.
Although the way to write it, which we show here for illustration, is a bit heavy because one must
ensure the global symmetry of the JPDF, the concept of this decoupling is quite simple. Defining $a_{i}=\sum_{j<i}m_{\ell_{j}}$, and recalling that $N=\sum_{i=1}^{\mathcal{M}}m_{\ell_{i}}$,
we can write 
\be  \label{P1} 
P\left(r_{1},\dots,r_{N}\right)=\frac{1}{N!}\sum_{\tau\in S_{N}}\prod_{i=1}^{\mathcal{M}}P_{\ell_{i}}\left[r_{\tau\left(a_{i}+1\right)},\dots,r_{\tau\left(a_{i}+m_{l_{i}}\right)}\right]
=
\frac{\prod_{i=1}^{\mathcal{M}}\left(m_{\ell_{i}}\right)!}{N!} 
\sum_{\substack{\cup_{i=1}^{\mathcal{M}}A_{\ell_{i}}=\{1,\dots,N\}\\
\forall i\ne j,\,A_{\ell_{i}}\cap A_{\ell_{j}}=\emptyset,\;\forall i,\,|A_{\ell_{i}}|=m_{\ell_{i}}
}}\;
\prod_{i=1}^{\mathcal{M}} P_{\ell_i}(\vec r_{A_{\ell_i}})
\ee
where $P_{\ell}\left(x_{1},\dots,x_{m_{\ell}}\right)$ is the joint PDF of the positions of $m_{\ell}$ noninteracting fermions in the 1d potential \eqref{VlSuppMat}. (and $S_N$ is the group of permutations of the set $\left\{ 1,\dots,N\right\}$). The last formula involves a sum over the partitions of the set $\left\{ 1,\dots,N\right\}$ into $\mathcal{M}$ subsets 
$\left\{ A_{\ell_{i}}\right\} _{i=1,\dots,\mathcal{M}}$. The last equality arises by regrouping
the terms in the first formula according to the values
of the sets $A_{\ell_{i}}=\left\{ \tau\left(a_{i}+1\right),\dots,\tau\left(a_{i}+m_{\ell_{i}}\right)\right\}$
and using the symmetry of each
$P_{\ell_{i}}\left(\left\{ r_{j}\right\} _{j\in A_{\ell_{i}}}\right)\equiv P_{\ell_{i}}\left(\vec{r}_{A_{\ell_{i}}}\right)$
when summing over permutations. 
Finally, Eq.~\eqrefMT{\eqsumofcumulantsangularsectors} in the text now follows from the fact that cumulants of sums of independent random variables are the sum of their cumulants.

\section{Free fermions in dimension $d$}
\label{sec:free} 

We give here some details about the derivation of the variance for free fermions, i.e., for $V(r)=0$
in dimension $d$ given in the text, together with an alternative method, and compare with known results.

\subsection{Free fermions, using the decoupling and the 1D inverse square potential}

Using the results of section \ref{sec:decoupling}, the case of free fermions $\hat H=\frac{{\bf p}^2}{2}$ 
in dimension $d$ can be studied using the 1d Hamiltonian $\hat H_\ell= - {\frac{1}{2}}\frac{\partial^2}{\partial r^2} + V_\ell(r)$ with $V_\ell(r) = \frac{\alpha (\alpha-1)}{2 r^2}$ with $\alpha = \ell + \frac{d-1}{2}$. 
Consider the number ${\cal N}_R$ of fermions in the sphere of radius $R$. From 
\eqrefMT{\eqsumofcumulantsangularsectors} in the text, its average
is given by 
\be
\label{eq:NRasSumOverEll}
\left\langle {\cal N}_{R}\right\rangle =\sum_{\ell=0}^{+\infty}g_{d}\left(\ell\right)\left\langle {\cal N}_{\left[0,R\right]}\right\rangle _{\ell}.
\ee
Note that since the potential is not confining
the sum over $\ell$ extends to infinity. However, within the sector of angular momentum $\ell$,
the support of the density at large $\mu$ is $[r^{-}(\ell), +\infty[$ with $r^{-}(\ell) \simeq \ell/\sqrt{2 \mu}$ for
$\ell \gg 1$. Hence, for a fixed $R$, the sum is effectively cutoff at $\ell=\ell_c(\mu,R)= k_F R$
with $k_F=\sqrt{2 \mu}$. Using the 1d bulk density we obtain, by replacing the sum in \eqref{eq:NRasSumOverEll} by an integral (which is justified for large $\mu$)
\be 
\label{av} 
\left\langle {\cal N}_{R}\right\rangle \simeq\int_{0}^{\sqrt{2\mu}R}d\ell\frac{2\ell^{d-2}}{\Gamma(d-1)}\int_{0}^{R}dr\frac{1}{\pi}\sqrt{\left(2\mu-\frac{\ell^{2}}{r^{2}}\right)_{+}} 
 = S_d \int_0^R dr \, r^{d-1} \frac{\mu^{d/2} }{(2 \pi)^{d/2} \Gamma(1+d/2)} 
 = \frac{(k_F R)^d}{2^d  \Gamma(1+ d/2)^2} 
\ee
where $(x)_+=\max(x,0)$, $S_d=2 \pi^{d/2}/\Gamma(d/2)$ is the area of the unit sphere embedded in dimension $d$, and the sum is dominated by values of $\ell \gg 1$, with $g_d(\ell) \simeq \frac{2 \ell^{d-2}}{\Gamma(d-1)}$ for $d>1$. This agrees with the standard result for free fermions
for any $d \geq 1$. Furthermore, the first two equalities in \eqref{av} also hold for an arbitrary potential $V(r)$ upon substituting 
$\mu \to \mu - V(r)$ (and $\mu^{d/2} \to (\mu - V(r))_+^{d/2})$, recovering the known result for the density
in the bulk Eq. (180) in \cite{DeanPLDReview}.
 This is a good test of the method.

Consider now the variance of ${\cal N}_R$. Using \eqrefMT{\eqsumofcumulantsangularsectors} in the text
we can proceed similarly as in \eqref{av} and use the asymptotic result for 
the variance of ${\cal N}_{[0,R]}$ in \eqref{1turning}, with the substitution 
$\tilde a \to R/\sqrt{2 \mu}$ and $\lambda = \frac{\alpha(\alpha-1)}{\mu^2}  \simeq \ell^2/\mu^2$ since the sum is again dominated by 
large values of $\ell$
\be
\label{vard}
\left\langle {\cal N}_{R}^{2}\right\rangle ^{c}=\sum_{\ell=0}^{\ell_{\max}(\mu)}g_{d}(\ell)\left\langle {\cal N}_{[0,R]}^{2}\right\rangle _{\ell}^{c}\simeq\int_{0}^{\sqrt{2\mu}R} \! d\ell \, \frac{2\ell^{d-2}}{\Gamma(d-1)}\frac{1}{2\pi^{2}}\left(\log\mu+\log4\tilde{a}\left(1-\frac{\lambda^{2}}{4\tilde{a}^{2}}\right)^{3/2}+c_{2}\right) .
\ee 
Performing the change of variable $\ell = \sqrt{2 \mu} \, R \Lambda$ and integrating over $\Lambda$, using 
$(1-d) \int_0^1 d\Lambda \, \Lambda^{d-2} \log(1- \Lambda^2) = \psi ^{(0)}\left(\frac{d+1}{2}\right)+\gamma_E$,
$\psi^{(0)}(x)$ being the di-gamma function,
one finds
\be
{\rm Var}{\cal N}_{R}\simeq\frac{(k_{F}R)^{d-1}}{\pi^{2}\Gamma(d)}\left[\log(k_{F}R)+1-\frac{1}{2}\gamma_{E}+2\log2-\frac{3}{2}\psi^{(0)}\left(\frac{d+1}{2}\right)\right]\quad,\quad k_{F}=\sqrt{2\mu} \; . \label{ff} 
 \ee
This formula gives the first two orders in an
expansion in the dimensionless parameter $k_F R \gg 1$  for any $d \geq 1$ \citep{footnote20}.

{\bf Remark} One can similarly calculate the variance of the number of fermions in a spherical shell $R_1 <r <R_2$ using \eqref{generalab1t} upon substituting 
$a \to R_1$, $b \to R_2$ and $\alpha(\alpha-1) \to \ell^2$ and 
using $\int_{x^-}^{a} \frac{dx'}{k_F(x')}= \frac{a}{2 \mu} k_F(a)$ and $k_F(x)=\sqrt{2 \mu- \frac{\ell^2}{x^2}}$. 

{\bf Comparison with known results}. The leading term is already known from various works, with quite different methods, 
as we now discuss. However the subleading term in 
\eqref{ff} is to our knowledge new. The term $\sim R^{d-1} \log R$
was explicitly computed for a $d$ dimensional sphere in Ref. \cite{Torquato}, Eq. (56) 
(in units such that the density is unity).

The variance of ${\cal N}_{\cal D}$ was given to leading order for an arbitrary domain ${\cal D}$ in \cite{Klitch,CalabreseMinchev1},
based on a conjecture of Widom \cite{Widom1,Widom2,Widom3}, for 
free fermions described by the kernel  $K_\mu({\bf x},{\bf y})=\int_{\Gamma(\mu)} \frac{d^d {\bf k}}{(2 \pi)^d} e^{i {\bf k} \cdot ({\bf x}-{\bf y})}$  where $\Gamma(\mu)$ is the Fermi volume. Let $\Omega$ be 
a fixed domain in $\mathbb{R}^d$ and ${\cal D}$ the domain obtained from $\Omega$ by rescaling space by $R$, then for large $R$
\be
{\rm Var} {\cal N}_{\cal D} \simeq \frac{1}{(2 \pi)^{d-1}} \frac{1}{4 \pi^2} 
R^{d-1} \log R \int_{\partial \Omega} d{\bf S}_x \int_{\partial \Gamma(\mu)} d{\bf S}_k
|{\bf n}_k \cdot {\bf n}_x| 
\ee 
where $\partial \Omega$ and $\partial \Gamma(\mu)$ are the boundaries of $\Omega$ 
and of the Fermi volume, and $n_x$ and $n_k$ the respective unit vectors. In our present case upon rescaling we can reduce to an integral over two unit spheres, which can be written as an integral over a single unit sphere
\be \label{widom2} 
{\rm Var} {\cal N}_{\cal D} \simeq \frac{1}{(2 \pi)^{d-1}} \frac{1}{4 \pi^2} 
(k_F R)^{d-1} \log (k_F R)  S_d^2 \frac{ \int_0^\pi d\theta (\sin \theta)^{d-2} |\cos \theta| }{ \int_0^\pi d\theta (\sin \theta)^{d-2}} \simeq \frac{1}{\pi^2 \Gamma(d)} (k_F R)^{d-1} \log (k_F R) 
\ee 
where $k_F^{d-1}$ arises from rescaling of the integral over $\partial \Gamma(\mu)$ and
we added a subleading term $\propto (k_F R)^{d-1} \log k_F$. We see that \eqref{widom2}
agrees with the leading term of our result \eqref{ff}.
Note also the recent work \cite{TanRyu2020} where a higher dimensional bosonisation method was used to recover the leading
order term of ${\rm Var}{\cal N}_{R}$ for free fermions in $d=2$. At this stage however this method 
does not predict the subleading term analytically. The authors of \cite{TanRyu2020} provided a 
numerical determination of this term which, as we checked, is in
excellent agreement with our analytical result Eq.~\eqref{ff}

\bigskip

\subsection{Derivation of the free fermion result from the $d$-dimensional kernel}

Here we provide a direct calculation of the variance of ${\cal N}_R$ for free
fermions (i.e., $V(r)=0$) in the infinite space in any dimension $d$. The exact kernel 
in that case, i.e., the $d$ dimensional analog of the sine kernel 
is given by \cite{Torquato,DeanPLDReview,DeanEPL2015}
\be \label{K_mu_free}
K_{\mu}(\x,\y)=\int_{k<k_{F}}\frac{d^{d}{\bf k}}{(2\pi)^{d}}e^{i{\bf k}\cdot(\x-\y)}=\left(\frac{k_{F}}{2\pi x}\right)^{d/2}J_{d/2}\left(k_{F}|\x-\y|\right)\;.
\ee
where $x = |\x|$ and $k_F=\sqrt{2 \mu}$ is related to the uniform density via $\rho(\x)=K_\mu(\x,\x) = \frac{k_F^d}{2^d \pi^{d/2} \Gamma(1+d/2)}$. The variance of ${\cal N}_R$ is given by
\bea \label{def_var}
{\rm Var} {\cal N}_R = W_1 - W_2 \quad , \quad W_1 = \int_{x < R} d^d\x \, \rho(\x)  \quad , \quad W_2= \int_{x<R} d^d \x \int_{y<R} d^d \y K_\mu(\x,\y)^2 .
\eea
One obtains $W_1= (k_F R)^d\frac{S_d}{2^d \pi^{d/2} d \, \Gamma(1+d/2)}
=
(k_F R)^d  \frac{1}{2^d \Gamma^2(1+d/2)}$ where we used the uniform density given above and the area of the unit sphere 
embedded in $d$ dimensions, $S_d=2 \pi^{d/2}/\Gamma(d/2)$.
The second term, $W_2$, in (\ref{def_var}) can be written as
\be
W_2
= \int_{k<k_F} \frac{d^d {\bf k}}{(2 \pi)^d} \int_{k'<k_F} \frac{d^d {\bf k}'}{(2 \pi)^d} A({\bf k}+{\bf k}') \quad , \quad
A(\p) := \int_{x<R} d^d \x \int_{y<R} d^d \y e^{i \p \cdot (\x-\y)} 
=  \left(\frac{2 \pi R}{p}\right)^{d} J^2_{d/2}(p R)  \, .
\ee 
One can rewrite $W_2$ as 
\be \label{W2_integral}
W_2 = \int d^d \p \, A(\p) B(\p)  ~~ , ~~
B(\p) :=  \int_{k<k_F} \frac{d^d {\bf k}}{(2 \pi)^d} \int_{k'<k_F} \frac{d^d {\bf k}'}{(2 \pi)^d} \delta^d(\p- ({\bf k}+{\bf k}') ) = \int \frac{d^d \z}{(2 \pi)^d} e^{i \z \cdot \p } \left( \frac{k_F}{2\pi z}\right)^{d} J^2_{d/2}(k_F z)
\ee
where we used the integral representation of the delta function over the $\z$ variable. In Eq. (\ref{W2_integral}) the integrals over $\p$ and $\z$ run over ${\mathbb R}^d$.
Hence rescaling $\z \to \z/k_F$ and $\p \to \p/R$ we obtain the following scaling form for the variance
\bea \label{scaling_variance}
{\rm Var} {\cal N}_R = {\cal U}_d(k_F R) \;, \; {\cal U}_d(x) = x^d \left( \frac{1}{2^d \Gamma^2(1+d/2)} - \frac{1}{(2 \pi)^{d}} \int {d^d \z} \int {d^d \p} \, e^{i \frac{\z \cdot \p}{x}} \frac{1}{z^d p^d} J^2_{d/2}(p) J^2_{d/2}(z)\right) 
\eea
where the scaling function generalizes the one obtained in $d=1$ below Eq.~\eqrefMT{\FTWO} in the main text, in the 
following sense
$\lim_{d \to 1}  {\cal U}_d(x) = \frac{1}{\pi^2} (U(0) - U(2 x))$, where $U(0)=c_2$. 

{\bf The case $d=2$}: in this case the double integral (\ref{scaling_variance}) reads
\bea
w_2&=& \frac{1}{(2 \pi)^2}  \int \frac{d^2 z}{|z|^2} J_1^2(|z|)\int_0^\infty \frac{p\,dp}{p^2} \int_0^{2 \pi} d\theta \, e^{i \frac{z p}{x} \cos{\theta}} J_1^2(p) =   (k_F R)^2\int_0^\infty \frac{d z}{z} J_1^2(z)\int_0^\infty \frac{dp}{p} J_0\left( \frac{z p}{k_F R}\right) J_1^2(p) \nn\\
&=& \frac{1}{\pi} \int_0^{2 x} \frac{d z}{z} J_1^2(z) \left[ \cos^{-1}\left( \frac{z}{2 x}\right) - \frac{z}{2 x}\sqrt{1 - \left(\frac{z}{2 x}\right)^2} \, \right] \;.
\eea
This leads to 
\be \label{V2}
{\cal U}_{2}(x)=\frac{x^{2}}{4}\left[\,_{2}F_{3}\left(\left\{ \frac{1}{2},\frac{1}{2}\right\} ,\left\{ 1,1,2\right\} ,-4x^{2}\right)+\frac{x^{2}}{4}\,_{2}F_{3}\left(\left\{ \frac{3}{2},\frac{3}{2}\right\} ,\left\{ 2,3,3\right\} ,-4x^{2}\right)\right]\;.
\ee
This function is plotted in Fig. \ref{Fig:U2_of_x_and_2d_HO} (b) in Section \ref{sec:numerics}. For large $x$ it behaves as ${\cal U}_2(x) = \frac{x}{\pi^2} \left( \ln x + {\gamma_E -2 + 5 \ln 2} + o(1) \right)  $, which agrees with Eq. \eqref{ff} for $d=2$ using that $\psi^{(0)}(3/2) = 2 - 2 \ln 2 - \gamma_E $. Note that the subleading terms are actually of order $O(1/\sqrt{x})$ and rapidly oscillating. 

{\bf The case $d>2$.} In general $d$ the computation is more complicated, and leads to the following expression
\be \label{scaling_gend}
{\cal U}_d(x) = x^d\left( \frac{1}{2^d \Gamma^2(1+d/2)}  - \frac{2^{1-d}}{\Gamma(d/2)} \int_0^1 \frac{du}{u} J^2_{d/2}(2 x u) \left[\frac{1}{\Gamma(1+d/2)} - \frac{2u}{\sqrt{\pi} \Gamma((1+d)/2)}  \,_2 F_1\left(\frac{1}{2}, \frac{1-d}{2}; \frac{3}{2}; u^2 \right) \right]\right)  \;.
\ee
In $d=3$ this integral can be performed explicitly 
{
\bea
\label{eq:U3}
{\cal U}_3(x) &=&\frac{1}{288\pi^{2}}\left\{ \left(12-144x^{2}\right)\text{Ci}(4x)-128x^{3}\text{Si}(4x)+8(8\pi x-9)x^{2}+12\gamma_E \left(12x^{2}-1\right)\right. \nn\\
&& \qquad\quad \left.-\left(32x^{2}+5\right)\cos(4x)-12\log(x)+4x\left[36x\log(4x)+7\sin(4x)\right]+5-24\log(2)\right\} ,
\eea
where $\text{Si}(z)=\int_{0}^{z}\left[\sin\left(t\right)/t\right]dt$ and $\text{Ci}(z)=-\int_{z}^{\infty}\left[\cos\left(t\right)/t\right]dt$ are the sine integral and cosine integral respectively.
For $x\gg1$, 
}
\bea \label{deq3}
{\cal U}_3(x) = \frac{x^2}{2 \pi^2} \left( \ln x + \gamma_E + 2 \ln 2 - \frac{1}{2}\right) - \frac{1}{24 \pi^2} \ln x + \frac{1}{288 \pi^2} (5 - 12 \gamma_E - 24 \ln 2) + o(1) \;.
\eea
${\cal U}_3(x)$ is plotted in Fig. \ref{Fig:U2_of_x_and_2d_HO} (c) in Section \ref{sec:numerics}.
Using $\psi^{(0)}(2) = 1 - \gamma_E$, one can check that the leading order in (\ref{deq3}) indeed coincides with (\ref{ff}) for $d=3$. We have checked the agreement for $d>3$.

For arbitrary dimension, at $k_F R \ll 1,$ $\cal{N}_R$ becomes a Bernoulli random variable, so ${\rm Var}{\cal N}_{R}\simeq\left\langle {\cal N}_{R}\right\rangle -\left\langle {\cal N}_{R}\right\rangle ^{2}$. Indeed, one can check that the corresponding approximate equality $W_2 \simeq W_1 ^ 2$ holds in this limit. This fact is also evident in the $x\ll1$ behavior of the functions ${\cal U}_d(x)$.

\section{General central potential in dimension $d$ and harmonic oscillator}

\label{sec:general} 

We now consider the case of noninteracting fermions in a general central 
potential in $d$ dimension with a single particle Hamiltonian $\hat H=\frac{{\bf p}^2}{2} + V(r)$.
Consider the number ${\cal N}_R$ of fermions in the sphere of radius $R$.
Using the results of section \ref{sec:decoupling}, we study its statistics using the 1d Hamiltonian $\hat H_\ell= - \frac{\partial^2}{\partial r^2} + V_\ell(r)$ with $V_\ell(r) = V(r) +\frac{\alpha (\alpha-1)}{2 r^2}$ with $\alpha = \ell + \frac{d-1}{2}$. We focus here on the large $\mu$ limit and we determine the cumulants of ${\cal N}_R$ using \eqrefMT{\eqsumofcumulantsangularsectors} in the text. In that limit the sum is dominated by values $\ell \gg 1$, hence we will approximate $V_\ell(r) \simeq V(r) +\frac{\ell^2}{2 r^2}$.

In dimension $d$ the bulk density is known to be given as $\rho^{\rm bulk}({\bf x})=\frac{1}{2^d \pi^{d/2} \Gamma(1+d/2)} k_F(r)^d$, where $k_F(r) = \sqrt{2 (\mu - V(r))_+}$. We first assume that (i) $V(r)$ is confining so that the bulk density is supported on the sphere of radius $r_e$, where $r_e$ is the unique root of $V(r_e)=\mu$. (ii) For $\ell>0$, $V_{\ell}(r)$ has either exactly two turning points, i.e., two roots $r^\pm(\ell)$ to the equation $V_\ell(r^\pm(\ell))=\mu$, or none. 
The bulk density of the associated 1d fermion problem, 
$\rho^{\rm bulk}_\ell(r)= \sqrt{2 (\mu - V_\ell(r))_+}/\pi = \sqrt{ ((r k_F(r))^2 - \ell^2)_+}/(\pi r)$ is thus non zero in the interval $r \in [r^-(\ell),r^+(\ell)]$ (where $V_\ell(r) \leq \mu$, equivalently $r k_F(r) \geq \ell$) and vanishes outside (where $V_\ell(r)>\mu$). These assumptions are equivalent to 
asking that the function $r \to r k_F(r)$ vanishes at $r=0$ and for $r \geq r_e$, 
and 
has a unique maximum at some $r=r^*$, see Figure \ref{Fig_kf}.
The equation $r k_F(r)= \ell$ has thus exactly two roots, $r^\pm(\ell)$, for $r < r^*$,
which annihilate at $\ell=\ell^*$.
The harmonic oscillator $V(r) \sim r^2$ satisfies these assumptions. We will
discuss later more general cases. 

\begin{figure}[ht]
\centering
\includegraphics[angle=0,width=0.6\linewidth]{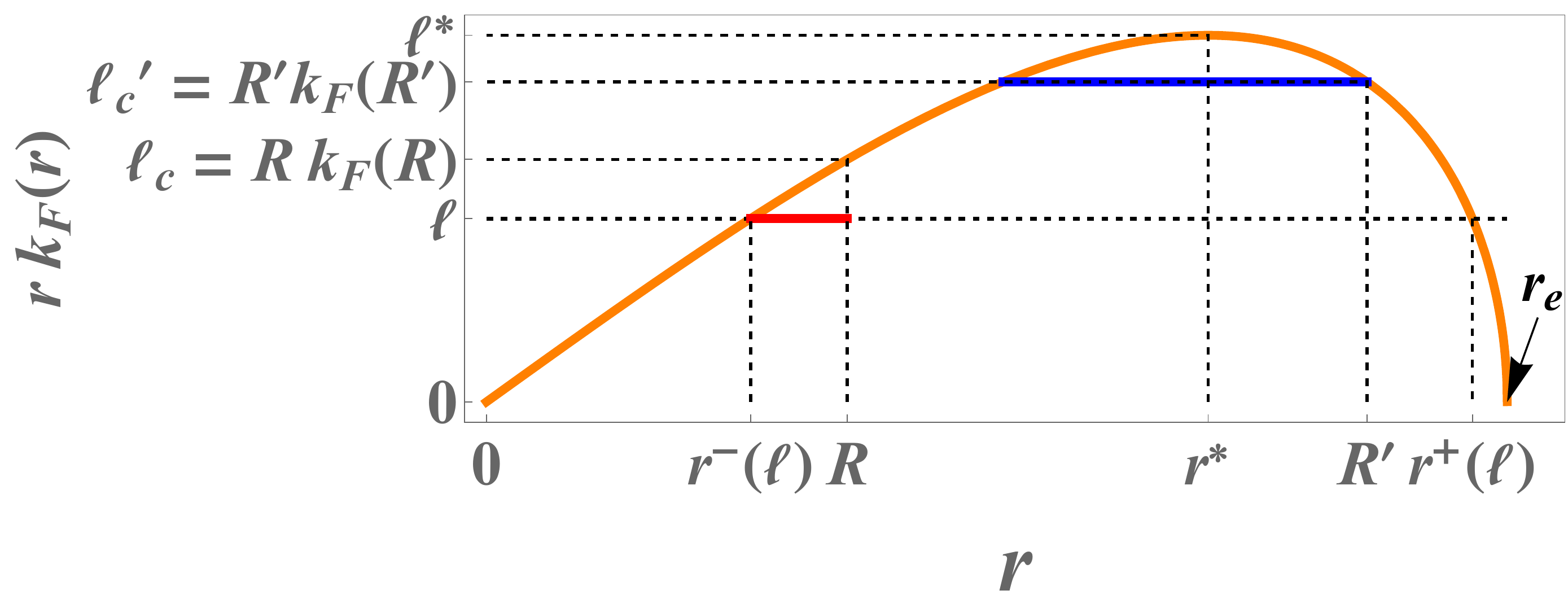}
\caption{{Plot of the function $r \to r k_F(r)=r \sqrt{2 \mu - V(r)}$ (for the HO for illustration). In this case for each $0<\ell<\ell^*$
there are only two semi-classical turning points for the 1d problem associated to $V_\ell(r)$, i.e., the two roots $r^-(\ell)<r^+(\ell)$ of $r k_F(r)=\ell$. The contribution 
{$\langle {\cal N}_{[0,R]}^2 \rangle^c_{\ell}$}
of the sector of angular momentum $\ell$ to
the variance of ${\cal N}_R$ in the sum in Eq. \eqrefMT{\eqsumofcumulantsangularsectors}
in the text is given by the variance of 
{${\cal N}_{[0,R]}$ which includes a macroscopic number of particles, concentrated on the subinterval}
$[r^{-}(\ell),R]$ highlighted in red. 
For $\ell \geq \ell_c = \ell_c = R k_F(R)$ this {sub}interval becomes empty 
hence {$\langle {\cal N}_{[0,R]}^2 \rangle^c_{\ell}\simeq 0$}. Similarly for $R'>r^*$, when 
$\ell \geq \ell'_c = R' k_F(R')$ the interval $[r^{-}(\ell),R]$ shown in blue,
becomes full (i.e., it contains $m_{\ell}$ fermions),
hence the contribution {$\langle {\cal N}_{[0,R]}^2 \rangle^c_{\ell}$} to the variance also vanishes for $\ell \geq \ell'_c$.} 
}
\label{Fig_kf}
\end{figure}

Using \eqrefMT{\eqsumofcumulantsangularsectors} in the text and proceeding as in \eqref{av}, the average number of fermions in a sphere of radius $R$ is obtained as a sum over the bulk densities of the 1d problems, i.e., $\langle {\cal N}_R \rangle \simeq \int_0^{+\infty} d\ell \frac{2 \ell^{d-2}}{\Gamma(d-1)} 
\int_0^R dr \rho^{\rm bulk}_\ell(r)$, where 
the densities $\rho^{\rm bulk}_\ell(r)$ vanish for $\ell > r k_F(r)$. From the remark below 
\eqref{av}, integration over $\ell$ recovers the 
result for $\rho^{\rm bulk}({\bf x})$ given above.

We now calculate the variance ${\rm Var} {\cal N}_R=\langle {\cal N}_R^2 \rangle^c$ using formula \eqrefMT{\eqsumofcumulantsangularsectors} in the text for $p=2$.
For the variance $\langle {\cal N}_{[0,R]}^2 \rangle^c_{\ell}$ of the 1d problem with potential $V_\ell$,  
we use the 
{formula \eqrefMT{\Haa} of the text, 
since for that potential ${\rm Var} {\cal N}_{[0,R[} = H(R,R)$,
and we recall that in this formula $\frac{d\mu}{dN} \simeq \frac{\pi}{\int_{x^-}^{x^+} dz/k_F(z)}$.}
One must substitute
$a \to R$, $x \to r$, $x^\pm \to r^\pm(\ell)$, $k_F(x) \to \sqrt{2(\mu - V_\ell(r))} = 
\frac{1}{r} \sqrt{(r k_F(r))^2 - \ell^2}$, leading to the general formula for the variance
\bea
&& {\rm Var} {\cal N}_R  \simeq \label{2td}  \\
&& \frac{1}{2\pi^{2}}\! \int_{0}^{Rk_{F}(R)}\frac{2\ell^{d-2}d\ell}{\Gamma(d-1)}\left[\log\left(\left(\left(Rk_{F}(R)\right)^{2}-\ell^{2}\right) \! \int_{r^{-}(\ell)}^{r^{+}(\ell)} \! \frac{rdr}{\pi R^{2}\sqrt{r^{2}k_{F}(r)^{2}-\ell^{2}}}\right)+\log2\left|\sin\pi\frac{\int_{r^{-}(\ell)}^{R}\frac{rdr}{\sqrt{r^{2}k_{F}(r)^{2}-\ell^{2}}}}{\int_{r^{-}(\ell)}^{r^{+}(\ell)}\frac{rdr}{\sqrt{r^{2}k_{F}(r)^{2}-\ell^{2}}}}\right|+c_{2}\right] \nn
\eea
where we recall that $k_F(r)=\sqrt{2 (\mu - V(r))}$.
Note that within the $\ell$ sector, the number of fermions ${\cal N}_{[0,R]} \simeq 0$ for $R < r^-(\ell)$ and 
${\cal N}_{[0,R]} \simeq m_\ell$ for $R > r^+(\ell)$, so the 1d variance $\langle {\cal N}^2_{[0,R]} \rangle_\ell^c$ is non zero
only when $R \in [r_-(\ell),r_+(\ell)]$ with 
$\langle {\cal N}^2_{[0,R]} \rangle^c_\ell
\simeq  \langle {\cal N}^2_{[r^-(\ell),R]} \rangle^c_\ell
=\langle {\cal N}^2_{[R,r^+(\ell)]} \rangle^c_\ell
$.
This explains the upper bound in \eqref{2td} 
since when $\ell$ reaches $\ell_c(\mu,R)=R k_F(R)$ then $R$ exits the interval 
$[r_-(\ell),r_+(\ell)]$ (on either sides, see Fig.~\ref{Fig_kf}).

As an example we consider the parametrization $V(r)=\mu v(r/r_e)$ where $v(\tilde r)$ is dimensionless
with $v(1)=1$. One defines $\lambda$ such that $\ell=\frac{1}{2} r_e \sqrt{2\mu} \lambda$ and one has $r^{\pm}(\ell)=r_e \tilde r^\pm(\lambda)$ where $\tilde r^\pm(\lambda)$ are the two roots of 
$\lambda=2 \tilde r \sqrt{1- v(\tilde r)}$. One then obtains in the limit where $k_F r_e= r_e\sqrt{2 \mu} \gg 1$ at
fixed ratio $\tilde R=R/r_e$ (i.e., in the bulk)
\be \label{resv} 
{\rm Var}{\cal N}_{R}\simeq\left(\frac{k_{F}r_{e}}{2}\right)^{d-1}\left[A_{d}\left(\tilde{R}\right)\log\left(\frac{k_{F}r_{e}}{2}\right)+B_{d}\left(\tilde{R}\right)+o(1)\right]\quad,\quad\tilde{R}=R/r_{e}\quad,\quad k_{F}=\sqrt{2\mu}
\ee 
with 
\bea && A_{d}\left(\tilde{R}\right)=\frac{1}{\pi^{2}\Gamma(d)}\left(2\tilde{R}\sqrt{1-v\left(\tilde{R}\right)}\right)^{d-1}\quad,\quad B_{d}\left(\tilde{R}\right)=\int_{0}^{2\tilde{R}\sqrt{1-v\left(\tilde{R}\right)}}\frac{d\lambda\,\lambda^{d-2}}{\pi^{2}\Gamma(d-1)}  \label{ABd} \\
&& \times\left[\log\left(\left(1-v\left(\tilde{R}\right)-\frac{\lambda^{2}}{4\tilde{R}^{2}}\right)\int_{\tilde{r}^{-}(\lambda)}^{\tilde{r}^{+}(\lambda)}\frac{\tilde{r}d\tilde{r}}{\pi\sqrt{\tilde{r}^{2}\left(1-v\left(\tilde{r}\right)\right)-\lambda^{2}/4}}\right)+\log4\left|\sin\pi\frac{\int_{\tilde{r}^{-}(\lambda)}^{\tilde{R}}\frac{\tilde{r}d\tilde{r}}{\sqrt{\tilde{r}^{2}\left(1-v\left(\tilde{r}\right)\right)-\lambda^{2}/4}}}{\int_{\tilde{r}^{-}(\lambda)}^{\tilde{r}^{+}(\lambda)}\frac{\tilde{r}d\tilde{r}}{\sqrt{\tilde{r}^{2}\left(1-v\left(\tilde{r}\right)\right)-\lambda^{2}/4}}}\right|+c_{2}\right] .
\eea
{\it Asymptotics near the edge} For $\tilde R \to 1$, i.e., as $R$ reaches the edge $r_e$, both $A_d$ and $B_d$ vanish. One has $v(\tilde R)\simeq1-v'(1)(1-\tilde R)$, hence 
$A_d(\tilde R) \simeq \frac{2^{d-1}}{\pi^2 \Gamma(d)} (v'(1) (1-\tilde R))^{\frac{d-1}{2}}$. The upper bound in the integral in \eqref{ABd} being small, one writes
$\lambda=2\left(v'(1)(1-\tilde{R})\right)^{1/2} \! u$, with $0<u<1$. In the integral in the numerator in the
last term one can replace the bounds $\int_{\tilde r^-(\lambda)}^{\tilde R} \to \int_{\tilde R}^{r^+(\lambda)}$,
and change variables to $\tilde r=1- (1- \tilde R) y$. The integral in the numerator becomes
$\simeq (\frac{1- \tilde R}{v'(1)})^{1/2} \int_{u^2}^1 \frac{dy}{\sqrt{y-u^2}}$. The argument of the sinus being small one can expand it and one finds that the integral $\int_{\tilde r^-(\lambda)}^{\tilde r^+(\lambda)}$
in the denominator exactly cancels the one in the first logarithm. One arrives at the asymptotics, for $\tilde R \to 1$
\be \label{Basymptgen} 
B_{d}(\tilde{R})\simeq\frac{2^{d-1}}{\pi^{2}\Gamma(d)}\left(v'(1)(1-\tilde{R})\right)^{\frac{d-1}{2}}\left(\frac{3}{2}\log(1-\tilde{R})+\frac{1}{2}\log v'(1)+c_{2}+3\log2+\frac{3}{2}(d-1)\int_{0}^{1}du \, u^{d-2}\log(1-u^{2})\right) \, .
\ee

{\it Harmonic oscillator}. We can now specify to the harmonic oscillator $V(r)=\frac{1}{2} r^2$, and we show that we recover the
result \eqrefMT{\varianceNddimHO} of the main text. One has $r_e=\sqrt{2 \mu}$, 
$v(\tilde r)=\tilde r^2$ and $\tilde r^\pm(\lambda)^2= \frac{1}{2} (1\pm\sqrt{1-\lambda^2})$. 
One thus immediately recovers the result for $A_d$ given in \eqrefMT{\Ad} in the text. To obtain the amplitiude $B_d$ 
we first compute the integral
\be
\int_{\tilde r^-(\lambda)}^{\tilde R} \frac{\tilde r d\tilde r}{\sqrt{\tilde r^2 - \tilde r^4 -\lambda^2/4}}
= \frac{1}{2} \phi(\tilde R) \quad , \quad \phi(\tilde R)= \frac{\pi}{2} - \arctan \frac{1- 2 \tilde R^2}{2 \sqrt{\tilde R^2-\tilde R^4-\lambda^2/4}}
\ee 
which equals $\pi/2$ for $\tilde R=r^+(\lambda)$. Hence the last logarithm in \eqref{ABd} becomes
$\log 4| \sin \phi(R)| = 3 \log 2 + \frac{1}{2} \log (R^2-R^4-\lambda^2/4) - \frac{1}{2} \log(1-\lambda^2)$
and we arrive at 
\be   
B_{d}(\tilde{R})=\int_{0}^{2\tilde{R}\sqrt{1-\tilde{R}^{2}}}\frac{d\lambda \, \lambda^{d-2}}{\pi^{2}\Gamma(d-1)}\left[\log\left(4\tilde{R}\frac{\left(1-\tilde{R}^{2}-\frac{\lambda^{2}}{4\tilde{R}^{2}}\right)^{3/2}}{\left(1-\lambda^{2}\right)^{1/2}}\right)+c_{2}\right] \, .
\ee 
This calculation is equivalent to the one sketched in the text in \eqrefMT{\integrate} 
where we used the result \eqrefMT{\varBessel} in the text for the variance ${\rm Var} {\cal N}^{\rm LUE}_{[0,R]}$ 
associated the 1d inverse square potential, related the LUE.

Integrating over $\lambda$, using the identity 
$(d-1) \int_0^1 d\lambda \, \lambda^{d-2} \log(1- z \lambda^2) = z \Phi \left(z,1,\frac{d+1}{2}\right)+\log (1-z)$,
where $\Phi(z,s,a)$ is the function Lerch transcendant, defined as $\Phi(z,s,a) = \sum_{k \geq 0} \frac{z^k}{(k+a)^s}$ we obtain an explicit expression for $B_d$ valid in any dimension $d \geq 1$ \cite{footnote20}
\bea \label{Bfull} 
 B_d(\tilde R) &=& 
\frac{1}{\pi^{2}\Gamma(d)}\left(2\tilde{R}\sqrt{1-\tilde{R}^{2}}\right)^{d-1}\bigg[\log\left(2\tilde{R}\sqrt{1-\tilde{R}^{2}}\right)+1-\frac{\gamma_{E}}{2}+2\log2-\frac{3}{2}\psi^{(0)}\left(\frac{d+1}{2}\right) \\
&+& \log\left(1-\tilde{R}^{2}\right)-2\tilde{R}^{2}\left(1-\tilde{R}^{2}\right)\Phi\left(4\tilde{R}^{2}\left(1-\tilde{R}^{2}\right),1,\frac{d+1}{2}\right)-\frac{1}{2}\log\left(1-4\tilde{R}^{2}\left(1-\tilde{R}^{2}\right)\right)\bigg]\,. \nonumber 
\eea 
In the limit $\tilde R \to 0$ it recovers the free fermion result \eqref{ff} using $k_F R = 2 \mu \tilde R$.
Specifying the formula  \eqref{Bfull} to $d=2,3$ one finds 
\be 
2\pi^{2}B_{2}\left(x\right)=2x\sqrt{1-x^{2}}\left\{ \log\left[\left(\frac{64x}{1-2x^{2}}\right)^{2}\left(1-x^{2}\right)^{3}\right]+2\gamma_E-2\right\} +\log\left(\frac{\left|1-2x\sqrt{1-x^{2}}\right|}{1+2x\sqrt{1-x^{2}}}\right)
\ee
and
\be 
2\pi^{2}B_{3}(x)=\left(1-2x^{2}\right)^{2}\log\left|1-2x^{2}\right|+4x^{2}\left(1-x^{2}\right)\left\{ \log\left[8x\left(1-x^{2}\right)^{3/2}\right]+\gamma_E\right\}  \, ,
\ee
which are also given in \eqref{B2} and \eqref{B3} in the main text.
The functions $A_d(x)$ and $B_d(x)$ for $d=2,3$ are plotted in Fig.~\ref{Fig_Ad_and_Bd}. 
The function $B_2(x)$ was compared with simulations in Fig.~1 in the text. 
These functions have a non trivial behavior with a maximum and a minimum. Near the edge, for 
$r=r_e$, i.e $x=\tilde R = 1$, the functions $B_2(x)$ and $B_3(x)$ vanish as
\bea
&& B_2(x) = \frac{1}{\pi^2} \sqrt{2-2 x} \left(3 \log (1-x) + 15 \log 2 + 2 \gamma_E  - 4 \right) 
+ O\left((1-x)^{3/2}\right) \, , \\
&& B_3(x) = \frac{2}{\pi^2} (1-x) (3 \log (1-x)+2 \gamma_E -1+ 9 \log 2)
+ O\left((x-1)^{2}\right)
\eea 
in agreement with the general formula \eqref{Basymptgen} using $v'(1)=2$.

\begin{figure}[ht]
\centering
\includegraphics[angle=0,width=0.5\linewidth]{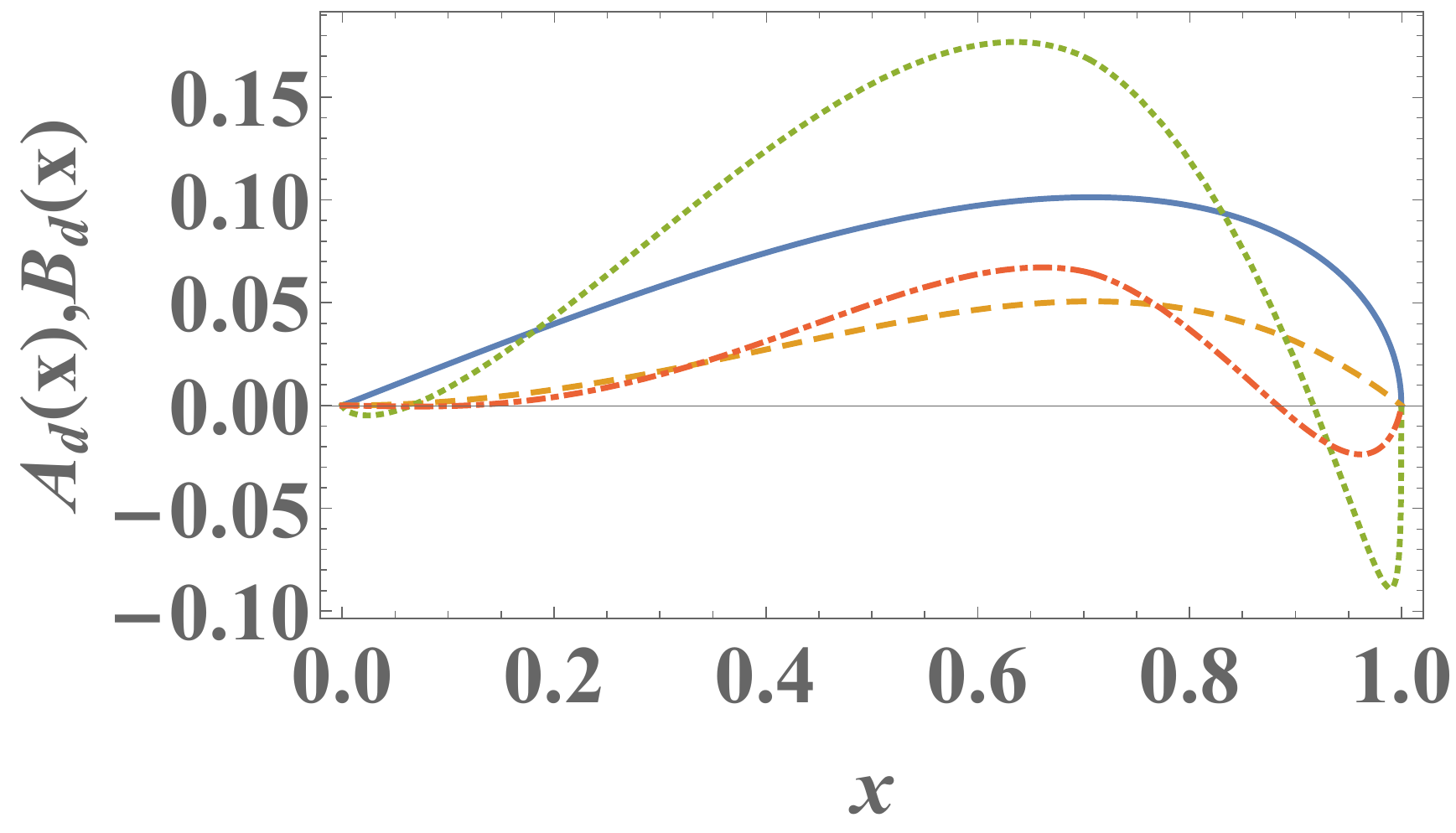}
\caption{The functions $A_{2}\left(x\right),A_{3}\left(x\right),B_{2}\left(x\right),B_{3}\left(x\right)$ (solid, dashed, dotted and dot-dashed lines respectively), that give the leading and subleading order terms of the variance of the number of particles inside a sphere for the harmonic oscillator for $d=2$ and $d=3$, see Eqs.~\eqrefMT{\varianceNddimHO}, \eqrefMT{\Ad}, \eqref{B2} and \eqref{B3} in the main text.
}
\label{Fig_Ad_and_Bd}
\end{figure}

{\bf Remark}. One can similarly calculate the variance of the number of fermions in a spherical shell $R_1 <r <R_2$ from Eq. \eqrefMT{\generalab} in the text upon similar substitutions as described above (e.g. 
$a \to R_1$, $b \to R_2$ and so on).\\


\bigskip

{\bf Other types of potentials}. Suppose now that the function $r \to r k_F(r)= r \sqrt{2 (\mu - V(r))}$ grows monotonically from $0$ to $+\infty$ for $r \in [0,+\infty[$. This is the case e.g. for a bounded and decreasing potential (non-confining).
In that case there is a single turning point for $V_\ell(r)$ for any $\ell$, i.e., a
single root $r^-(\ell)$ to $r k_F(r)=\ell$ and the situation is closer to the one for free fermions
(in which case $k_F(r)=\sqrt{2 \mu}$). Then we can repeat the above derivation using now the 1d formula \eqref{1turning} and
perform the same substitutions as above. Equivalently we can take the formal limit $r^+(\ell) \to +\infty$
in the $d$ dimensional formula \eqref{2td}. This leads to the simpler formula
\be
{\rm Var} {\cal N}_R  \simeq
\frac{1}{2\pi^{2}}\frac{2}{\Gamma(d-1)}\int_{0}^{Rk_{F}(R)}d\ell \, \ell^{d-2}\left\{ \log\left(\left[\left(Rk_{F}(R)\right)^{2}-\ell^{2}\right]\int_{r^{-}(\ell)}^{R}\frac{rdr}{R^{2}\sqrt{r^{2}k_{F}(r)^{2}-\ell^{2}}}\right)+c_{2}+\log2\right\} 
\ee
which reduces to the free fermion result \eqref{ff} when $k_F(r)=\sqrt{2\mu}$. 

There are other possible situations depending on the form of $V(r)$, 
and an exhaustive discussion goes beyond this Letter.
For instance for each $\ell$ there could be various number of roots (i.e., turning points), $r^\alpha(\ell)$ to the equation
$r k_F(r)=\ell$, with multiple interval supports $I_q(\ell)=[r^{(2 q-1)}(\ell),r^{(2 q)}(\ell)]$, $q=1,2,\dots$.
For a given $R$ only the interval containing $R$ contributes to the variance, since the other intervals are either full or empty of fermions.

\bigskip

\section{Edge behavior and matching with the bulk, in $d=1$ and $d>1$}

Here we give some more details about the matching of the number variance formulae
obtained in the bulk, when entering the region of the edge of the Fermi gas.

We start with $d=1$ and the discussion below \eqrefMT{\Haa} in the text, {and we focus on the inverse square potential, $V(x) = \frac{x^2}{2} + \frac{\alpha(\alpha-1)}{2x^2}$ which corresponds to the LUE}. For any smooth confining potential,
as discussed in \cite{DeanPLDReview}, for $x,y$ near the right edge $x^+$ (and similarly for $x^-$)
the kernel takes the universal scaling form 
$K_\mu\left(x,y\right)\simeq\frac{1}{w_{N}}K_{\text{Ai}}\left(\frac{x-x^+}{w_{N}},\frac{y-x^+}{w_{N}}\right)$ where $w_N$ is the width of the edge region 
$w_{N}=\left[2V'\left(x^+\right)\right]^{-1/3}$. Here $K_{\rm Ai}$ is the
Airy kernel given by $K_{\rm Ai}(x,y)=\frac{{\rm Ai}(x) {\rm Ai}'(y)- 
{\rm Ai}'(x) {\rm Ai}(y)}{x-y}$. Using this scaling form one obtains, for an arbitrary smooth potential and
$a$ in the edge region, as was 
done in \cite{MMSV14, MMSV16} for the case of the harmonic oscillator/GUE,
\be \label{edgeV2} 
\text{Var} {\cal N} _{\left[0,a\right]}  =  \int_{a}^{+\infty}dx\int_{0}^{a}dy\,K_{\mu}^{2}\left(x,y\right) 
\simeq \frac{1}{2} {\cal V}_{2} 
\left(\frac{a-x^+}{w_{N}}\right) 
\quad , \quad {\cal V}_2(\hat a) := {2}
\int_{\hat a}^{+\infty}du\int_{-\infty}^{\hat a}dv K_{\text{Ai}}^{2}\left(u,v\right)
\ee 
where the scaling function ${\cal V}_2(\hat a)$, defined in \cite{MMSV14, MMSV16}, 
is universal.
For a general potential the scaling variable is defined as $\hat a = \frac{a-x^+}{w_{N}}$ 
and \eqref{edgeV2} holds for $\hat a = O(1)$. We expect that the limit $\hat a \to - \infty$ of the edge result 
\eqref{edgeV2} should match with the limit $a \to x^+$ of \eqrefMT{\Haa} in the text, as we now check explicitly.
In the limit $a \to x^+$, $a< x^+$, one has $k_F(a) \simeq \sqrt{2(x^+-a) V'(x^+)}$ hence
$\int_a^{x^+} \frac{dz}{k_F(z)} \simeq 
 \frac{\sqrt{2 (x^+-a)}}{V'(x^+)^{1/2}}$. Evaluating the terms in \eqrefMT{\Haa} in the text
 we have $ \frac{1}{d \mu/dN} \sin \theta_a \simeq \frac{\pi}{d \mu/dN}  \frac{ \int_a^{x^+} \frac{dz}{k_F(z)}}{\int_{x^-}^{x^+} \frac{dz}{k_F(z)}} \simeq   \frac{1}{V'(x^+)^{1/2}} \sqrt{2 (x^+-a)}$ which leads to the
 asymptotic behavior for $a \to x^+$ coming from the bulk
\bea \label{asympt2}
\text{Var}{\cal N}_{\left[0,a\right]}\simeq\frac{1}{2\pi^{2}}\left[c_{2}+\log\left(4\left(x^{+}-a\right)^{3/2}\left(2V'\left(x^{+}\right)\right)^{1/2}\right)\right]=\frac{1}{2\pi^{2}}\left[\frac{3}{2}\log(-\hat{a})+c_{2}+2\log2\right]
\eea 
which can thus be nicely recast as a function of the edge scaling variable
$\hat a$. Our result \eqref{asympt2} can be compared with the known results for $\frac{1}{2}
{\cal V}_2(\hat a)$ for $\hat a \to - \infty$. The leading order (the logarithm) agrees with 
the known result for the HO/GUE obtained in \cite{MMSV14,Gustavsson}.
The $O(1)$ constant can further be extracted from the mathematical work of
Bothner and Buckingham using Riemann Hilbert techniques in \cite{BK18} (from the coefficient of $v^2/2$ in the Taylor expansion in $v$ of
their Eq. 1.11 with their $s \equiv \hat a$). Here we have obtained this asymptotics by
a completely different method and both results agree. 
Finally, as stated in the text, one can show that when $a$ is fixed in the bulk and $b$ reaches the edge, the heights at these locations decorelate and $H(a,b) = o(1)$, equivalently from Eq.~\eqrefMT{\defNab} 
in the main text,
$\text{Var}\left(\mathcal{N}_{\left[a,b\right]}\right)\simeq\text{Var}\left(\mathcal{N}_{\left[a,\infty\right[}\right)+\text{Var}\left(\mathcal{N}_{\left[b,\infty\right[}\right)$.\\

For a confining central potential $V(r)$ in dimension $d$, we use again \eqrefMT{\eqsumofcumulantsangularsectors} in the text for $p=2$, substituting the $d=1$ edge 
scaling form discussed above for each angular momentum sector $\ell$
associated to $m_\ell$ fermions in the potential $V_\ell$, hence
\begin{equation} \label{sumvar} 
\left\langle {\cal N}_{R}^{2}\right\rangle ^{c}=\sum_{\ell\geq0}^{\ell_{\max}(\mu)}g_{d}(\ell)\left\langle {\cal N}_{R}^{2}\right\rangle _{\ell}^{c}\simeq\sum_{\ell\geq0}^{\ell_{\max}(\mu)}\frac{2\ell^{d-2}}{\Gamma\left(d-1\right)}\frac{1}{2}{\cal V}_{2}\left(\frac{R-r^{+}(\ell)}{w_{m_{\ell}}}\right)
\end{equation}
where $w_{m_{\ell}}=\left[2V_{\ell}'\left(r^{+}(\ell)\right)\right]^{-1/3}=\left[2V'\left(r^{+}(\ell)\right)-\frac{\ell^{2}}{r^{+}(\ell)^{3}}\right]^{-1/3}$. 
By definition $r^+(\ell)$ is solution of (for $\ell \gg 1$)
\be
V(r^+(\ell)) - V(r_e) + \frac{\ell^2}{2 r^+(\ell)^2} = 0 .
\ee 
One can tentatively expand in powers of $\ell$, recalling that $r^+(\ell \simeq 0)=r_e$
\begin{equation}
r^+(\ell) - r_e = -\frac{\ell^{2}}{2 r_e^2 V'\left(r_e\right)}
- \frac{4 V'(r_e) + r_e V''(r_e)}{ 8 r_e^5 V'(r_e)^3} \ell^4 
+ O(\ell^6)
= - w_N^3 \frac{\ell^{2}}{r_e^2}+ 
O\left(\frac{w_{N}^{6}}{r_{e}^{5}}\ell^{4}\right)
\label{eq0} 
\end{equation}
where we introduced $w_N=(2 V'(r_e))^{-1/3}$ the width of the
edge region in any dimension (for a central potential) \cite{DeanPLDReview}.
It is natural to expect (and consistent as we find below) that the values of $\ell$
which contribute most in the sum in \eqref{sumvar} are such that 
$r^+(\ell) - r_e = O(w_N)$. From the first term in the expansion in \eqref{eq0}, we see that they
are of order $\ell = O(r_e/w_N)$. It is consistent since the subleading term 
$O(\ell^4)$ is then smaller than the $O(\ell^2)$ term by a factor
$w_N/r_e$ (and similarly for the higher orders). Furthermore, one can check that replacing 
$w_{m_\ell}$ by $w_N$ in the scaling variable in \eqref{sumvar}, for $\ell =O(r_e/w_N)$, also
amounts to neglecting subdominant terms. Hence the scaling variable in \eqref{sumvar} 
can be replaced as
$\frac{R-r^{+}(\ell)}{w_{m_{\ell}}}\simeq\frac{1}{w_{N}}\left(R-r_{e}+w_{N}^{3}\frac{\ell^{2}}{r_{e}^{2}}\right)$ and we obtain
\begin{equation} \label{final0} 
\left\langle {\cal N}_{R}^{2}\right\rangle ^{c}\simeq\int_{0}^{\ell_{\max}(\mu)}\frac{2\ell^{d-2}d\ell}{\Gamma\left(d-1\right)}\frac{1}{2}{\cal V}_{2}\left(\hat{R}+w_{N}^{2}\frac{\ell^{2}}{r_{e}^{2}}\right)\simeq\left(\frac{r_{e}}{w_{N}}\right)^{d-1}\int_{0}^{\infty}\frac{d\xi\,\xi^{\frac{d-3}{2}}}{2\Gamma(d-1)}{\cal V}_{2}\left(\hat{R}+\xi\right)\quad,\quad\hat{R}=\frac{R-r_{e}}{w_{N}}
\end{equation}
upon introducing the $O(1)$ variable $\xi = (w_N \ell/r_e)^2$, which is the
formula \eqrefMT{\final} in the text.
Note that we have also obtained
this formula by a direct calculation \cite{TBP}, without using the decoupling in 
\eqrefMT{\eqsumofcumulantsangularsectors} in the text, but using instead 
the exact form of the edge kernel in dimension $d$ obtained in \cite{DeanPLDReview}. 
{The formula \eqref{final0} is compared with numerical simulations in
$d=2$ in Fig. \ref{Fig:U2_of_x_and_2d_HO} in Section \ref{sec:numerics}.}

To see how this formula matches with our result for the bulk in dimension $d$, we now evaluate \eqref{final0} 
in the limit $\hat R \to -\infty$,
inserting the asymptotics \eqref{asympt2} of $\frac{1}{2} {\cal V}_2(\hat a)$ for large negative $\hat a$. 
The integral over $\xi$ in \eqref{final0} can be splitted into a first part $\int_0^{-\hat R} d\xi$ and
a second part $\int_{-\hat R}^{+\infty} d\xi = \int_0^{+\infty} dv$ writing $\xi=-\hat R+v$. Since 
${\cal V}_2(v)$ decays fast at $v \to +\infty$, this second part is $O((-\hat R)^{\frac{d-3}{2}})$
which is subdominant. In the first part we write $\xi=- \hat R u^2$ and obtain for $\hat R \to -\infty$
\be
\label{asympt3} 
\left\langle {\cal N}_{R}^{2}\right\rangle ^{c}\simeq\left(\frac{r_{e}}{w_{N}}\right)^{d-1}\frac{(-\hat{R})^{\frac{d-1}{2}}}{\pi^{2}\Gamma(d)}\left[\frac{3}{2}\log(-\hat{R})+c_{2}+2\log2+\frac{3}{2}(d-1)\int_{0}^{1}duu^{d-2}\log(1-u^{2})\right] \, .
\ee 
Upon using that $-\hat R=(1-\tilde R)r_e/w_N$ with $w_{N}=\left[r_{e}/(2\mu v'(1)\right]^{1/3}$ 
one can check that \eqref{asympt3} agrees with the result \eqref{resv} and the
asymptotics for $A_d$ and $B_d$ given below \eqref{ABd} and in \eqref{Basymptgen}
for a general smooth central potential of the form $V(r)=\mu v(r/r_e)$. This shows
that our results for the variance in the bulk and at the edge match smoothly in any $d$.

{Finally, different universality classes occur in the case of "hard edges".
For instance in $d=1$, the kernels for the hard box near $x=0$ or $x=L$,
or for the inverse square well near $x=x^-$, take universal scaling forms 
at a distance $1/k_F$ near the edge, studied in \cite{Hardwalls}.
We checked that our bulk results match these known cases
in $d=1$, when $a \to 0$.}

\section{Higher cumulants and entropy}

\subsection{Higher cumulants}
\label{sec:cum} 

In $d=1$, for the specific potentials and geometries related to RMT described in Section \ref{sec:special} 
the higher cumulants of the number of fermions ${\cal N_{\cal I}}$ in any interval ${\cal I}$ can be extracted 
from known results in random matrix theory,
using the mapping between the fermion positions $x_i$ and the eigenvalues $\lambda_i$.
In RMT the generating function of the cumulants is computed from Fisher-Hartwig type asymptotics
of Hankel and Toeplitz determinants, using Riemann Hilbert methods
\cite{DIK2009,Charlier_hankel,Charlier1,CharlierJacobi,CharlierSine2019}.
Let us focus on Ref. \cite{CharlierJacobi} where the GUE, LUE and JUE are studied from
which we obtain the results for the HO, inverse square well and the Jacobi box respectively.

Using standard methods for determinantal processes, such as the Cauchy Binet formula, the generating function of the cumulants of ${\cal N_{\cal I}}$ for the HO/GUE with $V(x)=\frac{1}{2} x^2$ (where $x_i = \lambda_i$) is given by the
Hankel determinant
\be \label{GF_GUE}
\left\langle e^{-s{\cal N}_{{\cal I}}}\right\rangle =\left\langle \prod_{i=1}^{N}\left(1-\left(1-e^{-s}\right)\chi_{{\cal I}}(x_{i})\right)\right\rangle =\dfrac{\det_{0\leq k,l\leq N-1}\left(\int_{-\infty}^{\infty}e^{-x^{2}}x^{k+l}\left(1-(1-e^{-s})\chi_{{\cal I}}(x)\right)dx\right)}{\det_{0\leq k,l\leq N-1}\left(\int_{-\infty}^{\infty}e^{-x^{2}}x^{k+l}dx\right)}=\frac{G_{N}(\vec{0},\vec{\beta},V,0)}{G_{N}(\vec{0},\vec{0},V,0)}
\ee  
where $\chi_{\cal I}(x)$ is the indicator function of the interval ${\cal I}$, i.e., $\chi_{{\cal I}}(x)=1$ if $x\in {\cal I}$ and $\chi_{{\cal I}}(x)=0$ otherwise. The last identity connects to the
notations in Ref. \cite{CharlierJacobi}, where $V$ denotes the potential $V(x)=2x^2$ 
in (1.1)-(1.2) there, with $W=0$ and $\psi(x)=2/\pi$, and the charges $\alpha_i$ in (1.6) there are all zero for the GUE.
The choice of the $m$ non zero parameters $\vec \beta$ entering (1.6)-(1.7) depends on the type of
interval ${\cal I}$. For ${\cal I}=[a,b]$ with $a=\tilde a \sqrt{2 N}$ and 
$b=\tilde b \sqrt{2 N}$ (where we are interested in $-1<\tilde a \neq \tilde b <1$ in the bulk) 
the choice is $\vec{\beta}= (\beta_1 = p/(2 i \pi), \beta_2 = - s/(2 i \pi))$, $t_1 = \tilde a$  and $t_2 = \tilde b$, so that from (1.7) there, $\omega(x)=1$ outside the interval $[t_1 = \tilde a,t_2 = \tilde b]$
and $e^{-s}$ inside. For the semi-infinite interval ${\cal I}=[-\infty,a]$
one must choose 
$\vec{\beta}= (\beta_1 = s/(2 i \pi))$ and $t_1=\tilde a$. The case of a collection
of intervals is similarly obtained considering $m \geq 3$. Using the Theorem 1.1 in Ref. \cite{CharlierJacobi} one obtains for
large $N$, up to terms of order $O(\log N/N)$
\be \label{barnes}
\log\left\langle e^{-s{\cal N}_{{\cal I}}}\right\rangle =\sum_{j=1}^{m}\log G\left(1+\beta_{j}\right)G\left(1-\beta_{j}\right)+\dots\quad=\begin{cases}
2\log G\left(1+\frac{is}{2\pi}\right)G\left(1-\frac{is}{2\pi}\right)+\dots\quad, & {\cal I}=[a,b]\\
\log G\left(1+\frac{is}{2\pi}\right)G\left(1-\frac{is}{2\pi}\right)+\dots\quad, & {\cal I}=[-\infty,a]
\end{cases}
\ee 
where $\dots$ denote terms proportional to $\beta_j$ and $\beta_j^2$, i.e., to $s$, $s^2$.  These
terms have been discussed above and reproduce respectively the first and and second cumulant
of ${\cal N}_{\cal I}$. Expanding the Barnes function in \eqref{barnes} in powers of $s$ one obtains the formula
\eqrefMT{\cumulevenGUEgen} of the text for the higher order cumulants of ${\cal N}_{[a,b]}$ for the harmonic oscillator, with $a,b$ in the bulk on macroscopic scales. 
The odd cumulants are zero and the $2n$-th cumulants are given by $\kappa_{2n}$ defined in \eqrefMT{\cumulevenGUEgen} in the text, which, remarkably is independent of $a,b$.
The higher cumulants of ${\cal N}_{[-\infty,a]}$ are smaller by a factor $1/2$. 

For the inverse square potential $V(x) = \frac{x^2}{2} + \frac{\alpha(\alpha-1)}{2 x^2}$ we can use Theorem 1.2 of Ref. \cite{CharlierJacobi} ($G_N$ is now denoted $L_N$) with $V(x)=2(x+1)$ in (1.1)-(1.2) there, with $W=0$ and $\psi(x)=1/\pi$, and the charges $\alpha_i$ in (1.6) all zero except $\alpha_0=\alpha-\frac{1}{2}$. The parameter $\beta_i$ with
$1 \leq i \leq m$ are the same as in the previous case. For the interval ${\cal I}= [2 \sqrt{N} \tilde a,
2 \sqrt{N} \tilde b]$ one has $t_1= 2 \tilde a^2-1$ and $t_2=2 \tilde b^2-1$. One sees that the result \eqref{barnes}, leading to \eqrefMT{\cumulevenGUEgen} of the text also holds for the inverse square potential.
Note that the analog of the semi-infinite interval $]-\infty,a]$ is now $[0,a]$. From Theorem 1.3 of Ref. \cite{CharlierJacobi} 
we see that the same result holds for the Jacobi box potential. Finally the same result holds for the higher cumulants for ${\cal N}_{[a,b]}$ for the CUE, hence for the free fermions on the circle 
for which the generating function can be written as a Toeplitz determinant 
\cite{ForresterFrankel2004, DIK2009,AbanovIvanovQian2011}.

Furthermore the same result holds for microscopic scales, e.g. for an interval $[a,b]$ 
with $|a-b| \ll 1$, in the limit where $|a-b| k_F(a)$ is fixed but large, see Theorem 
1.1 of Ref. \cite{CharlierSine2019} as is shown from similar analysis using the sine-kernel.
This leads to the conjecture given in the text
that it arises from microscopic scales and holds for an arbitrary smooth potential $V(x)$. \\

In space dimension $d>1$, for a central potential $V(r)$ and a spherical domain ${\cal D}$
of radius $R$, the formula 
\eqrefMT{\eqsumofcumulantsangularsectors} of the text determines the $p$-th order cumulant 
of ${\cal N}_{{\cal D}}$ from the knowledge of the $p$-th order cumulant
of ${\cal N}_{[0,R]}$ for the set of 1d potentials $V_\ell(r)= V(r) + \frac{\alpha(\alpha-1)}{2 r^2}$ 
on $[0,r]$ with $\alpha=\ell + \frac{d-1}{2}$, where $\ell$ labels the angular momentum sector.
In the large $\mu$ limit, we know from Section \ref{sec:free} that this sum is dominated 
by values of $\ell \gg 1$, such that $g_d(\ell) \simeq \frac{2 \ell^{d-2}}{\Gamma(d-1)}$. 

Let us start with free fermions $V(r)=0$. In that case one needs the higher cumulants for
$V_\ell(r) \simeq \frac{\ell^2}{2 r^2}$ for $\ell \gg 1$. These are known from a very recent
work in mathematics \cite{Charlier3} on the Bessel process, which yields again the same formula \eqref{barnes}, recalling that $[0,R]$ corresponds to the second line in this formula, hence
leading to $1/2$ times the result in \eqrefMT{\cumulevenGUEgen} of the text. Using 
\eqrefMT{\eqsumofcumulantsangularsectors} in the text we obtain, for $k_F R \gg 1$ and $n \geq 2$
\be \label{aga} 
\left\langle {\cal N}_{R}^{2n}\right\rangle ^{c}\simeq\frac{2}{\Gamma(d-1)}\int_{0}^{k_{F}R}d\ell\,\ell^{d-2}\times\frac{1}{2}\kappa_{2n}=\frac{(k_{F}R)^{d-1}}{\Gamma(d)}\left(\kappa_{2n}+o(1)\right)\quad,\quad k_{F}=\sqrt{2\mu}
\ee
where the integration upper bound comes from the constraint 
$V_\ell(r) \simeq \frac{\ell^2}{2 r^2} \leq \mu$. Note that since the result
of \cite{Charlier3} is rigorous, our formula for the higher cumulants for
free fermions is expected to be exact independently of the conjecture stated
above and in the main text.

For the harmonic oscillator $V(r)=\frac{1}{2} r^2$, it turns out that the result for the higher cumulants discussed above for the LUE holds also
for large $\ell = O(\mu)$ \cite{thank_Charlier}. Therefore, in this case, we arrive at the following formula
%
\be \label{cumV} 
\left\langle {\cal N}_{R}^{2n}\right\rangle ^{c}=\frac{\left(k_{F}(R)R\right)^{d-1}}{\Gamma(d)}\left(\kappa_{2n}+o(1)\right)\quad,\quad k_{F}(R)=\sqrt{2\left(\mu-V(R)\right)}
\ee
since the upper bound in the integral in \eqref{aga} is now $\ell=\ell_c(\mu,R) \simeq
k_F(R)R$, from the constraint that $V_\ell(r) \simeq V(r) + \frac{\ell^2}{2 r^2} \leq \mu$. 
In the text below Eq. \eqrefMT{\cumulfree} in the main text, we conjecture that this result in Eq. \eqref{cumV}, 
which is close to being rigorous for $V(r) = \frac{1}{2}r^2$,
holds for a more general smooth central potential.


\subsection{Entanglement entropy} 

We now apply our results to the calculation of the bipartite R\'enyi entanglement entropy of a $d$-dimensional domain ${\cal D}$ with its complement $\overline{\cal D}$. It is defined for $q \geq 0$ as
$S_q({\cal D})=\frac{1}{1-q}\ln\Tr[\rho_{\cal D}^q]$, where
$\rho_{\cal D}=\Tr_{\overline{\cal D}}[\rho]$ is obtained by tracing out 
the density matrix $\rho$ of the system over $\overline{\cal D}$. For noninteracting fermions \cite{Kli06,KL09,CalabreseMinchev2,Hur11} the entropy can be expressed in any dimension $d$ as a series involving the cumulants of ${\cal N}_{\cal D}$
 \bea \label{entropymoments} 
S_{q}({\cal D})=\frac{\pi^{2}}{6}\left(1+\frac{1}{q}\right){\rm Var}{\cal N}_{{\cal D}}+\sum_{n\geq2}s_{n}^{(q)}\langle{\cal N}_{{\cal D}}^{2n}\rangle^{c}\quad,\quad s_{n}^{(q)}=\frac{(-1)^{n}(2\pi)^{2n}2\zeta(-2n,\frac{1+q}{2})}{(q-1)q^{2n}(2n)!}
\eea
where the coefficient $s_n^{(q)}$ are given in \cite{CalabreseMinchev2}
and $\zeta(s,a)=\sum_{k=0}^\infty (k+a)^{-s}$ is the generalized Riemann zeta function. 
An immediate consequence of this property, together with \eqrefMT{\eqsumofcumulantsangularsectors} of the text, is that, for any central 
potential $V(r)$ and any domain ${\cal D}$ with spherical symmetry in dimension $d$, the entropy
can be written as a sum of entropies of corresponding one-dimensional domains associated to 
an angular momentum sector. For instance, for ${\cal D}={\cal B}_R$ the sphere of radius $R$ centered at 
the origin, one has 
\be \label{entropysum} 
S_q({\cal D}) =  \sum_{\ell=0}^{\ell_{\max}(\mu)} g_d(\ell)  S^\ell_q([0,R]) 
\ee 
where $S^\ell_q([0,R])$ is the bipartite entanglement entropy of the interval
$[0,R]$ for the 1d problem with potential $V_\ell(r)$ for $r \in \mathbb{R}^+$.
Although \eqref{entropysum} can be useful, we will instead directly substitute our result \eqref{cumV} 
for the cumulants in dimension $d$ into \eqref{entropymoments}, leading to
\bea \label{entropyBR} 
S_{q}({\cal B}_{R})=\frac{\pi^{2}}{6}\left(1+\frac{1}{q}\right){\rm Var}{\cal N}_{R}+\frac{(k_{F}(R)R)^{d-1}}{\Gamma(d)}\sum_{n\geq2}s_{n}^{(q)}(\kappa_{2n}+o(1))
\eea
which is valid for any central potential $V(r)$, with $k_F(R)=\sqrt{2 (\mu - V(R))}$,
in the limit where $R k_F(R) \gg 1$. 
The analogous formula in $d=1$ for the interval $[a,b]$ and a smooth potential $V(x)$ is, in the limit of large $\mu$
\bea
S_{q}\left([a,b]\right)=\frac{\pi^{2}}{6}\left(1+\frac{1}{q}\right){\rm Var}{\cal N}_{[a,b]}+\sum_{n\geq2}s_{n}^{(q)}(\kappa_{2n}+o(1)) \, . 
\eea
Note that in the case of a semi-infinite interval $[c,a]$ with $c=-\infty$, $c=0$, for fermions
on $\mathbb{R}$ and $\mathbb{R}^+$ respectively, there is a factor $1/2$ in front 
of the last sum. 

To evaluate the sum $\sum_{n \geq 2} s_n^{(q)} \kappa_{2n}$, we take advantage of the fact that it is independent both of the potential and the dimension $d$, and is already known
for free fermions in $d=1$. In $d=1$ it is well known that the entanglement entropy of free fermions
for an interval ${\cal D}=[a,b]$ is given by (for $k_F |a-b| \gg 1$) 
\be
S_{q}^{{\rm ff}}([a,b])=\frac{q+1}{6q}\log\left(2k_{F}|a-b|\right)+E_{q}+o(1)~,~E_{q}=\frac{q+1}{q} \! \int_{0}^{+\infty}\frac{dt}{t}\left[\frac{1}{1-q^{-2}}\left(\frac{1}{q\sinh(t/q)}-\frac{1}{\sinh t}\right)\frac{1}{\sinh t}-\frac{e^{-2t}}{6}\right]
\ee 
with $k_F=\sqrt{2 \mu}$ and 
where $E_q$ was obtained in \cite{JinKorepin2004}, see also Eq. (11) in \cite{CalabreseEntropyFreeFermions}. 
Since we also know the variance
of ${\cal N}_{[a,b]}$ for free fermions, ${\rm Var} {\cal N}_{[a,b]} = \frac{1}{\pi^2} (\log k_F |a-b| + c_2 + o(1))$,
(see the text), we obtain that $\sum_{n \geq 2} s_n^{(q)} \kappa_{2n} = \tilde E_q=E_q -  \frac{q+1}{6 q} (1 + \gamma_E)$, using that $c_2=\log 2 + 1 + \gamma_E$. This leads to the 
equation 
\eqrefMT{\entropy} given in the text
\be \label{entropy3}
S_{q}({\cal D})=\frac{\pi^{2}}{6}\frac{q+1}{q}{\rm Var}{\cal N}_{{\cal D}}+\eta_{d}\frac{\left(k_{F}(R)R\right)^{d-1}}{\Gamma(d)}\left(\tilde{E}_{q}+o(1)\right)
\ee 
for $k_F(R) R \gg 1$,
where we introduced the parameter $\eta_d$ to discuss the various cases. For a central potential
in dimension $d>1$ and ${\cal D}= {\cal B}_R$ the sphere of radius $R$, Eq. \eqref{entropy3} holds
with $\eta_d=1$, as a consequence of \eqref{entropyBR}. 
For a spherical shell $r \in [R_1,R_2]$, \eqref{entropy3} can be generalized by replacing
$\left(k_{F}\left(R\right)R\right)^{d-1}\to2\left(k_{F}\left(R_{1}\right)R_{1}\right)^{d-1}+\left[\left(k_{F}\left(R_{2}\right)R_{2}\right)^{d-1}-\left(k_{F}\left(R_{1}\right)R_{1}\right)^{d-1}\right]$
(i.e., using $\eta_d=2$ and $\eta_d=1$ for different $\ell$). 
In dimension $d=1$, if ${\cal D}=[a,b]$ where
$a,b$ are both in the bulk, the Eq. \eqref{entropy3} holds in the limit of large $\mu$, with $\eta_d=1$, setting
$d=1$ in the equation, i.e., replacing
$\frac{(k_F(R) R)^{d-1} }{\Gamma(d)} \to 1$ in Eq. \eqref{entropy3}. For the
semi-infinite interval $[c,a]$ with $c=-\infty$, $c=0$, for fermions
on $\mathbb{R}$ and $\mathbb{R}^+$ respectively, one has $\eta_d=\frac{1}{2}$.\\

{\bf Remark}: for $q=1$ there is a simpler formula which relates the 
generating function of the cumulants and the entanglement entropy \cite{Caux2019},
i.e.,  $S_{1}({\cal D})=\frac{1}{4}\int_{-\infty}^{\infty}\frac{\ln\left\langle e^{-s{\cal N}_{{\cal D}}}\right\rangle }{\sinh^{2}(s/2)}\,ds$. This allows to obtain the previous results without considering explicitly the sum over the cumulants, using the results for the generation function, see Eq. \eqref{barnes}, and
the identity $E_1= \frac{1}{2} \int_{-\infty}^\infty \ln \left[G(1+\frac{p}{2 i \pi}) G(1 - \frac{p}{2 i \pi}) \right] \frac{1}{\sinh^2(p/2)} \, dp=0.495018 \ldots$.\\

Our main result for the entanglement entropy is thus Eq. \eqref{entropy3}. Combined with
the explicit expressions for the variance that we have obtained in this paper, Eq. \eqref{entropy3} provides
explicit expressions for the entropy in the limit of large $\mu$, or large $k_F(R) R$, in the following cases. 

In $d=1$ for the potentials $V(x)$ related to RMT, the HO, the inverse square well, the 
Jacobi box, the formula \eqref{entropy3} is based only on the available rigorous results for all cumulants, see Section \ref{sec:cum}, and thus does not rely on any conjecture. 
The leading term $O(\log \mu)$ was known in previous works for the HO, and some formula were proposed for the $O(1)$ subleading term \cite{CalabresePLDEntropy,DubailStephanVitiCalabrese2017,V12}.
Our general formula leads to the following result for the entanglement entropy for the
HO, $V(x)=\frac{1}{2} x^2$, for any interval $[a,b]$ in the bulk in the large $N$ limit
\be \label{eq17_1}
S_{q}\left(\left[\tilde{a}\sqrt{2N},\tilde{b}\sqrt{2N}\right]\right)=
 \frac{q+1}{6 q} 
\left[\ln \mu + \frac{3}{4} \ln(1-\tilde a^2)(1-\tilde b^2) +  \ln \frac{4|\tilde a - \tilde b|}{1 - \tilde a \tilde b + \sqrt{(1-\tilde a^2)(1-\tilde b^2)}} + c_2 \right] + \tilde E_q  + o(1)
\ee
with $\mu=N$, and we recall that $\tilde E_q = E_q - \frac{(q+1)(1 +\gamma_E)}{6 q}$. 
Note that this formula is quite similar to Eq. (18) in \cite{DubailStephanVitiCalabrese2017}
which however is only valid up to a $O(1)$ constant (independent of $\tilde a,\tilde b$),
which is obtained here. 
We can obtain similar results for the semi-infinite interval of the HO, and for the inverse square well. In the case of the hard 
box, i.e., $V(x) = 0$ for $x \in [0,L]$ and Dirichlet boundary conditions in $x=0, L$ (which is related to the JUE), one can use the the formula \eqref{entropy3} together with the result obtained for the variance in Eq. \eqref{hardboxresults} to compute the entanglement entropy of the segment $[0,a]$, in which case $\eta_d = 1/2$ as explained below Eq. \eqref{entropy3}. One obtains, for large $\mu$ (equivalently for large $N$)
\bea \label{entropy_box}
S_q\left([0,a]\right) = \frac{1}{12} \frac{q+1}{q} \ln \left[ 4 N \sin \left(\frac{\pi a}{L} \right)\right] + \frac{E_q}{2} + o(1) \;,
\eea
which coincides with the known result (see Eq. (65) in Ref.~\cite{CalabreseEntropyFreeFermions}).

In $d=1$ for an arbitrary smooth potential $V(x)$ the formula \eqref{entropy3} 
can be combined with our results \eqrefMT{\generalab} and \eqrefMT{\Haa} in the text for the variance to obtain the 
explicit expression of the entanglement entropy for an interval $[a,b]$ in the
bulk, and a semi infinite interval. It relies on our conjecture of the universality of the higher order 
cumulants. In Ref. \cite{DubailStephanVitiCalabrese2017} some formula were also proposed
based on a very different approach using field theory and did not attempt to determine the
$O(1)$ terms. 

In $d>1$ our formula \eqref{entropy3} for a spherical domain is exact for free fermions
based on rigorous results in the math literature, 
see Section \ref{sec:cum}, and thus does not rely on any conjecture. 
The leading term $S_q({\cal D})  \propto R^{d-1} \log R$ at large $R$
agrees with the Widom conjecture for a spherical domain \cite{Widom1,Widom2,Widom3,Klitch,CalabreseMinchev1},
proved in \cite{ProofKlitch}. In addition, we obtain here the first
correction $O(R^{d-1})$. 

In $d>1$ for a general central potential our formula \eqref{entropy3} for a spherical domain
using our results for the variance, displayed in Section \ref{sec:general}, and 
\eqrefMT{\varianceNddimHO} and \eqrefMT{\Ad} in the text for the HO, 
is completely new. It has the same validity as formula
\eqref{cumV} for the higher cumulants, i.e., it relies on two very natural conjectures
detailed in Section \ref{sec:cum}.

\section{Numerical calculations}
\label{sec:numerics} 

In this Section we briefly discuss some of the methods to compute numerically the number variances such as those displayed in Fig.~1 of the main text and in the figures below.
The first one amounts to evaluate numerically the double integral \eqref{HK} that gives the height
field and the variance from \eqref{HH1}, which however is delicate because of the fast oscillations of the integrand.
An alternative method, which we found to be more efficient in most cases, is to generate a set of realizations of the positions of the fermions drawn from the quantum JPDF $|\Psi_0|^2$, and compute the empirical variance from it. For some particular potentials, such realizations can be generated very efficiently using the known connections to RMT, see Section \ref{sec:special}. This is the method that we used 
to generate all of the numerical data presented in this work \cite{footnote13},
{except for the hard-box potential where a numerical evaluation of the integral \eqref{HK} was performed.}

Consider the HO. In $d=1$, we generate the eigenvalues of a random $N\times N$ GUE matrix.
In $d>1$, to compute the variance of the number of particles ${\cal N}_R$ inside a sphere of radius $R$, we generate realizations of the radial coordinates $r_1, \dots, r_N$. By exploiting the decoupling properties
between the different angular sectors discussed in Section \ref{sec:decoupling}, we generate samples 
of radial coordinates within each angular sector separately. This amounts to generate fermion positions in the effective $1d$ potential $V_\ell(r)= \frac{r^2}{2} + \frac{(\ell + \frac{d-3}{2})(\ell + \frac{d-1}{2})}{2 r^2}$.
This corresponds to the inverse square well with $\alpha= \ell + \frac{d-1}{2}$.
This is conveniently done by exploiting the mapping to the eigenvalues of a random matrix from the
LUE matrix (as described in Section \ref{sec:special}). We used the tridiagonal matrix representations of GUE and LUE matrices \citep{Dumitriu2002} in order to efficiently generate their eigenvalues.


We start by testing our numerical methods on two cases related to GUE and LUE which were studied previously in \cite{MMSV14,MMSV16}. 
In Fig.~\ref{Fig_1d_HO} we give our results for $\text{Var}\left(\mathcal{N}_{\left[-a,a\right]}\right)$ and $\text{Var}\left(\mathcal{N}_{\left[a,\infty\right[}\right)$ for $N=100$ fermions in a harmonic trap in $d=1$. Each of the data points was generated by simulating $5\times {10}^4$ GUE matrices. 
Similarly, in Fig.~\ref{Fig:hard_box_and_LUE} (a) we plot results for $\text{Var}\left(\mathcal{N}_{\left[0,a\right]}\right)$ for $N=100$ fermions in an inverse square potential with $\alpha=5/2$, where each of the data points was generated by simulating $5\times {10}^4$ LUE matrices with $N=100$ and $M=102$.
Finally, in Fig.~\ref{Fig:hard_box_and_LUE} (b) we plot results for $\text{Var}\left(\mathcal{N}_{\left[0,a\right]}\right)$ for $N=100$ fermions in a hard-box potential on the interval $[0,L]$, where a numerical evaluation of the integral \eqref{HK} was performed, using the explicitly known kernel \citep{Hardwalls}.
In all cases,
the data shows excellent agreement with our theoretical predictions, as described in the figures' captions.

\begin{figure}[ht]
\centering
\includegraphics[angle=0,width=0.49\linewidth]{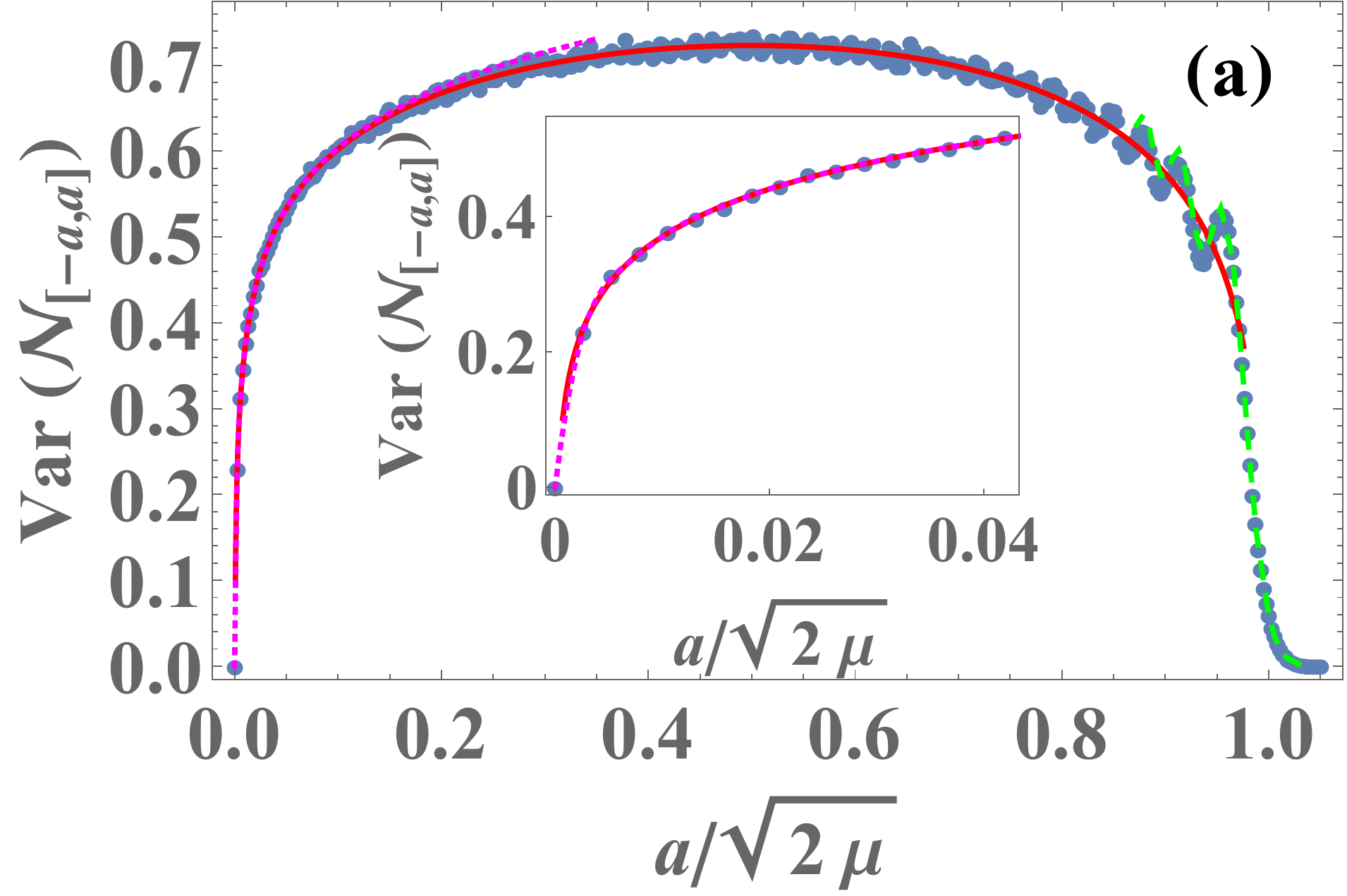}
\includegraphics[angle=0,width=0.48\linewidth]{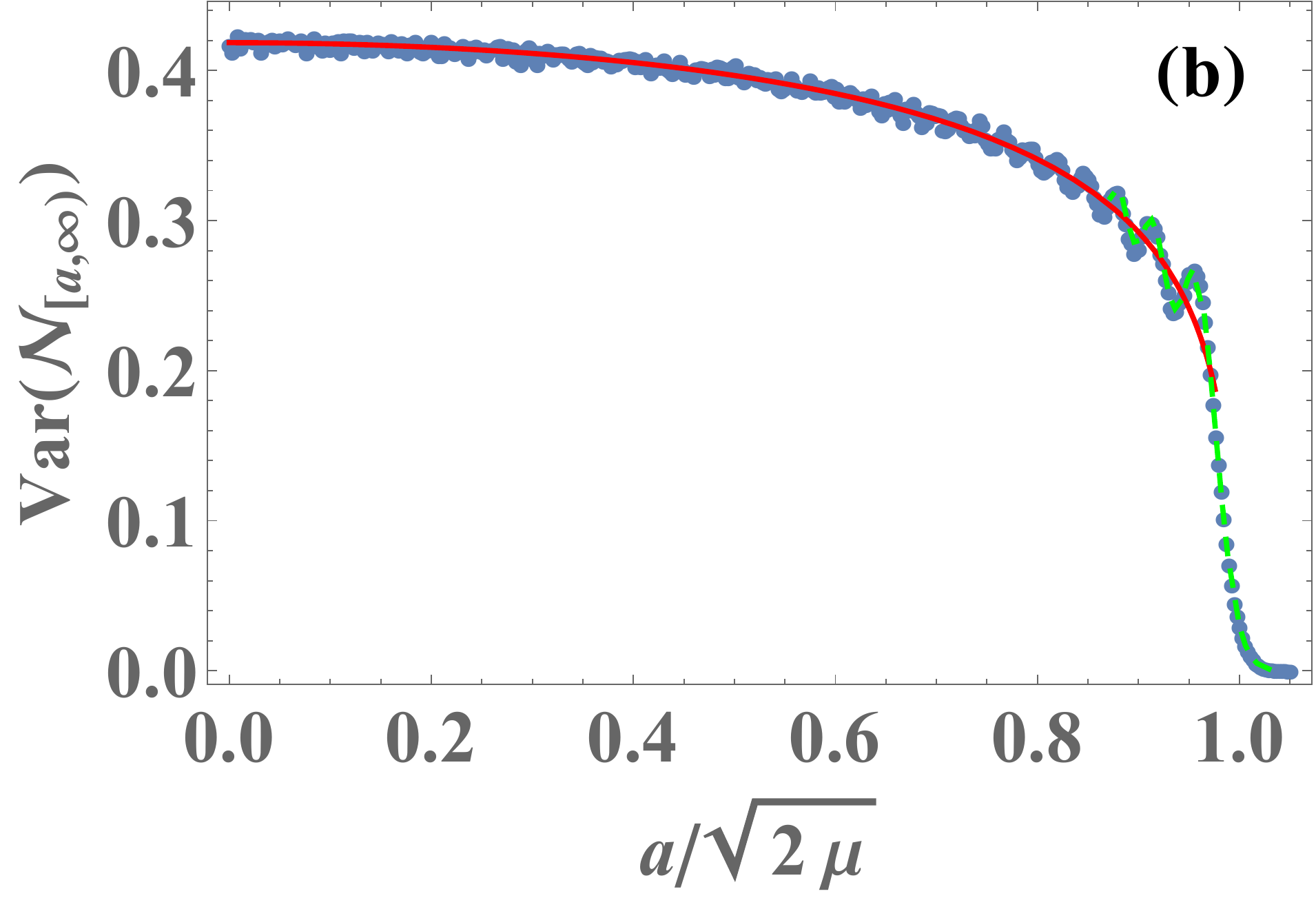}
\caption{Variance of $\mathcal{N}_{\left[-a,a\right]}$ (a) and of $\mathcal{N}_{[a,\infty)}$ (b) for the $1d$ harmonic oscillator
$V(x) = x^2 / 2$, plotted vs $\tilde a = a/\sqrt{2 \mu}$ for $\mu = 100$ (we recall that $\mu \simeq N$). 
The simulations (symbols) show excellent agreement with our predictions: In the bulk, with 
Eq.~\eqref{HOHO} in (a) and Eq.~\eqref{aaa} in (b), and near the edge $\tilde a=1$, with the scaling form \eqref{edgeV2}.
The dotted line in (a) (see the {\bf inset} for a zoom on this region) is the prediction for the variance for a 
microscopic interval, $\pi^2 {\rm Var} {\cal N}_{[-a,a]} \simeq 
U(0)-U(2 a k_F(0))$, as described around Eq.~\eqrefMT{\FTWO} in the main text: it is expected to
be valid in the regime $a \sim 1/\sqrt{\mu}$, i.e., $\tilde a \sim 1/\mu \ll 1$.
}
\label{Fig_1d_HO}
\end{figure}

We now give some additional details for the HO in $d=2$, $V(r)=\frac{1}{2} r^2$, with $\mu=100$, corresponding to $N=\mu\left(\mu+1\right)/2=5050$. The numerical data that is plotted in Fig.~1 of the main text was computed over $2\times {10}^6$ realizations of sets $r_1, \dots , r_N$ of the fermions' radial coordinates using the method that we described above. 
{For clarity, we display in Fig.~\ref{Fig:U2_of_x_and_2d_HO} (a) a close-up of the edge regime $\tilde{R}\simeq1$ of Fig.~1 of the main text, showing excellent agreement between the numerical data and the theoretical edge prediction \eqrefMT{\final} in the text.}
Although we did not report it in Fig.~1 in the text we find excellent agreement for small values
of $R$ with the free fermion result \eqref{ff}.

\begin{figure}[ht]
\centering
\includegraphics[angle=0,width=0.47\linewidth]{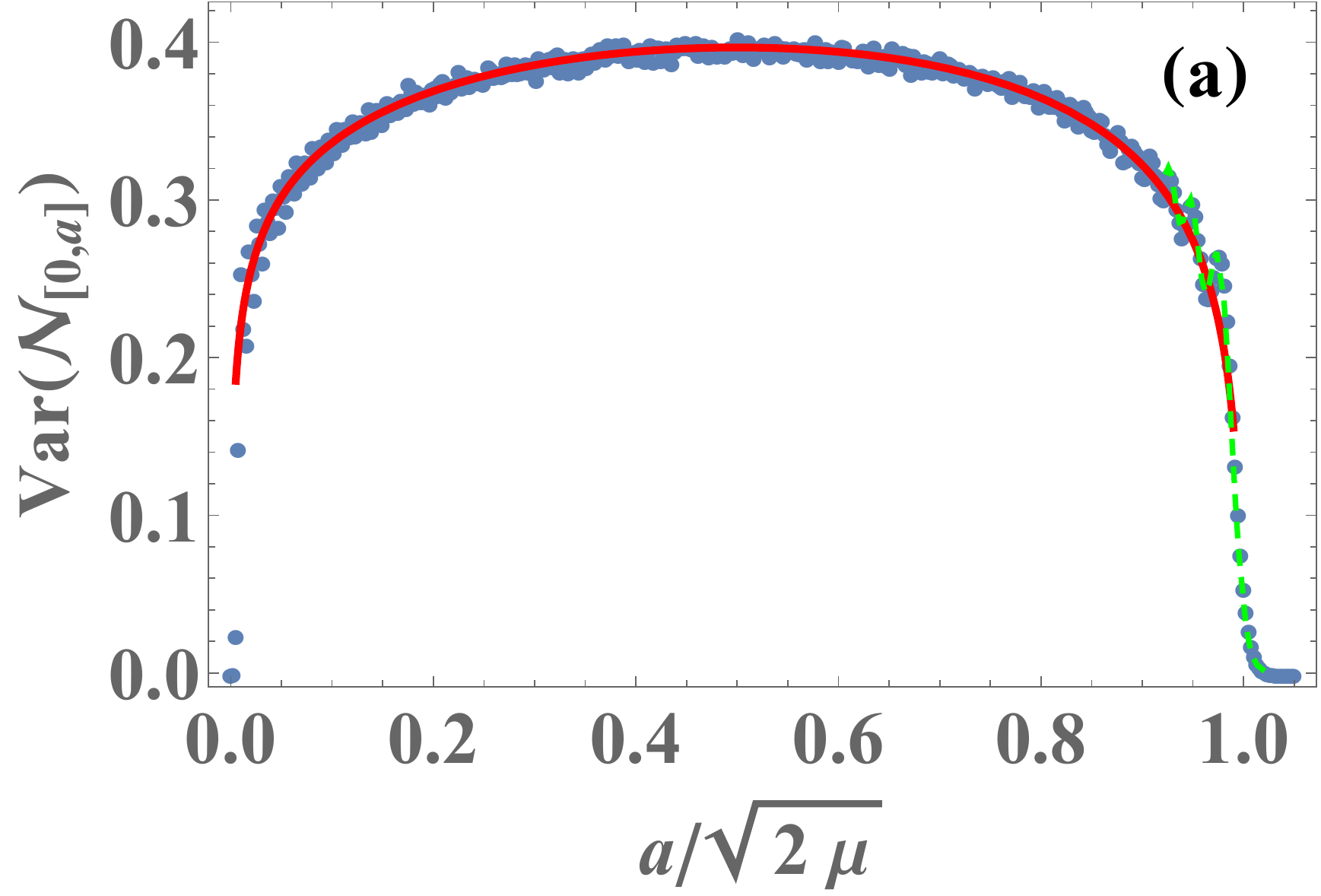}
\includegraphics[angle=0,width=0.48\linewidth]{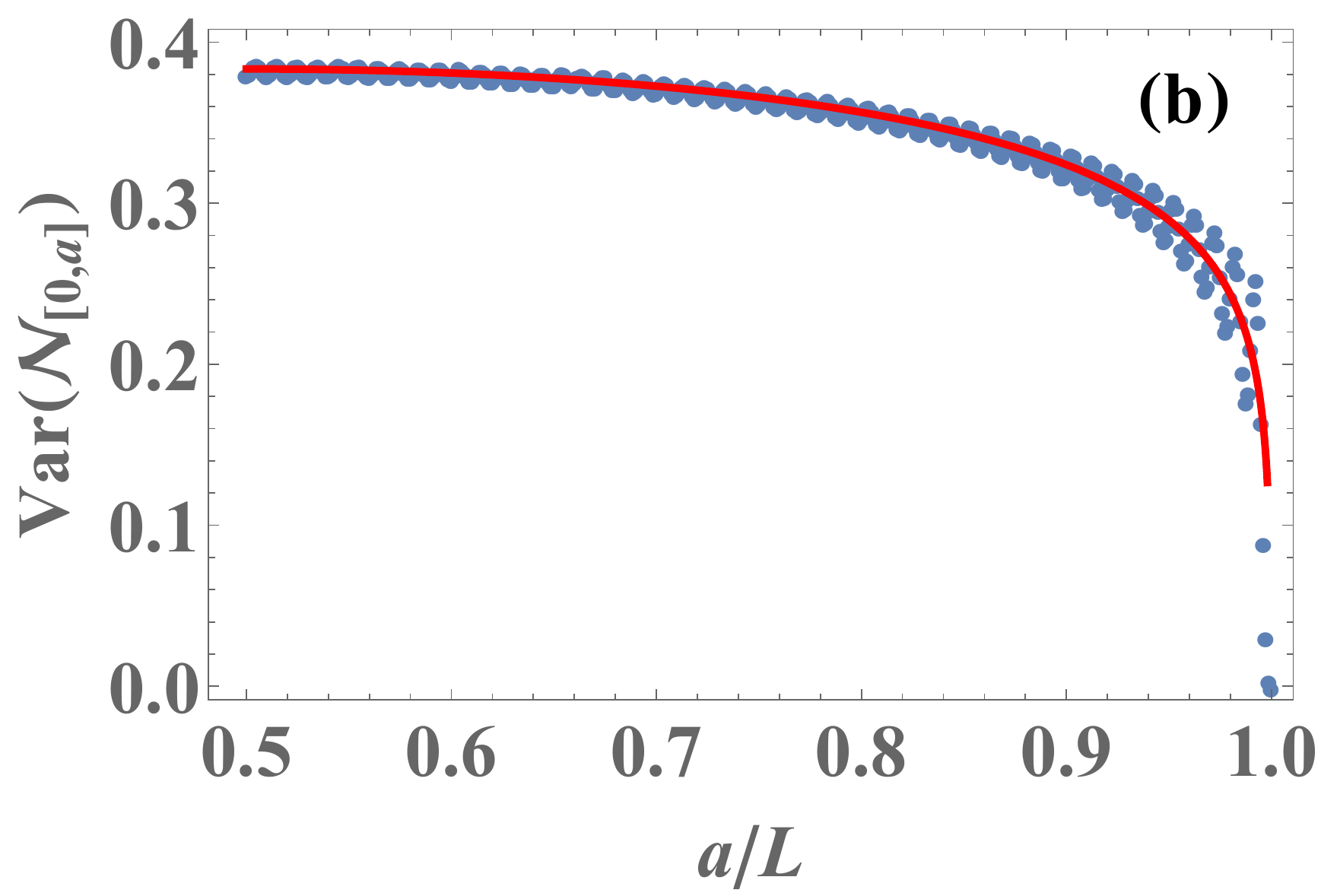}
\caption{Variance of $\mathcal{N}_{\left[0,a\right]}$ for $N=100$ fermions in (a) the inverse square potential $V(x) = \frac{x^2}{2} + \frac{\alpha(\alpha-1)}{2x^2}$ with $\alpha = 5/2$, and (b) the hard-box potential on the interval $[0,L]$.
The symbols, corresponding to numerical simulations in (a) and a numerical evaluations of the integral \eqref{HK} in (b), show excellent agreement with our predictions: for (a), with equation \eqrefMT{\varBessel} in the main text
with $\lambda = 0$ and $\tilde a=a/\sqrt{2 \mu}$ in the bulk, and the scaling form \eqref{edgeV2} near the edge, 
 and for (b), with Eq.~\eqref{hardboxresults}.
}
\label{Fig:hard_box_and_LUE}
\end{figure}

\begin{figure}[ht]
\centering
\includegraphics[angle=0,width=0.34\linewidth]{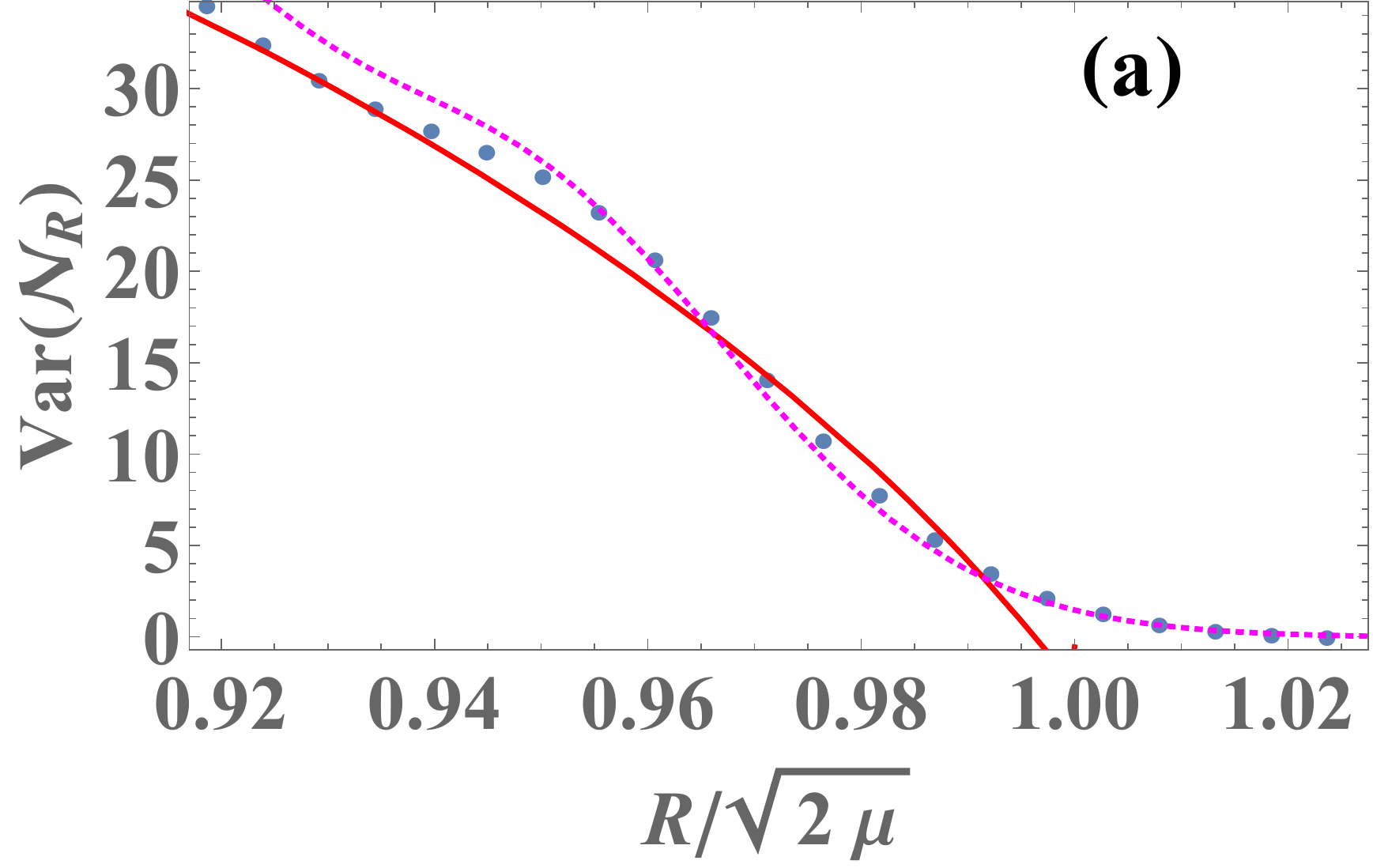}
\includegraphics[angle=0,width=0.32\linewidth]{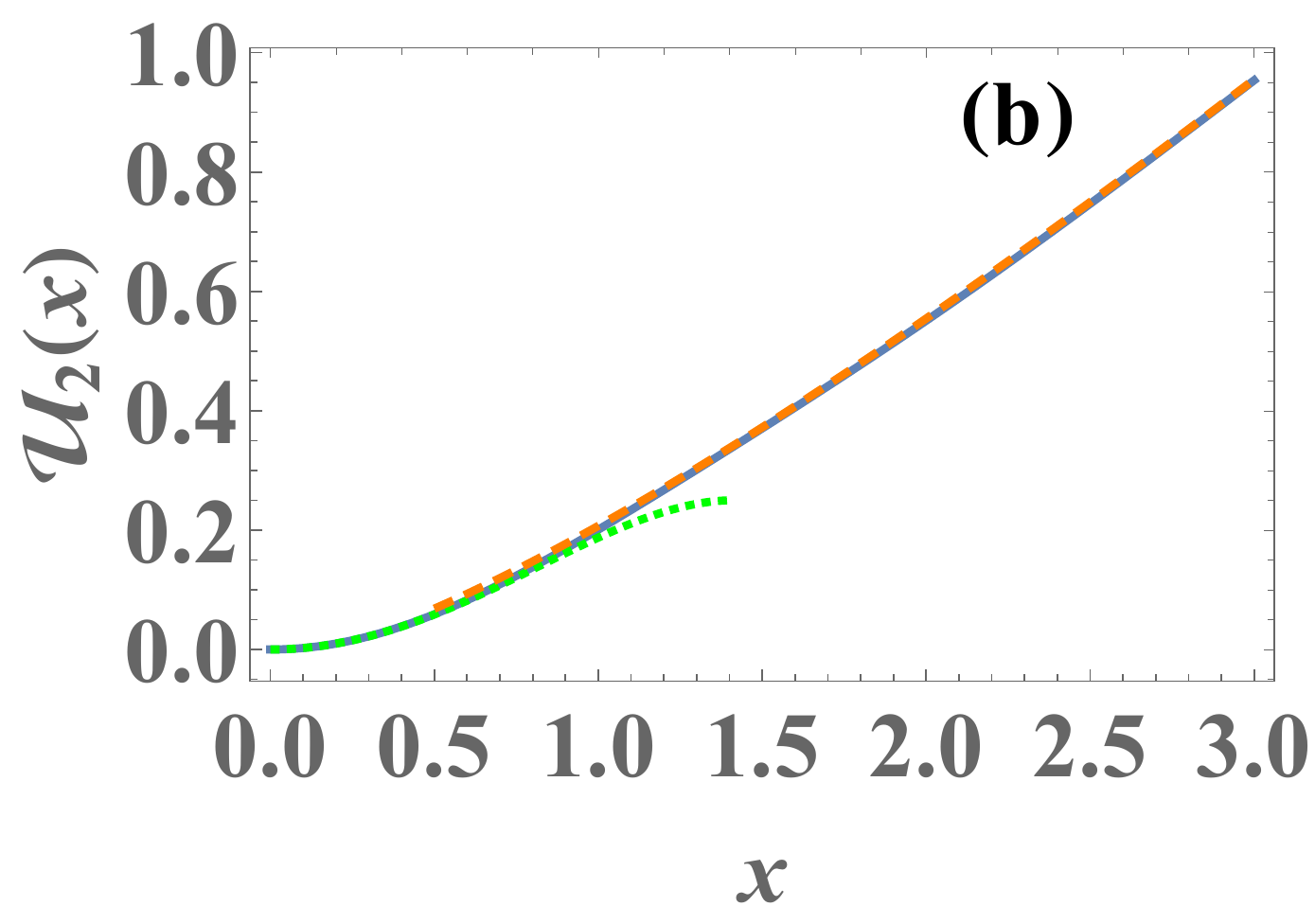}
\includegraphics[angle=0,width=0.32\linewidth]{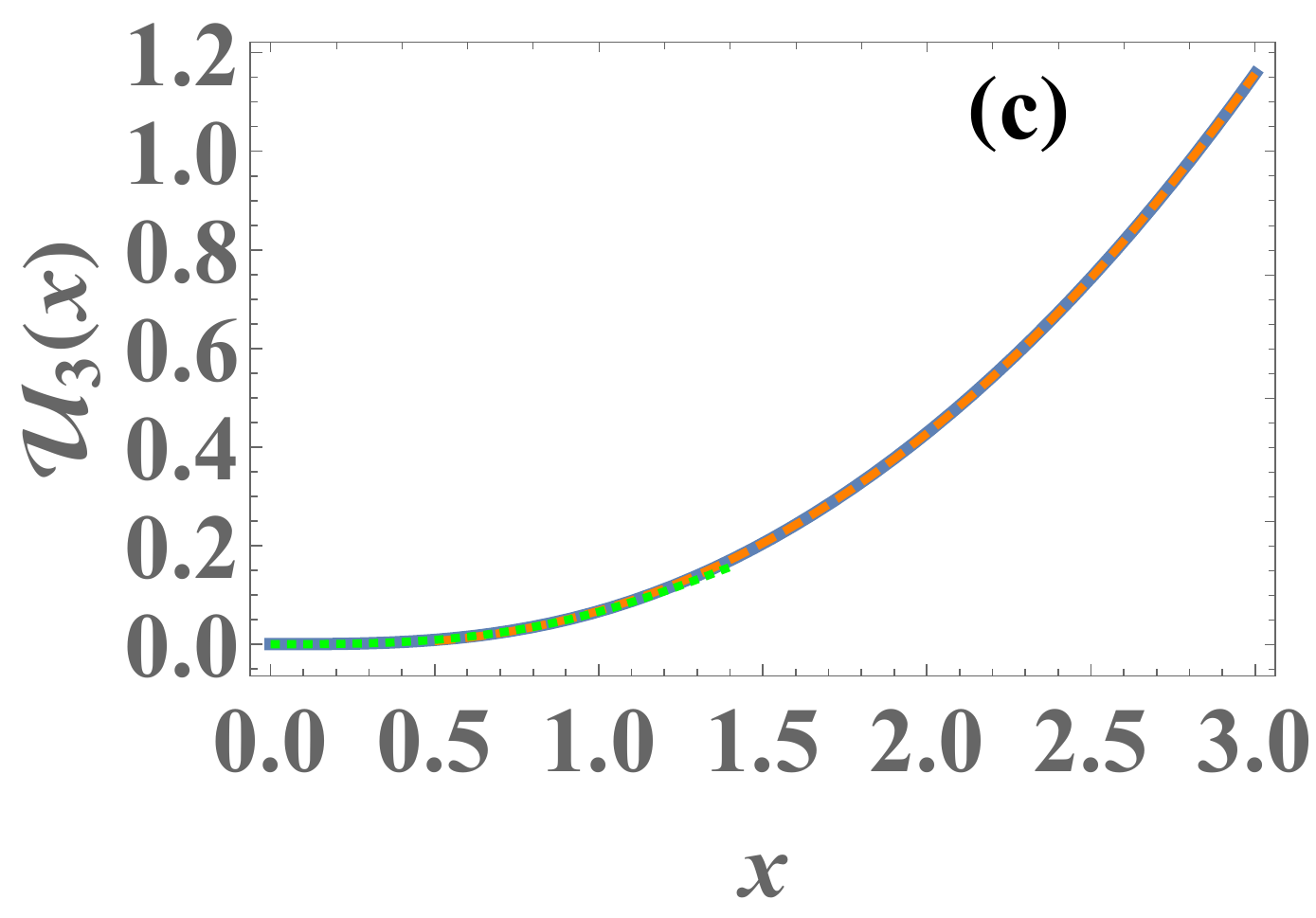}
\caption{ 
(a) A zoom in on the edge regime $\tilde{R}\simeq1$ of Fig.~1 of the main text, showing excellent agreement between the numerical data (symbol) and the theoretical edge prediction, 
\eqrefMT{\final} in the text  (dotted line).
The data was computed over $2\times {10}^6$ simulated realizations of the fermions' radial coordinates $r_1,\dots,r_N$. All data points were generated over the same set of simulations. 
(b) and (c) The scaling functions ${\cal U}_2(x)$ and ${\cal U}_3(x)$ respectively, that describe the variance of the number of free fermions inside a disc in $d=2$ and a sphere in $d=3$, see Eqs.~\eqref{V2} and \eqref{eq:U3} respectively (blue solid lines respectively). Also plotted are their asymptotic behaviors ${\cal U}_{2}\left(x\ll1\right)\simeq x^{2}/4-x^{4}/16$ and ${\cal U}_{3}\left(x\ll1\right)\simeq2x^{3}/\left(9\pi\right)-4x^{6}/\left(81\pi^{2}\right)$ (green dotted lines) which describe a Bernoulli distribution of $\mathcal{N}_{\mathcal{D}}$, and ${\cal U}_{2}\left(x\gg1\right)\simeq x\left(\ln x+\gamma_E-2+5\ln2\right)/\pi^{2}$ and ${\cal U}_{3}\left(x\gg1\right)$ given in Eq.~\eqref{deq3} (orange dashed lines) which correspond to the macroscopic limit.
}
\label{Fig:U2_of_x_and_2d_HO}
\end{figure}

\bigskip

Finally we checked numerically our formula for $K_{\mu}\left(x,y\right)^{2}$ \eqrefMT{\aTWO} 
in the main text, see discussion in Section \ref{sec:K2}, for the HO in $d=1$
(in which case $\theta_x = \arccos(-x/\sqrt{2 N})$). The prediction reads $K_{\mu}\left(x,y\right)^{2}\simeq A_{\mu}\left(x,y\right)^{2} \! /2$, up to rapidly oscillating terms which average to zero on scales larger than microscopic, with, for the HO
\be
\label{eq:Amu_def}
A_{\mu}(x,y)=\frac{\sqrt{1-\frac{xy}{2N}}}{\pi\left(x-y\right)\left(1-\frac{x^{2}}{2N}\right)^{1/4}\left(1-\frac{y^{2}}{2N}\right)^{1/4}} \, .
\ee
{Eq.~\eqref{eq:Amu_def} has been obtained before in the context of GUE random matrices \cite{FrenchMellowPandey1978, BrezinZee1993, Beenakker1993}.}
In Fig.~\ref{Fig:Kernel} the kernel $K_\mu$, rescaled by the amplitude $A_\mu$, is plotted 
as a function of $y$, for $N=200$ and $x=-10$. The result oscillates rapidly as a function of $y$, with an amplitude very close to unity. The quality of this approximation improves as $N$ is increased (not shown).
At $\left|x-y\right| \ll \sqrt{N}$ the amplitude \eqref{eq:Amu_def} matches smoothly with the amplitude $\left[\pi\left(x-y\right)\right]^{-1}$ of the oscillations predicted by the sine kernel 
\be
\label{sinekernelSM}
K_{\mu}(x,y)\simeq\!\frac{\sin\left(k_{F}(x)|x-y|\right)}{\pi|x-y|}.
\ee

\begin{figure}[ht]
\centering
\includegraphics[angle=0,width=0.97\linewidth]{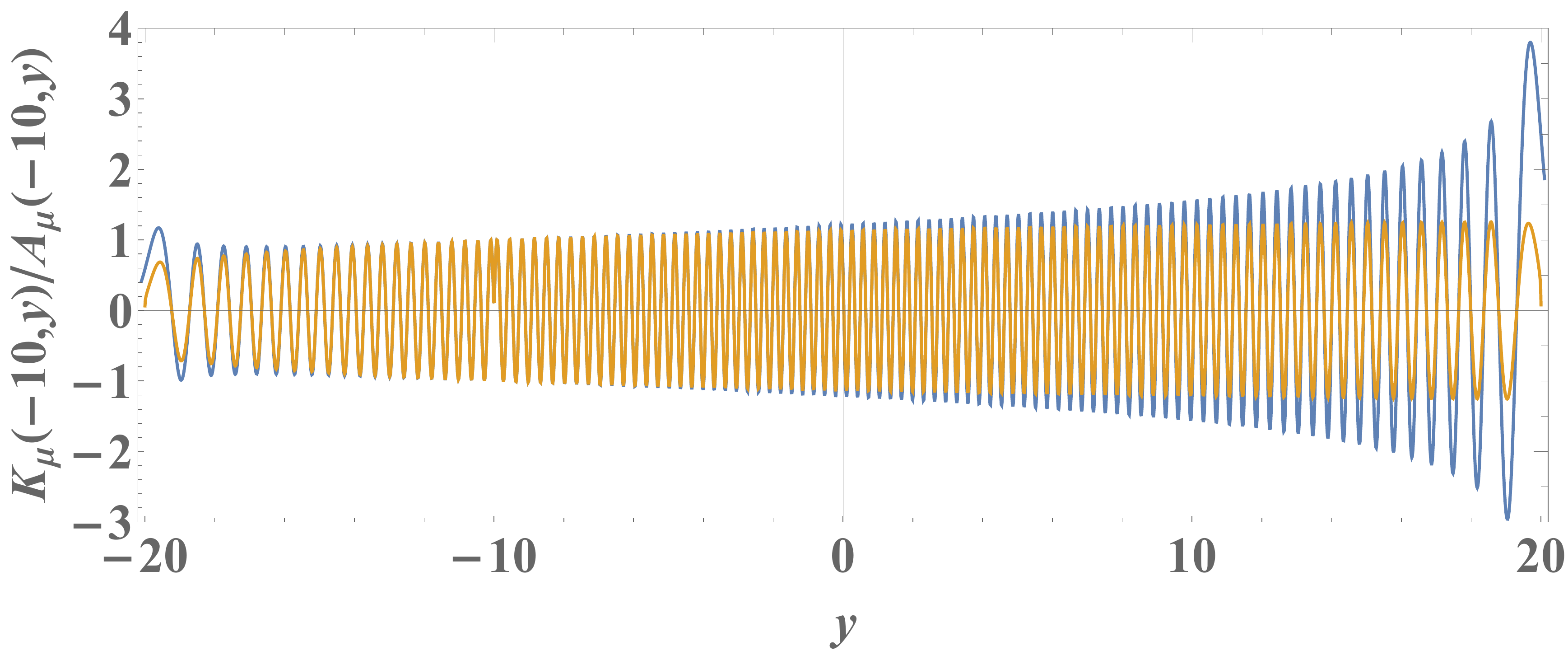}
\caption{Orange line: The kernel $K_{\mu}\left(x,y\right)$ of the harmonic oscillator in $d=1$, see Eq.~\eqref{eq:KHarmonic_exact}, divided by the amplitude
in \eqref{eq:Amu_def}, as a function of $y$, for $N=200$ and $x=-10$. The result oscillates rapidly with an amplitude very close to unity, for all $y$ in the bulk $\left|y\right|<\sqrt{2\mu}$. At $y\simeq x$ it agrees with $K_{\mu}\left(x,y\right) / A_{\text{sine}}\left(x,y\right)$ where $A_{\text{sine}}\left(x,y\right)=\left[\pi\left(x-y\right)\right]^{-1}$ is the amplitude of the oscillations predicted by the sine kernel,  Eq.~\eqref{sinekernelSM} (blue line).}
\label{Fig:Kernel}
\end{figure}


\end{widetext} 

\begin{thebibliography}{}
\bibitem{footnote1} 
For any union of disjoint intervals one has
${\rm Var} {\cal N}_{\cup_i [a_i,b_i]} = \sum_{ij} H(a_i,a_j)+H(b_i,b_j)-2 H(a_i,b_j)$
\bibitem{footnote:large_k}
{What is meant is that the only states which are not
accurately described by our WKB calculation are the few very deep low energy states, 
which however are not relevant in the large $N$ limit.}
\bibitem{footnote10} 
The lower bound of the integral in the definition of $\phi_k(x)$ depends
in general on $k$, but is unimportant in the argument \cite{SM}.
\bibitem{footnote3} 
In that regime the sum over $m$ can 
be approximated by a continuous integral.
\bibitem{footnote4}
In the edge regime it takes the form obtained in
\cite{AKPLDSP,CharlierAiry}.
\bibitem{footnote5}
The $O(1)$ constant can further be extracted from \cite{BK18}
\bibitem{Charlier2}
It seems that the constant term can also be proved rigorously, see \cite{Charlier3}.
\bibitem{footnote6}
With half that result for a semi-infinite interval.
\bibitem{footnote7}
Note that \eqref{cumuleven_GUE_gen} also holds at the edge
from the side of the bulk, 
for $- \hat a \gg 1$, \cite{BK18,CharlierAiry}.
\bibitem{footnote9}
Note that for $d=1$ the dependence on $R$ 
in \eqref{entropy} drops out.
\bibitem{footnote16} 
There may be exceptional points
when a symmetry is present, e.g. for the 
HO and $y=-x$ one has 
$\tilde \phi_N(x) = {(-1)^N} \tilde \phi_N(y)$ 
and one of these term does not oscillate. However the
weight of these configurations is subdominant when computing
the variance.
\bibitem{footnote11} 
For $V(x) \neq 0$ the exact energy levels are not always doubly degenerate, e.g. a delta
impurity at $x=0$ breaks the degeneracy between odd (sinus) and even (cosinus)
wavefunctions. Note that the ground state is never degenerate. However this effect
goes beyond the semi-classical approximation.
\bibitem{footnote12}
Note that $\theta_x=\pi - \tilde \theta_x$ where $\tilde \theta_x=
\pi  \frac{\int_{x}^{x^+} dz/k_F(z) }{\int_{x^-}^{x^+} dz/k_F(z) }$
and our main formula are equivalently valid with $\theta_x \to \tilde \theta_x$.
\end{thebibliography}
\end{document}